\documentclass[12pt,a4paper,final]{iopart}
\pdfoutput=1
\usepackage{graphicx}
\usepackage{cite}
\usepackage{bigfoot}
\usepackage{appendix}
\usepackage[breaklinks=true,colorlinks=true,linkcolor=black,urlcolor=black,citecolor=black,pdfencoding=auto]{hyperref}
\usepackage{amsfonts, amsmath, dsfont}
\usepackage{empheq}

\usepackage{bookmark}

\def\contentsname{\empty}

\newcommand{\blue}{ }

\newcommand{\bea}{\begin{eqnarray}}
\newcommand{\eea}{\end{eqnarray}}

\newcommand{\be}{\begin{equation}}
\newcommand{\ee}{\end{equation}}

\def\Ai{\text{Ai}}

\newcommand{\nn}{{\nonumber}}
\def\be{\begin{equation}}
\def\ee{\end{equation}}

\def\l{\lambda}

\def\l{\lambda}

\makeatletter
\def\@fnsymbol#1{\ifcase#1\or \dagger\or \ddagger\or \S\or
   \|\or \P\or ^{+}\or ^{\tsty *}\or \sharp
   \or \dagger\dagger \or \mbox{$\clubsuit$} \else\@ctrerr\fi\relax}
\makeatother

\begin{document}

\title[Tail of the two-time height distribution
for 1D KPZ growth]{Tail of the two-time height distribution for KPZ growth in one dimension}

\author{Jacopo de Nardis$^1$ and Pierre Le Doussal$^1$}
\address{$^1$ CNRS-Laboratoire de Physique Th\'eorique de l'Ecole Normale Sup\'erieure, 24 rue Lhomond, 75231 Paris Cedex, France}
\ead{jacopo.de.nardis@phys.ens.fr}
\ead{ledou@lpt.ens.fr}

\begin{abstract}
Obtaining the exact multi-time correlations for one-dimensional growth models described by the Kardar-Parisi-Zhang (KPZ)
universality class is presently an outstanding open problem. Here, we
study the joint probability distribution function (JPDF) of the height
of the KPZ equation with droplet initial conditions, at two different times $t_1<t_2$, in the limit where both times are
large and their ratio $t_2/t_1$ is fixed.
This maps to the JPDF of the free energies of two directed polymers with two different lengths and in the same random potential.
Using the replica Bethe ansatz (RBA) method, we obtain the exact
tail of the JPDF when one of its argument (the KPZ height at the earlier time $t_1$)
is large and positive.
Our formula interpolates between two limits where the JPDF decouples: (i) for $t_2/t_1 \to +\infty$ into a
product of two GUE Tracy-Widom (TW) distributions, and (ii) for $t_2/t_1 \to 1^+$ into a product of a
GUE-TW distribution and a Baik-Rains distribution (associated to stationary KPZ evolution). The lowest cumulants
of the height at time $t_2$, conditioned on the one at time $t_1$, are expressed analytically as expansions
around these limits, and computed numerically for arbitrary $t_2/t_1$. Moreover we compute the connected two-time correlation, conditioned to a large enough value at $t_1$, providing a quantitative prediction for the so-called persistence of correlations (or ergodicity breaking) in the time evolution from the droplet initial condition. Our RBA results are then compared
with arguments based on Airy processes, with satisfactory agreement. These predictions are
universal for all models in the KPZ class and should be testable in experiments and numerical simulations.
\end{abstract}



\tableofcontents

\section{Introduction}

Numerous exact results have been obtained in recent years for discrete growth models in the so-called
Kardar-Parisi-Zhang (KPZ) universality class in one dimension. This includes discrete models for the stochastic growth of an interface, such as the PNG model \cite{png,spohn2000}, but also, more broadly, particle transport models, such as asymmetric exclusion processes (TASEP, ASEP and $q$-TASEP)
\cite{spohnTASEP,ProlhacTASEP,BorodinMacdo,corwinsmallreview},
directed polymers (DP) in random potentials (last passage percolation \cite{Johansson2000}),
combinatorial problems, and others, where the
analogous of a height field evolving under a local stochastic rule can be defined. The KPZ behavior was also observed in non-linear fluctuating hydrodynamics \cite{Spohnfluct,Spohnfluct2} and equilibrium and non-equilibrium properties of Bose-Einstein condensates \cite{Lamacraft,Altman1,Altman2}.
At the center of this universality class lies the continuum KPZ equation \cite{KPZ}, for which exact results were
obtained more recently, for the three main types of initial conditions (droplet, flat, stationary)
\cite{we,dotsenko,spohnKPZEdge,corwinDP,sineG,we-flat,we-flatlong,cl-14,dotsenkoGOE,Quastelflat,SasamotoStationary,BCFV}.
The remarkable aspect of these results was to show how the probability distribution function (PDF) of the fluctuations of the growing height function at one space point, $h(X,t)$, is described at large time by the Tracy-Widom distributions, describing also the statistical behavior of
 the largest eigenvalue of the Gaussian ensembles of random matrices\cite{TW1994}. In particular for droplet initial conditions the PDF of the height converges to the Tracy-Widom distribution associated to the Gaussian Unitary Ensemble (GUE), while for flat initial conditions it converges to the Tracy-Widom distribution associated to the Gaussian orthogonal ensemble (GOE) of random matrices. Remarkably, the PDF of the height with Brownian initial conditions (stationary class) converges to the so called Baik-Rains distribution \cite{png}, which has no equivalent in the random matrices theory.
Furthermore it was found that the multi-space point joint PDF (JPDF) identifies at large time
\cite{ps-2point,ps-npoint,dotsenko2pt,Spohn2ptnew,corwinRG}, with the one of a well characterized
determinantal process, the Airy processes (\Ai$_2$/\Ai$_1$) \cite{spohn2000,ferrariAiry-a}. These predictions were confirmed in remarkable experiments on liquid crystals \cite{exp4,Takeuchi}. \\

All these results and progresses however deal with {\it one time quantities}, i.e. observing
the KPZ interface at some fixed time $t$ after the initial condition at $t=0$. However the
{\it two-time observables} are also of great interest, for instance the JPDF of the couple of random variables $(h(0,t_1), h(X,t_2))$ at two very large times $t_1,t_2$, the ratio $t_2/t_1$ being fixed, a limit believed to be universal across the KPZ class.
Remarkably little is known theoretically about such multi-time correlations, which encode the full time evolution of the KPZ equation. On the other hand the numerical results \cite{henkeltwT} and experiments are more advanced on the topic and there is a strong need for exact results to be compared with the recent remarkable experimental results
\cite{TakeuchiCrossover,TakeuchiPersistence,TakeuchiHHLReview}.

{ In order to make substantial theoretical advances on the KPZ equation and, more broadly, on the whole KPZ class,
the study of directed polymer models \cite{directedpoly} (also called last passage percolation
\cite{Johansson2000}) has often proved useful in the past.}
Indeed, in the continuum the two models are identical, i.e. the continuum KPZ equation is equivalent, under the Cole-Hopf mapping \cite{ColeHopfDeterminist,ColeHopfStochasticMath,HairerSolvingKPZ,reviewCorwin},
to a problem of a continuum
directed polymer (DP) in a random potential, with $h(x,t)=\ln Z(x,t)$ being minus the free-energy
and the time $t$ being the polymer length.

In the present context, Johansson recently
obtained a rigorous formula for the two-time JPDF of the free energies of
the zero-temperature Brownian {\it semi-discrete} DP model
\cite{Johansson2times}, known to be in the KPZ class.
However this formula is quite involved and until now it has not been possible to analyze it analytically or
even numerically, in order to produce predictions that can be compared with experiments.
Using Airy processes, some results were also recently obtained by Ferrari and Spohn, with an explicit formula valid only for the two-time height correlation (i.e. second moment), and only in the stationary case
\cite{FerrariSpohn2times}. Another notable
exception is the pioneering work of Dotsenko
\cite{dotsenko2times1,dotsenko2times2,dotsenko2times3}, who proposed a formula for the two-time JPDF of the free energies (i.e. of the KPZ heights). However we believe that this formula is not correct and we briefly explain our reasons in the text.

For the continuum directed polymer/KPZ equation, the replica Bethe ansatz method allows to write aesthetically appealing, formal equations for the joint integer moments of the partition sum of the directed polymer $Z(X,t) = e^{h(X,t)}$ at fixed temperature and for two different and  {\it finite} lengths. It involves summations over the known eigenstates of the attractive one-dimensional delta-Bose gas, the so-called Lieb-Liniger model \cite{ll}. Hence one could attempt a summation over these states, to obtain an explicit formula for these moments, and from them, extract the two-time JPDF. This program was indeed successfully carried through for the one-time problems \cite{we,dotsenko,we-flat}. Unfortunately, when more than one time is involved, the summand contains the so-called {\it form factors} of the bosonic creation and annihilation operators. Those form factors are notoriously difficult to calculate in full generality
\cite{Slavnov,piroli-cal}. In
Refs. \cite{dotsenko2times1,dotsenko2times2,dotsenko2times3} the above described summation was indeed attempted with a guess for the form factors which, in our opinion, is not correct, although it leads
to a simplified summation. As will be confirmed here, many terms of the exact sum are thus missing, leading then to an incomplete result for the JPDF. Although one could hope that that these terms are negligible in the large time limit, we show here explicitly that this is
not the case. Hence the problem is still mostly unsolved.\\

The aim of the present paper is to study the joint probability distribution function (JPDF) of the two heights
$(h(0,t_1),h(X,t_2))$ of the
Kardar-Parisi-Zhang equation at two different times $t_1<t_2$ and with droplet initial conditions. We do not obtain a full solution of the problem, namely the full JPDF, which still remains an open problem. However, we are able to obtain an exact expression for
the JPDF in the limit where $h(0,t_1)$ (i.e. its fluctuations) is positive and large, i.e. the exact tail of the JPDF when one of the two arguments
is large. The dependence in the second argument, describing the fluctuations of $h(X,t_2)$, remains non-trivial.
Although we do also obtain an expression valid for finite time, we focus here on the limit where both times are large while their ratio $t_2/t_1$ is fixed.
As this ratio $t_2/t_1$ is varied, our formula for the JPDF exhibits a crossover between two limits
where it decouples into a product of two distributions, each of them describing the fluctuations at one time. In the limit where the two times become equal $t_2/t_1 \to 1^+$ we recover the product of a GUE Tracy-Widom PDF and a Baik-Rains distribution. By independent arguments, based on Airy processes, we can show that this decoupling should
hold exactly in that limit (beyond the tail), which provides an important check of our method.
The opposite limit, where the ratio $t_2/t_1$ is very large, leads to a factorized JPDF, into a
product of two GUE Tracy-Widom distributions. The subleading corrections decay in time as $(t_1/t_2)^{1/3}$,
which leads to a non-trivial two-time correlations also in the limit where $t_2 \gg t_1$,
a phenomenon known as persistence of correlations \cite{Takeuchi,TakeuchiPersistence,FerrariSpohn2times,GueudreUnpub}
and that we show here show from our exact expression for the tail of the JPDF. We obtain exact expressions for the
cumulants of $h(X,t_2)$ conditioned to a fixed large value of $h(0,t_1)$, and their expansion
in the limit of large $t_1/t_2$. Finally, our predictions are universal for all models in the KPZ class
and should be testable in experiments and numerical simulations.

As is detailed below, our method is based on a partial summation over the Bethe ansatz states:
we include only a single string state (bound state of all the particles describing the different replicas) propagating in the first time interval
$[0,t_1]$, which can be written explicitly, and, as we argue, gives the exact tail of the JPDF when
$h(0,t_1)$ is large. The resulting non-trivial summation over the other many-string states related
to the second variable can then be performed exactly at arbitrary finite times and with
no approximation.
\\

The outline of the paper is as follows. In Section \ref{sec:review} we review the
known results for the one-time PDF of the height of the KPZ equation, introduce units and
notations, and define the tail approximation. In Section \ref{sec_twotimeproblem}
we define the JPDF that we aim to calculate, and
we present the main results of the paper, some in explicit form, and for others
we refer to formula later in the text. In Section \ref{sec:RBA} we present
the method of the replica Bethe ansatz (RBA) for the two-time problem.
We recall the quantum mechanical approach based on the attractive delta-Bose gas, i.e.
the Lieb Liniger (LL) model. We use it to express the two-time joint moments of the DP partition sums
as sums over two sets of eigenstates of the LL model, the so-called strings. We define and present a formula for
the generating function. We define the tail approximation used in this paper, were
the summation is restricted to 1-string states for the first set, and unrestricted for the second.
In Section \ref{ff} we { derive} the expression of the 1-string form factor, namely the form factor of a generic power of the annihilation operator between a state containing only one single string (bound state) and a generic Bethe state. The explicit formula allows to perform
the summation explicitly in Section \ref{sec:calc}. After a number of manipulations, we
obtain our main result for the tail generating function, and the JPDF. We
analyze this JPDF in three limits, and perform expansions around these
limits (the $t_2/t_1 \to 1$ limit, the large $t_2/t_1$ limit and the large $h_1$ limit).
In Appendix \ref{app:airy} we analyze the problem using Airy processes, and
compare with the results of the RBA method. Finally Section \ref{sec:conclusion}
contains the Conclusion.

Further technical details are given in the Appendices.
In  Appendix \ref{app:dots} we detail the comparison between our work and
the papers \cite{dotsenko2times1,dotsenko2times2,dotsenko2times3}.
In Appendix \ref{app:ff} we present the calculation of the form factor used in the
text. In  Appendix \ref{sec:Z11} we present an independent calculation of the
``double tail" in both height variables. In Appendix \ref{app:w} we show a
more general result for the $t_2/t_1\to 1^+$ limit. In Appendix
\ref{app:BR} we give more details about extended Baik-Rains distributions
and derive some new expansion formula.
In Appendix \ref{app:sums} we explain a contour integral method for constrained
summations needed in the text. In Appendix \ref{sec:useful} we give useful identities,
such as expressions of derivatives of the GUE-TW distribution. Finally,
in Appendix \ref{app:higher} we calculate higher orders in the large $t_2/t_1$ expansion.


\section{KPZ equation, directed polymer and review of the one-time results}
\label{sec:review}

\subsection{Units and the Cole-Hopf mapping}

In this paper we study the one-dimensional Kardar-Parisi-Zhang (KPZ) equation \cite{KPZ}. It describes,
in the continuum, the stochastic growth of an interface, of height $h(x,t)$ at point $x \in \mathbb{R}$, as a function of time $t$
\be \label{kpzeq}
\partial_t h(x,t) = \nu \partial_x^2 h(x,t) + \frac{\lambda_0}{2}  (\partial_x h(x,t))^2 + \sqrt{D} ~ \xi(x,t)
\ee
driven by a unit white noise $\overline{\xi(x,t) \xi(x',t')}=\delta(x-x') \delta(t-t')$. We will use the following scales as
units
\begin{align}
x^* = \frac{(2 \nu)^3}{D \lambda_0^2} \quad , \quad t^* = \frac{ 2 (2 \nu)^5}{D^2 \lambda_0^4}
\quad , \quad   h^* = \frac{2 \nu}{\lambda_0}
\end{align}
i.e. we set $x \to x^* $, $t \to t^* $ and $h \to h^* $
so that from now on $x,t,h$ are in dimensionless units and the KPZ equation becomes
Eq. (\ref{kpzeq}) with $\nu=1$, $\lambda_0=2$ and $D=2$.

The Cole-Hopf mapping solves the KPZ equation from an arbitrary
initial condition $h(x,0)$ as follows: the solution at time $t$ can be written as:
\be
e^{h(x,t)} = Z(x,t) \equiv  \int dy Z_\eta(x,t|y,0) e^{h(y,t=0)}.
\ee
Here $Z_\eta(x,t|y,0)$ is the partition function of the continuum directed polymer in the random potential
$- \sqrt{2} ~ \eta(x,t)$ with fixed endpoints at $(x,t)$ and $(y,0)$:
\be \label{zdef}
Z_\eta(x,t|y,0) = \int_{x(0)=y}^{x(t)=x}  Dx e^{-  \int_0^t d\tau [ \frac{1}{4}  (\frac{d x}{d\tau})^2  - \sqrt{2} ~ \eta(x(\tau),\tau) ]}
\ee
which solves the (multiplicative) stochastic heat equation (SHE):
\begin{align} \label{dp1}
\partial_t Z = \nabla^2 Z + \sqrt{2} ~ \eta Z
\end{align}
with Ito convention and initial condition $Z_\eta(x=0,t|y,0)= \delta(x-y)$. Equivalently, $Z(x,t)$ is the solution of
(\ref{dp1}) with initial conditions $Z(x,t=0)=e^{h(x,t=0)}$. We will sometimes omit the ``environment" index $\eta$.
Here and below overbars denote averages over the white noise $\eta$.

\subsection{Large time results}

{ Let us briefly review the known results on the one-time fluctuations of the KPZ field. We focus
here on the droplet initial condition. To define it for arbitrary time one usually considers the
wedge initial condition defined as
\begin{equation}
h_{w_0}(x,0) =  - w_0  |x| + \ln(\frac{w_0}{2})
\end{equation}
and the ``hard wedge" limit $w_0 \to +\infty$, so that $\exp(h_{w_0}(x,0)) \to \delta(x)$. In that limit
$h_{w_0}(x,t) \to h(x,t)$ for any $t>0$, and one has
\bea \label{drop0}
h(x,t)= \ln Z_\eta(x,t|0,0)
\eea
where $Z_\eta(x,t|0,0)$ is the partition function of the DP defined in the previous Section
\footnote{Note that if one is interested only in the large time limit, any finite $w_0>0$ leads to
the same result, up to a constant shift.}.
Hence the droplet height profile $h(x,t)$ corresponds to the free energy of a DP in a random medium going from the point $(0,0)$ to $(x,t)$ (see Fig. \ref{fig:twotimefig}).}


At large time the KPZ field grows linearly in time with $O(t^{1/3})$ fluctuations.
At the level of the fluctuations in one space point (choosing here $x=0$) these are
governed by the GUE Tracy Widom of PDF, $f_2(\sigma) = F_2'(\sigma)$, i.e. one has at large $t$ \cite{spohn2000,we}
\bea
h(x=0,t) = - \frac{t}{12} + t^{1/3} \chi_2 + o(t^{1/3}) \quad , \quad {\rm Prob}(\chi_2 < \sigma) = F_2(\sigma)
\eea
where $F_2(\sigma)$ is given by a Fredholm determinant
\bea \label{airyK1}
&& F_2(\sigma) = {\rm Det}[I - P_\sigma K_{\rm Ai} P_\sigma]
\eea
involving the Airy Kernel $K_{\Ai}$:
\bea \label{airyK2}
\fl   K_{\rm \Ai}(v,v')= \int_{0}^{+\infty} dy \Ai(y+v) \Ai(y+v')
= \frac{\Ai(v) \Ai'(v')-\Ai'(v) \Ai(v')}{v-v'}
\eea
and where $P_\sigma(v)=\theta(v-\sigma)$ is the projector on $[\sigma,+\infty[$.
To get rid of the part linear in
time we will, from now on, redefine the KPZ field, and the DP partition sum, at all times, as
\bea \label{redef}
h(x,t) = - \frac{t}{12} + \tilde h(x,t) \quad , \quad Z(x,t) = e^{- t/12} \tilde Z(x,t)
\eea
and for notational simplicity, we will omit the tilde in what follow.

\subsection{Tail approximation}
\begin{figure}
\centering
\includegraphics[scale=1.3]{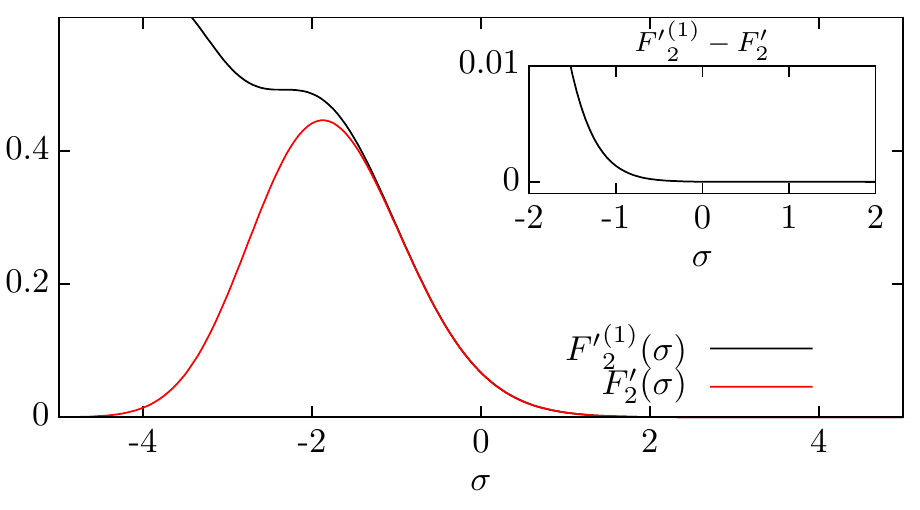}
\caption{Plot of the Tracy-Widom distribution $F'_2(\sigma)$ (red line) compared with its tail for positive $\sigma$ given by $F'{}_2^{(1)}(\sigma)$ (black line). Inset: difference between the Tracy-Widom distribution $F'_2(\sigma)$ and its tail $F'{}_2^{(1)}(\sigma)$, given in \eqref{TWtail}.   }
\label{fig:gue_approx}
\end{figure}
Since we obtain below some exact results for the tail of the two-time JPDF, it is useful to introduce here the tail of the single time GUE-TW distribution. We define
the tail approximations of the CDF of the GUE-TW distribution as the function 
\begin{equation} \label{TWtail}
F^{(1)}_2(\sigma)  \equiv  1 - \text{\text{Tr}} [P_\sigma K_{\Ai}] = 1 - \int_\sigma^{+\infty} dv K_{\rm Ai}(v,v)
\end{equation}
It corresponds to keeping only the first term in the series expansion for the determinant of a matrix (here an operator) around the identity: ${\rm Det}[I + A] = 1 + \text{Tr}[A] + O(A^2)$.
Hence, compared to the exact
result, it neglects terms of order $O(K_{\rm Ai}^2)$
containing products of four Airy functions and more. This tail function captures the leading (stretched) exponential behavior for large and positive $\sigma$,  $F^{(1)}_2(\sigma) -1= O(e^{- \frac{4}{3} \sigma^{3/2}})$ for $\sigma \to +\infty$ and
the corrections are of higher (stretched) exponential order
\bea\label{eq:tail_approx}
F_2(\sigma) = F^{(1)}_2(\sigma)  + O(e^{- \frac{8}{3} \sigma^{3/2}})
\eea
As can be seen on Fig. \ref{fig:gue_approx}, this approximation is reasonably good (with error less than $10^{-3}$) for any $\sigma > -1$. More generally below we will always use the superscript $(1)$
to indicate the tail approximation {\eqref{eq:tail_approx}}. It is important to note that
keeping the full trace in (\ref{TWtail}) is a considerably better approximation that
its leading large $\sigma$ limit. As discussed below, this approximation also
consists in keeping only single string states in the RBA method, leading
to huge simplifications in the application of the method.

%

\section{Two-time problem: summary of the results} \label{sec_twotimeproblem}

\subsection{Notations}
\begin{figure}[ht]
\centering
\includegraphics[scale=0.6]{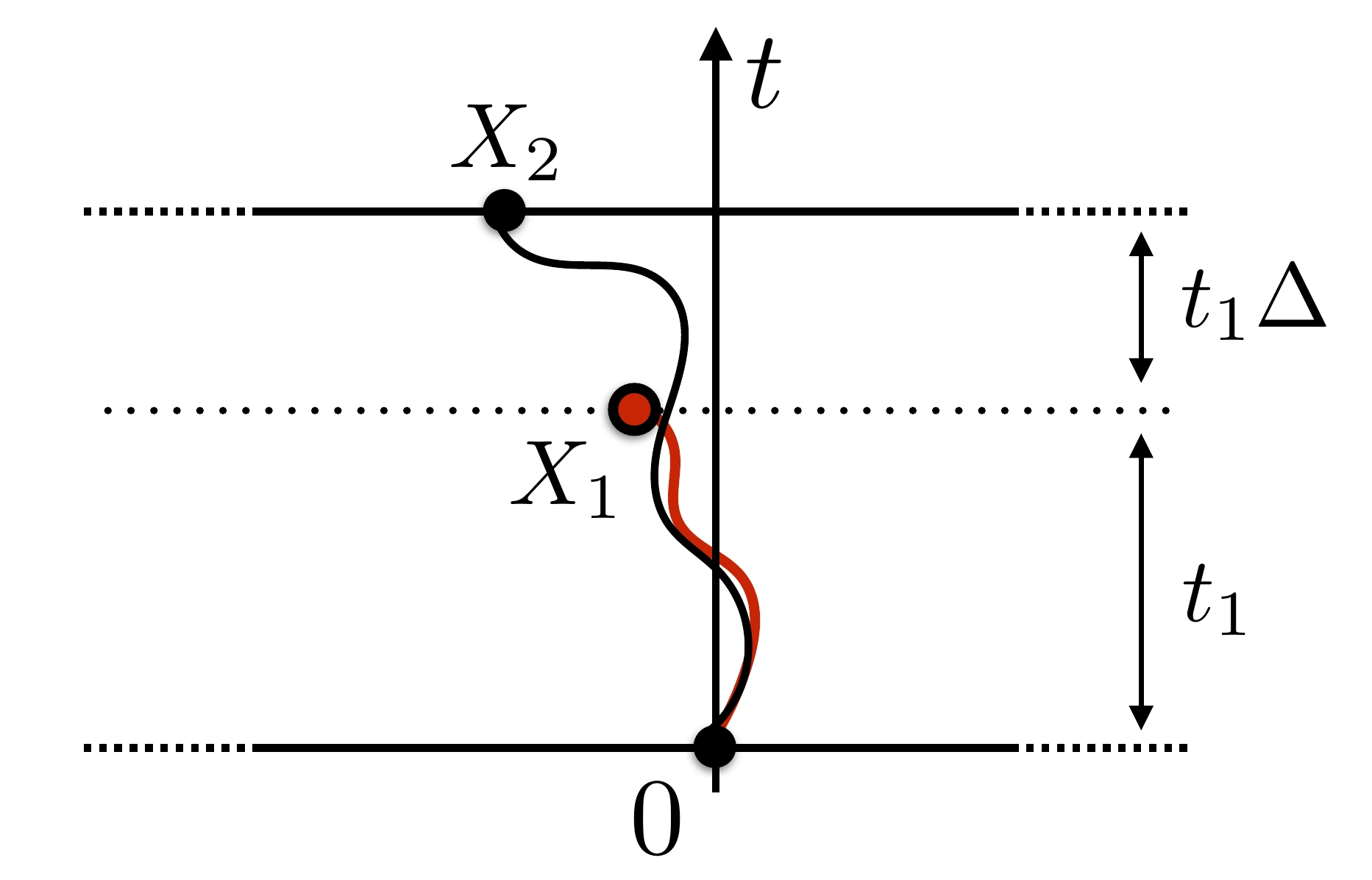}
\caption{Schematic representation of the two-time setting: we consider the joint probability distribution JPDF of the free energy of a DP in a random medium starting at $(0,0)$ and ending at $(X_1,t_1)$ and another polymer  in the same random medium starting at the same position and ending at  $(X_2,t_2)$. { In the following,
we focus on the case $X_1=0$ and $X_2=X$.}}
\label{fig:twotimefig}
\end{figure}
Here we study the KPZ height field $h(x,t)$ with droplet initial condition
at two different times $t=t_1$ and $t=t_2 > t_1$,
and two different space points $x=0$ and $x=X$, and we denote
\bea  \label{defH12}
&& H_1 \equiv  h(0,t_1) = \ln Z_1 \quad , \quad Z_1\equiv Z_\eta(0,t_1| 0,0)  \\
&& H_2 \equiv  h(X,t_2) =  \ln Z_2 \quad , \quad Z_2\equiv  Z_\eta(X,t_2| 0,0)
\eea
In particular we will be also interested in the difference of the two height functions
\bea
H\equiv  H_{21} \equiv  H_2-H_1
\eea
built over the time difference $t_{21}\equiv  t_2-t_1$. In the large time limit, namely when both $t_1$ and $t_2$ are sent to $+\infty$, the relevant parameter characterizing the JPDF of the two heights
will be the time difference rescaled by the earlier time, denoted as
\begin{equation}
\Delta = \frac{t_2 - t_1}{t_1} >0
\end{equation}

Let us start with $X=0$, the general case being discussed below.
From the previous Section, we know that the two heights grow in time as $H_1 \sim t_1^{1/3}$ and $H_2 \sim t_2^{1/3}$. It is
thus convenient to introduce the rescaled heights $h_1$, $h_2$ and $h$, through\\
\bea \label{defh1}
\fl && ~~~~~~  H_1 = t_1^{1/3} h_1 
\quad , \quad H_2 = t_2^{1/3} h_2 \quad , \quad H_2  - H_1    = (t_2-t_1)^{1/3}  h  = \Delta^{1/3}  t_1^{1/3}  h  \label{defh}
\eea
Here $ h_1,  h_2,  h$ are random variables whose joint distributions we want to determine.

From the previous Section we know that at large time each individual height is distributed according to the GUE-TW
distribution
\bea
&& \! \! \! \! \! \!\! \! \!   \lim_{t_1 \to +\infty} \text{Prob}(h_1 < \sigma_1) = F_2(\sigma_1) \quad , \quad  \lim_{t_2 \to +\infty} \text{Prob}(h_2 < \sigma_2) = F_2(\sigma_2)
\eea

We define the joint probability distribution function (JPDF) at two arbitrary finite times $0< t_1<t_2$ of
the rescaled KPZ field $h_1$ and height difference $h$ as
\begin{equation} \label{defP}
P_{\Delta, \lambda}( \sigma_1, \sigma) \equiv  \overline{ \delta(h_1 - \sigma_1) \delta(h - \sigma) }
\end{equation}
where $\sigma_1,\sigma$ are two real numbers and $\lambda$ parametrizes the time $t_1$ as
$\lambda = (t_1/4)^{1/3}$.
We will be particularly interested in the limit of all times being large
keeping $\Delta$ finite, such that $t_1 \Delta \sim t_1$, which we denote as {
\begin{equation} \label{defPinfty}
P_{\Delta}(  \sigma_1,    \sigma) =  \lim_{t_1 \to + \infty} P_{\Delta, \lambda}(  \sigma_1,    \sigma) \equiv  \lim_{\lambda \to + \infty} P_{\Delta, \lambda}(  \sigma_1,    \sigma)
\end{equation} }

{
\subsection{The two-time joint PDF}
}

As mentioned in the introduction the full JPDF, $P_{\Delta}(\sigma_1,\sigma)$, is very difficult to compute exactly, and here
we obtain this function only in the region of large positive $\sigma_1$, and for arbitrary $\sigma$. More precisely we find that
\begin{equation} \label{defP1}
P_{\Delta}(  \sigma_1,    \sigma) = P^{(1)}_{\Delta}(\sigma_1, \sigma) + O(e^{- \frac{8}{3} \sigma_1^{3/2}})
\end{equation}
where $P^{(1)}_{\Delta}(\sigma_1,\sigma)$ is of order $O(e^{- \frac{4}{3} \sigma_1^{3/2}})$ uniformly in $\sigma$. { We are able to derive a relatively simple expression for the tail approximation $P^{(1)}_{\Delta}(\sigma_1, \sigma)$. We report the result as follows:}{\blue
\begin{empheq}[box=\fbox]{equation} \label{jpdf0}
P^{(1)}_{\Delta}(\sigma_1,\sigma)  = \left( \partial_{\sigma_1} \partial_{\sigma} -  {\Delta^{- 1/3}} \partial^{2}_{\sigma} \right) \left(  {F_2(\sigma)}  \text{\text{Tr}} \left[   \Delta^{1/3}  P_\sigma K^{\Delta}_{\sigma_1}   P_\sigma   (I-P_{\sigma} K_{\Ai}P_{\sigma} )^{-1}   - P_{\sigma_1} K_\Ai  \right] \right)
\end{empheq}}
expressed in terms of the Airy kernel, as well as of a novel kernel
\begin{empheq}[box=\fbox]{equation}
K^{\Delta}_{\sigma_1}(u,v)   =  \int_0^\infty dy_1 dy_2 \Ai\left(-  {y_1}{ }   +  u \right) K_{\Ai}(y_1 \Delta^{1/3} + \sigma_1, y_2\Delta^{1/3} + \sigma_1)  \Ai \left(-  {y_2}{}  +  v \right)
\label{K4def0}
\end{empheq}
where we recall that $F_2(\sigma)$ and $F^{(1)}_2(\sigma_1)$ are respectively the GUE-TW CDF and its tail
approximation, given respectively
by (\ref{airyK1}) and (\ref{TWtail}). { Below, the function $f_\Delta(\sigma_1,\sigma)$
in \eqref{jpdf0} is traded for the function called $\hat g^{(1)}_{\Delta}(s_1,s)$,
see \eqref{deriv3}, and
another equivalent expression is given
in terms of the difference of two Fredholm determinants
in (\ref{finalfinal}).}



Our exact result for the JPDF satisfies two important properties.
In the limit of infinite time difference $t_2/t_1 \to +\infty$ (which corresponds to the limit $\Delta \to + \infty$), it converges to the product of two GUE-TW
distributions
\begin{align} \label{prodGUE}
 \lim_{\Delta \to \infty}  P^{(1)}_{\Delta}(\sigma_1,\sigma)   =  F_2^{(1) \prime}(\sigma_1)  \ F_2'(\sigma)
\end{align}
and in the limit of small (scaled) time separation $(t_2-t_1)/t_1 \ll 1$ (which translates into $\Delta \to 0^+$) it also decouples as follows
\begin{equation} \label{decoupleBR}
\lim_{ \Delta \to 0} P^{(1)}_{\Delta}(\sigma_1,\sigma)  =     F_2^{(1) \prime}(\sigma_1)    \   F_0'(\sigma)
\end{equation}
where $F_0'(\sigma)$ is the Baik-Rains { (BR)} probability distribution \cite{png,SasamotoStationary} which
governs the stationary growth profile in the infinite time limit, and whose explicit expression is recalled in
\eqref{BR1}.

In fact, reexamining the problem in terms of Airy processes in Section \ref{app:airy}, we argue that in these two limits our result is valid for
any value of $\sigma_1$, i.e. we predict the exact property of the full JPDF
\bea
&& \lim_{\Delta \to \infty}  P_{\Delta , \infty}(\sigma_1,\sigma)   =  F_2'(\sigma_1)  \ F_2'(\sigma) \\
&& \lim_{ \Delta \to 0} P_{\Delta , \infty}(\sigma_1,\sigma)  =   F _2'(\sigma_1)    \   F_0'(\sigma) \label{prod2}
\eea
hence our exact result is able to reproduce this property. Note that the
second property has not been found previously. It is non-trivial since
here the large time limits $t_1,t_2 \to \infty$ have been taken.

A systematic expansion in the large $\Delta$ limit is performed in
Section \ref{sec:largeDelta}, and the results are obtained in
an expansion in $1/\Delta^{1/3}$. The small $\Delta$ limit is
analyzed in Section
\ref{sec:smallDelta} where (\ref{decoupleBR}) is shown,
and the leading correction (in $\Delta^{1/3}$) are obtained.

All the above results can be extended to an arbitrary endpoint $X$, in terms of the scaling variable
\bea \label{endpoint_def}
\hat X = \frac{X}{2 (\Delta t_1)^{2/3}}
\eea
In such a case the definition of the scaled height difference variable $h$ becomes
\bea \label{defhX}
H_2-H_1= (\Delta t_1)^{1/3} ( h - \hat X^2) = \frac{X^2}{4 \Delta t_1} + (\Delta t_1)^{1/3} h
\eea
where we recall $\Delta t_1=t_2-t_1$.
Then definition (\ref{defP}) and formula (\ref{jpdf0}) and (\ref{jpdf0}) still hold, with a more general expression
for the kernel $K^{\Delta}_{\sigma_1}$, given in (\ref{K4def}). The property
(\ref{prodGUE}) also holds. In the property (\ref{decoupleBR}),
$F_0(\sigma)$ is replaced by $F_0(\sigma-\hat X^2;\hat X)\equiv H(\sigma;\frac{\hat X}{2},-\frac{\hat X}{2})$,
the extended Baik-Rains distribution (i.e. its CDF). The function $H$ is
given in \cite{png}. It describes the one point PDF of the (scaled) height of
the KPZ equation with tilted stationary initial conditions, in the large time limit,
as recalled in \eqref{BR1BR2} and in \eqref{BaikRainsExplicit}.\\

\subsection{{ Conditional probability and moments:} numerical evaluations, and
series expansions}

\subsubsection{Conditional probability and { its moments}.} \label{sec:condmom}

In addition to the JPDF, of great interest is the conditional PDF (CPDF)
for $h=\sigma$ given that $h_1=\sigma_1$, together with its moments {
\bea
&& P(\sigma|\sigma_1) \equiv  \frac{P_{\Delta}(\sigma_1 , \sigma)}{F_2{}'(\sigma_1)}
\quad , \quad \langle h^p \rangle_{\sigma_1} \equiv  \int_{-\infty}^{+\infty} d\sigma ~ \sigma^p  P(\sigma|\sigma_1)
\eea}
From what we discussed above as $\Delta$ increases from zero to infinity it is predicted to
crossover from the Baik-Rains distribution $F_0$ to the GUE-TW distribution $F_2$ (and their moments).
It is useful to recall the known values of the lowest cumulants, as well as the skewness
and kurtosis of the two limiting cases
\bea \label{cumGUE}
&& \langle h \rangle_{\mbox{\tiny GUE}}  = -1.771086807411 \quad , \quad \langle h^2 \rangle^c_{\mbox{\tiny GUE}}  = 0.8131947928329 \\
&&  {\rm Sk}_{\mbox{\tiny GUE}} \equiv  \frac{\langle h^3 \rangle^c_{\mbox{\tiny GUE}}}{(\langle h^2 \rangle^c_{\mbox{\tiny GUE}})^{3/2}} = 0.224084203610 \\
&& {\rm Ku}_{\mbox{\tiny GUE}} \equiv  \frac{\langle h^4 \rangle^c_{\mbox{\tiny GUE}}}{(\langle h^2 \rangle^c_{\mbox{\tiny GUE}})^{2}}
= 0.0934480876
\eea
as well as \cite{HalpinStat2014}
\bea \label{cumBR}
&& \langle h \rangle_{\mbox{\tiny BR}}  = 0 \quad , \quad  \langle h^2 \rangle^c_{\mbox{\tiny BR}}  = 1.15039 \\
&&  {\rm Sk}_{\mbox{\tiny BR}}  = 0.35941 \quad , \quad  {\rm Ku}_{\mbox{\tiny BR}}  = 0.28916
\eea
At the two limit points $\Delta=0$ and $\Delta=+\infty$ the CPDF and the conditional
moments (cumulants) are independent of $\sigma_1$ since the JPDF factorizes. For intermediate values of
$\Delta$ these depend on $\sigma_1$ and the result that we obtain here is their
value in the limit of large positive $\sigma_1$. More precisely we
calculate
\bea
&& P^{(1)}(\sigma|\sigma_1) \equiv  \frac{P^{(1)}_{\Delta}(\sigma_1 , \sigma)}{F^{(1)}_2 {}'(\sigma_1)}
\quad , \quad \langle h^p \rangle^{(1)}_{\sigma_1} \equiv  \int_{-\infty}^{+\infty}
d\sigma ~ \sigma^p  P^{(1)}(\sigma|\sigma_1) \label{condcum}
\eea
which deviate from their exact values with corrections that are exponentially suppressed in $\sigma_1$
at large positive $\sigma_1$
\bea
\langle h^p \rangle_{\sigma_1} = \langle h^p \rangle^{(1)}_{\sigma_1} + O(e^{- \frac{4}{3} \sigma_1^{3/2}})
\eea

In particular we are able to obtain analytical expressions for the moments of the conditional probability in the limit $\Delta \to 0$ { (here given for $\hat X=0$)}
\bea
&& \langle h^p \rangle_{\sigma_1}^{(1)} =
\langle h^p \rangle_{\mbox{\tiny BR}}  + \Delta^{1/3} \frac{\Ai(\sigma_1)^2}{K_\Ai(\sigma_1,\sigma_1) } A_p
+ O(\Delta^{2/3})
\eea
{ where the first three coefficients $\{A_{p}\}_{p=1}^{3}$ are calculated as $\{-0.5751, -0.2214,-2.1755 \}$ and the general formula is given in \eqref{small_delta_expansion_final}.}

In the limit $\Delta \to \infty$ they converge to the moments of the GUE-TW distribution, and we have obtained
the corrections as
\begin{equation}\label{large_delta_pers}
\langle h^p \rangle_{\sigma_1} ^{(1)}
= \langle h^p \rangle_{\mbox{\tiny GUE}} +  R_{1/3}(\sigma_1) \frac{p }{\Delta^{1/3}}
 \langle h^{p-1} \rangle_{\mbox{\tiny GUE}}  + O(\Delta^{-2/3})
\end{equation}
with
\begin{equation}
R_{1/3}(\sigma_1) = \left( \frac{  \left[\int_{\sigma_1}^\infty dy \Ai(y )\right]^2 - \int_{\sigma_1}^\infty dy K_{\Ai}(y,y) }{K_\Ai(\sigma_1,\sigma_1)}\right)
\end{equation}
{ where the next two orders in the $\Delta^{-1/3}$ expansion
are reported and analyzed, together with the cumulants, in
\eqref{condmomexp}  and the equations below it. Note that it is shown there that
these coefficients are independent of $\hat X$.} Indeed the dependence of the cumulants on the endpoint $\hat{X}$ vanishes exponentially in $\Delta$, see Fig. \ref{fig:correlations_integrated_X}, therefore the coefficients of the large $\Delta$ expansion of the first cumulants are all independent of the final endpoint. { As discussed
below this is a manifestation of the persistence of correlations and ergodicity breaking during the time evolution
with droplet initial condition.}

In order to have an easier comparison with numerics or experiments
it may be more advantageous to define a cumulative conditioning, i.e.
\bea
\fl && P^{(1)}(\sigma| h_1>\sigma_c) \equiv  \frac{\int_{\sigma_c}^{+\infty} d \sigma_1 P^{(1)}_{\Delta}(\sigma_1 , \sigma)}{1 - F^{(1)}_2(\sigma_c)}
\quad , \quad \langle h^p \rangle_{h_1>\sigma_c}^{(1)} \equiv  \int_{-\infty}^{+\infty} d\sigma ~ \sigma^p  P^{(1)}(\sigma|
h_1 > \sigma_c) \label{condcumint}
\eea
for which more statistics can be accumulated, making it easier to compare with numerical or experimental results, see Fig.  \ref{fig:correlations_integrated}. On the other hand the integrated cumulants have
properties very similar to the conditional cumulants.\\

Note that an important sum rule satisfied by the exact JPDF is
\begin{equation}
\int_{-\infty}^{+\infty} d\sigma P_{\Delta}(\sigma_1 , \sigma) = F'_2(\sigma_1)
\end{equation}
Similarly, our result for the tail function of the JPDF, $P^{(1)}_{\Delta}$, satisfies the sum rule
(shown in \ref{sumrule_appendix})
\begin{equation}
\int_{-\infty}^{+\infty} d\sigma P^{(1)}_{\Delta}(\sigma_1 , \sigma) = F_2^{(1)}{}'(\sigma_1)
\end{equation}
Therefore we can use the difference $ F_2(\sigma_c) -  F_2^{(1)} (\sigma_c)$ as a good measure of the error on the cumulative distribution \eqref{condcumint} due the approximating the JPDF with its tail for large $\sigma_1$. Encouraged by Fig. \ref{fig:gue_approx}, we surmise
that the predictions remain quite precise for a quite broad range of values of $\sigma_c$ and especially in the domain $\sigma_c >0$.

\begin{figure}[!ht]
\centering
\includegraphics[scale=1.2]{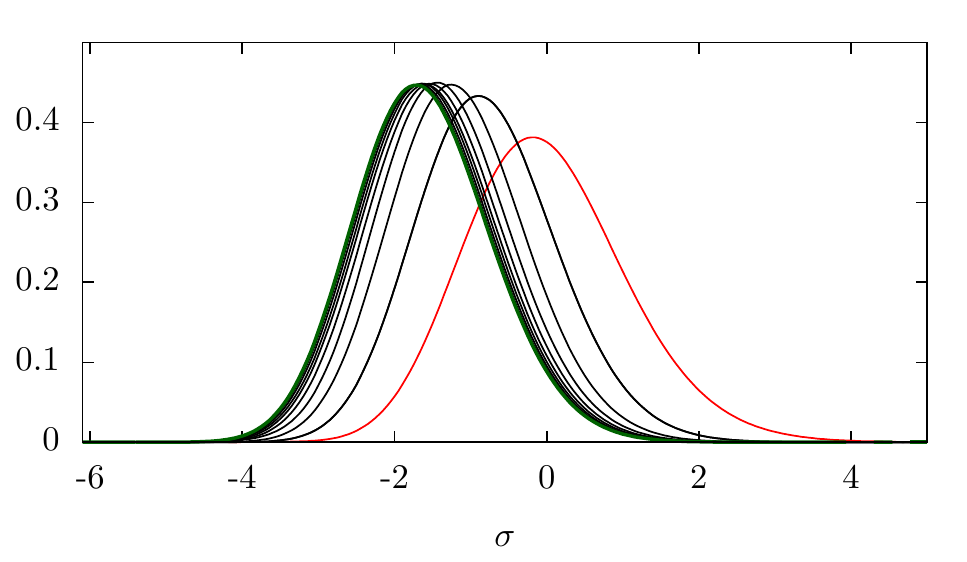}
\caption{Plot of the conditional probability distribution $P^{(1)}(\sigma|\sigma_1)$, defined in (\ref{condcum}),
of the scaled height difference $h \equiv  (H_2-H_1)/t_1^{1/3}=\sigma$ for a fixed value
of the height at the earlier time $h_1\equiv H_1/t_1^{1/3} = \sigma_1 = 0$,
as a function of $\sigma$. The various curves correspond to increasing values of $\Delta^{1/3} = (0.7 k)  $ with $k = 0, \ldots , 10$ (from right to left). The functions interpolate between the $\Delta =0$ point (red line) which coincides with the Baik-Rains probability distribution $F_0'(\sigma)$, and the $\Delta \to \infty$ (green line) which corresponds to the GUE Tracy-Widom probability distribution $F'_2(\sigma)$.   }
\label{fig:pdf_evolution}
\end{figure}

{
\subsubsection{Unconditioned mean value}

While here we obtain results for the conditional first moment of $h$, it is useful to
point out that the {\it unconditionned first moment} can be calculated exactly for arbitrary
fixed $\Delta$ and $t_1 \to +\infty$, from pure scaling. Indeed, using that at large time
\bea
&& \langle H_1 \rangle = \langle \sigma_1 \rangle_{\mbox{\tiny GUE}}  t_1^{1/3} + o(t_1^{1/3}) \\
&& \langle H_2 \rangle = \langle \sigma_1 \rangle_{\mbox{\tiny GUE}} (t_1+\Delta t_1)^{1/3} + o(t_1^{1/3})
\eea
one immediately obtains from the definitions (\ref{defh}) that
\bea \label{mean1tot0}
&&  \langle h \rangle =   \frac{\langle H_2 \rangle - \langle H_1 \rangle}{(\Delta t_1)^{1/3} }
= \langle \sigma \rangle_{\mbox{\tiny GUE}} \frac{(1+ \Delta)^{1/3} - 1}{\Delta^{1/3}}
\\
&& =  \langle \sigma \rangle_{\mbox{\tiny GUE}} \left(1 - \frac{1}{\Delta^{1/3}} + (1+ \frac{1}{\Delta})^{1/3} - 1 \right)
\label{exactmean}
\eea
where the second form is adapted to obtaining readily the series expansion at large $\Delta$.
Note that this behavior in $1/\Delta$ holds equally well for a larger class of initial conditions, e.g. for
flat or stationary initial condition (replacing
$\langle \sigma \rangle_{\mbox{\tiny GUE}}  \to \langle \sigma \rangle_{\mbox{\tiny GOE}}$
and $\langle \sigma \rangle_{\mbox{\tiny GUE}}  \to \langle \sigma \rangle_{\mbox{\tiny BR}}$
respectively).

Formula \eqref{mean1tot0} immediately leads to the following asymptotic behaviors:
\bea
&& \fl \langle h \rangle_{\Delta \to 0}  \simeq \langle \sigma \rangle_{\mbox{\tiny GUE}}   \frac{\Delta^{2/3}}{3}
(1+ O(\Delta)) ~, \quad
\langle h \rangle_{\Delta \to +\infty}  \simeq \langle \sigma \rangle_{\mbox{\tiny GUE}} (1 - \frac{1}{\Delta^{1/3}}
+ \frac{1}{3 \Delta} ) + O(\frac{1}{\Delta^2} )  \label{asympt1}
\eea
(see also Appendix \ref{app:airy} for Airy processes arguments). Interestingly,
similar behavior for the unconditioned cumulants at small and large $\Delta$ was
found numerically and experimentally, not for the droplet geometry, but for the flat initial condition in
\cite{TakeuchiCrossover}, although no determination of the prefactors (predicted by the present
arguments) was discussed there. \\
}

\subsubsection{Persistence of correlations and conditional two-time correlations}
The first order correction $O(\Delta^{-1/3})$ of the conditional moments to their respective GUE values,
i.e. the term $\Delta^{-1/3} R_{1/3}(\sigma_1)$ in (\ref{large_delta_pers}), is
at the origin of the property of ``persistence of correlations''  for droplet initial conditions.
This property, also called breaking of ergodicity in the temporal evolution, states that
the two-time connected correlation of the KPZ heights $H_1,H_2$ at two large and well separated times
$t_1,t_2$ do not decay to zero in the limit $t_2/t_1 \to +\infty$ \cite{Takeuchi,TakeuchiPersistence,FerrariSpohn2times,GueudreUnpub},
i.e. the following dimensionless connected correlation has a finite limit
\begin{equation}
 \lim_{\Delta \to +\infty} \frac{\langle H_1 H_2 \rangle^c }{\langle H_1^2 \rangle^c }  = c
 \end{equation}
In the present work we define a conditional version of the above function
\bea\label{eq:conn_corr_eq}
c_{\Delta, \sigma_c} = \frac{\langle H_1 H_2 \rangle^c_{h_1 > \sigma_c}}{\langle H_1^2 \rangle^c_{h_1 >\sigma_c}}  \quad , \quad c_{\sigma_c} =  \lim_{\Delta \to +\infty} c_{\Delta, \sigma_c}
\eea
involving correlations restricted to realizations such that $h_1 > \sigma_c$.
Our main result for the JPDF \eqref{jpdf0} allows to numerically evaluate this function for
general $\Delta$, see Fig. \ref{fig:conn_corr}. In Section \ref{sec:cond} we furthermore
obtain an analytic expression for its large $\Delta$ limit, $c_{\sigma_c}$, which is exact
for large positive $\sigma_c$ and expected to be accurate up to $\sigma_c \approx -1.5$
(see Fig. \ref{fig:coeff4} there). An extrapolation to $\sigma_c \to - \infty$
provides an estimate for { $c = \lim_{\sigma_c \to -\infty} c_{\sigma_c} \approx 0.58 \pm 0.05$
close to the observation
in Fig. 12 of \cite{Takeuchi}. Note that our result provides a proof by explicit calculation
that the large $\Delta$ limit of this ratio is finite. A complementary study is made using
Airy processes in Section \ref{app:airy} with satisfactory agreement. Both methods show that there is
no dependence of this ratio in the endpoint position $\hat X$.

Note that since this property does not occur for flat initial conditions
\cite{Takeuchi,TakeuchiPersistence,FerrariSpohn2times,GueudreUnpub}
we surmise that in that case $\langle h \rangle_{\sigma_1}  \Big|_{\text{flat}}
= \langle h \rangle_{\mbox{\tiny GOE}} +  o(\Delta^{-1/3})$.

 }

\subsubsection{Numerical evaluations}

\begin{figure}[!ht]
\centering
\includegraphics[scale=1.4]{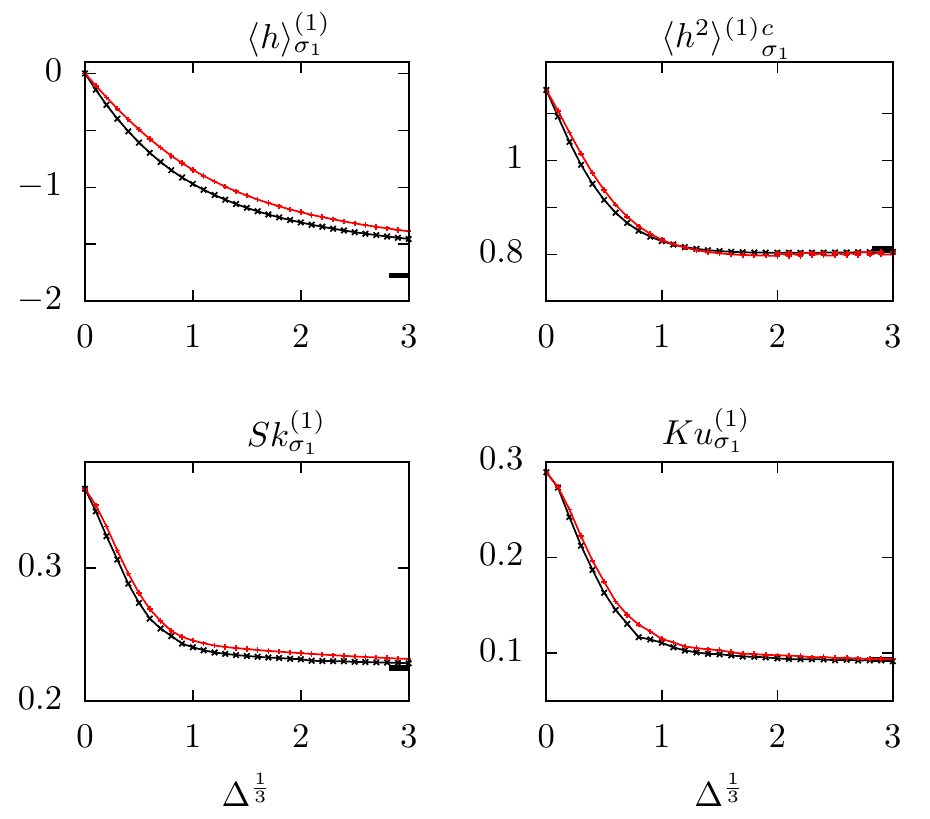}
\caption{Plot of the lowest cumulants (mean, variance, skewness and kurtosis) of the conditional probability distribution $P^{(1)}(\sigma|\sigma_1)$, defined in (\ref{condcum}), at fixed $\sigma_1 = 0$ (red lines) and $\sigma_1 = 1$ (black lines) as a function of $\Delta^{1/3}$. The small horizontal lines on the right in each plot indicate the asymptotic values for infinite $\Delta$, namely the GUE expectation values reported in \eqref{cumGUE}.  Note the weak non-monotonicity of the variance which is a genuine effect. }
\label{fig:correlations}
\end{figure}

 \begin{figure}[!ht]
\centering
\includegraphics[scale=1.4]{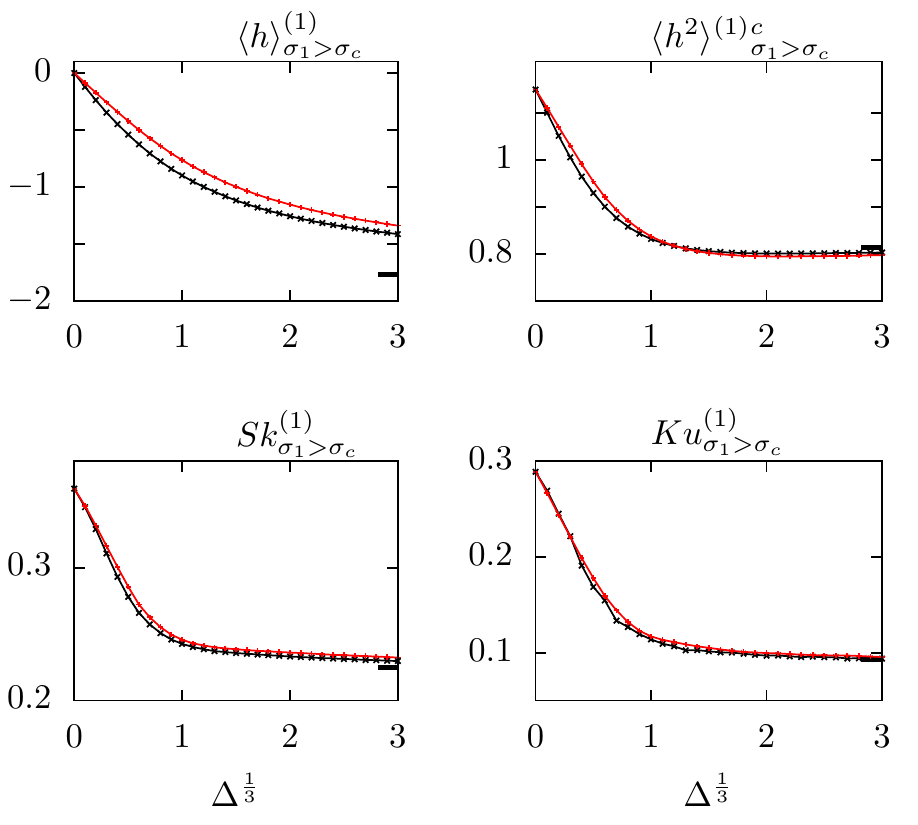}
\caption{Plot of the lowest cumulants (mean, variance, skewness and kurtosis) of the integrated conditional probability distribution $P^{(1)}(\sigma| h_1>\sigma_c)$, defined in \eqref{condcumint}, at fixed $\sigma_{c} = -1$ (red lines) and $\sigma_{c} = 0$ (black lines) as a function of $\Delta^{1/3}$. The small horizontal lines on the right in each plot indicate the asymptotic values for infinite $\Delta$, namely the GUE expectation values reported in \eqref{cumGUE}. The small deviation from its asymptotic value of the kurtosis at $\sigma_c = -1$ is due to numerical error.}
\label{fig:correlations_integrated}
\end{figure}

\begin{figure}[!ht]
\centering
\includegraphics[scale=1.4]{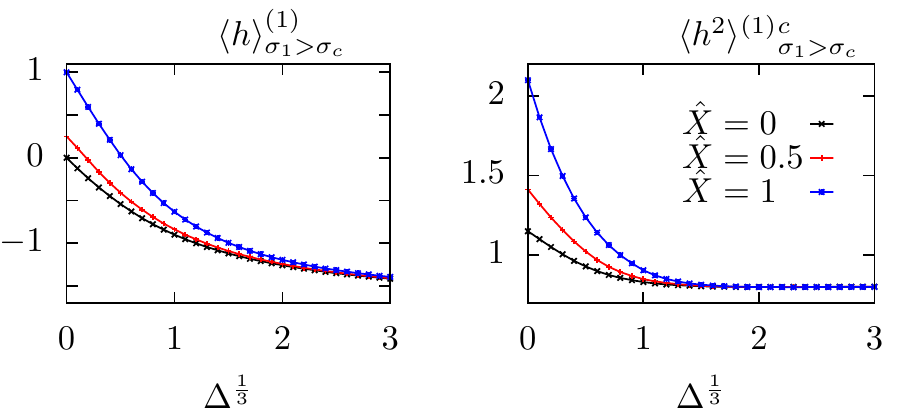}
\caption{Plot of the mean and variance of the integrated conditional probability distribution $P^{(1)}(\sigma| h_1>\sigma_c)$ at fixed $\sigma_{c} = 0$ and endpoint position \eqref{endpoint_def} $\hat{X}=0$ (black lines),  $\hat{X}=0.5$ (red lines),  $\hat{X}=1$ (blue lines) as a function of $\Delta^{1/3}$. At $\Delta=0$ the cumulants converge to the mean and variance of the Baik-Rains distribution for with a displacement $\hat{X}$ (see Fig. \ref{fig:BR_X}). In the other limit, for large $\Delta$, the dependence on $\hat{X}$ is exponentially (as $\sim e^{ - \text{const } \Delta^{1/3}}$) vanishing.   }
\label{fig:correlations_integrated_X}
\end{figure}

\begin{figure}[!ht]
\centering
\includegraphics[scale=1.1]{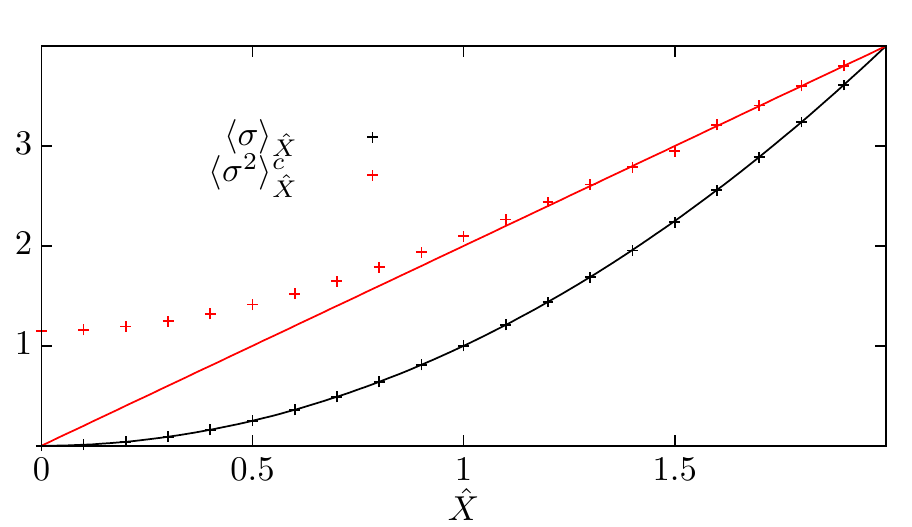}
\caption{ Plot of first two cumulants of the variable $\sigma$ in the Baik-Rains probability distribution formula \eqref{BaikRainsExplicit}, as a function of the displacement $\hat{X}$. The mean value $\langle \sigma \rangle_{\hat{X}} $ (black points) is exactly given by the function $\hat{X}^2$ (black line) while the variance $\langle \sigma^2 \rangle^c_{\hat{X}}$ (red points) converges to the line $2 |\hat{X}|$ (red line) for large $\hat{X}$, { in agreement with Brownian motion statistics.}}
\label{fig:BR_X}
\end{figure}

\begin{figure}[!ht]
\centering
\includegraphics[scale=1.1]{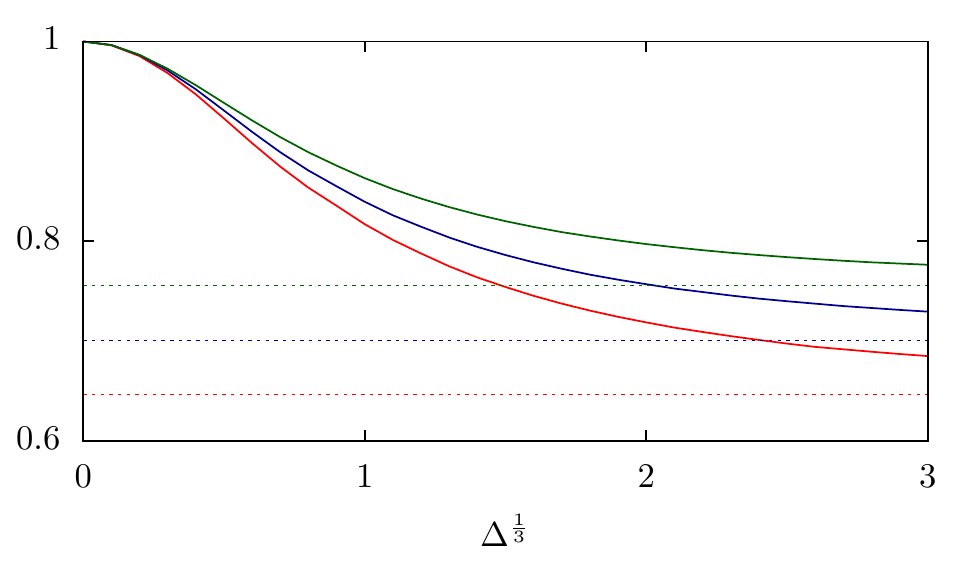}
\caption{ Plot of the conditional correlator $c_{\Delta, \sigma_c}$ defined in \eqref{eq:conn_corr_eq} as a function of $\Delta^{1/3}$ and for three different values of $\sigma_c$, from top to bottom: $\sigma_c= -0.5$ (green), $\sigma_c= -1$ (blue), $\sigma_c= -1.5$ (red). The dashed lines are their respective asymptotic values at infinite $\Delta$. The latter are computed from the large $\Delta$ expansion of the JPDF \eqref{jpdf0}, in equation \eqref{large_delta_pers}. For any value of $\sigma_c$ the data approach their asymptotic values with corrections of order $\Delta^{-2/3}$ (see Section \ref{sec:cond} for details).     }
\label{fig:conn_corr}
\end{figure}

In Fig. \ref{fig:pdf_evolution} we report the results for the conditional probability distribution function
obtained by numerical evaluations of equation \eqref{defP1}
as a function of $\Delta$ and  at fixed $\sigma_1$. In Fig. \ref{fig:correlations}) we plot the lowest
conditional cumulants of the scaled height difference $h=\sigma$, i.e. the mean value, standard deviation, skewness and kurtosis,
for a fixed value of $h_1=\sigma_1$ (in particular $\sigma_1 = 0,1$) where the approximation \eqref{eq:tail_approx} is believed to be correct  (see Fig. \ref{fig:correlations}) up to deviation of $10^{-4}$ (an estimate based on the approximation of the GUE probability distribution with its tail, see Fig. \ref{fig:gue_approx}). We note the genuine feature that the variance is slightly non-monotonous, even if the deviations from a monotonic behavior are very small (at least for the values of $\sigma_1$ considered here). On the other hand the other cumulants are monotonous decreasing functions of $\Delta$. In Fig. \ref{fig:correlations_integrated_X} we study the effect of the endpoint displacement $\hat{X}$ on the first two cumulants as a function of $\Delta$. Since at $\Delta = 0 $ we recover the BR distribution with an arbitrary endpoint $\hat{X}$ (see Appendix \ref{app:BR}) we also plot here for completeness its first two cumulants as a function of $\hat{X}$ in Fig. \ref{fig:BR_X}.
\\

The numerical absolute error on the cumulants is estimated (for both cases of conditional and integrated probability) to be around $10^{-6}$ for the mean value, $10^{-5}$ for the variance, $10^{-4}$ for the skewness and $10^{-3}$ for the kurtosis. This decrease of precision for higher cumulants is due to the finite size effects induced by the discretization of the kernels  and which are larger along the tails of the distribution. For more details on the numerical evaluations see Appendix \ref{app:numerics}.

\section{Replica Bethe ansatz approach to the two-time correlations} \label{sec:RBA}

\subsection{Joint integer moments from the \texorpdfstring{$\delta-$}{delta-}Bose gas}

As for the calculation of the one-time observables, the RBA method starts
with the calculation of the integer moments of the DP partition sums, from
which one aims to extract the joint PDF of KPZ height fields.
Here we consider the joint moments
\begin{equation} \label{defjmoments}
Z_{n_1,n_2}(t_1,t_2) = \overline{ Z_1^{n_1} Z_2^{n_2} } =
\overline{ Z_\eta(X_1,t_1| 0,0)^{n_1} Z_\eta(X_2,t_2| 0,0)^{n_2} }
\end{equation}
with $n_1,n_2 \geq 0$, as defined in (\ref{defH12}). Here
we have allowed for arbitrary endpoints $X_1,X_2$, however
thanks to the so-called statistical tilt symmetry, one can choose
with no loss of generality, $X_1=0$ and $X_2=X$, which we
will do later on \footnote{As compared to the result for $(X_1,X_2)=(0,X)$,
a general $X_1$ only results in a simultaneous deterministic
shift of $H_1$ by $- X_1^2/(4 t_1)$ and of $H_2$ by $(X^2- (X-X_1)^2)/(4 (t_2-t_1))$.}

For calculational convenience, we will consider a slightly more general partition sum $Z_2$
defined as
\bea \label{defZ2}
Z_2= Z^{w,X_1}_\eta(X_2,t_2| 0,0) \equiv  \int dy e^{- w |y-X_1|} Z_{\eta}(X_2,t_2|y,t_1) Z_{\eta}(y,t_1|0,0)
\eea
with $w>0$. From the definition of the partition sums as path integrals (\ref{zdef}) it is
easy to see that in the limit $w \to 0$ one recovers
\bea
\lim_{w \to 0} Z_2= Z_\eta(X_2,t_2| 0,0)
\eea
hence one obtains (\ref{defjmoments}). We will perform the calculation with
$w>0$ (implicitly, for notational simplicity) and perform the limit $w \to 0$ at the end.

It turns out that all the $n-$th joint moments of the DP can be mapped to quantum expectation values
{ in a model} of $n$ attractive 
bosons on the line \cite{kardareplica,bb-00}. This is known as the attractive Lieb-Liniger model \cite{ll} for $n$ bosons with Hamiltonian
\bea \label{HLL}
 H_n = -  \sum_{\alpha=1}^n \frac{\partial^2}{\partial x_\alpha^2} - 2 \bar c
\sum_{1 \leq \alpha < \beta \leq n} \delta(x_\alpha - x_\beta)
\eea
on the line $x \in \mathbb{R}$. In our chosen unit $\bar c=1$, which we will use from now on.
For definiteness one fixes periodic boundary conditions with period $L$,
where $L \to +\infty$ is considered afterwards. The joints moments of the DP can then be expressed as the matrix elements of the quantum evolution operator with imaginary time $t$
\bea
\fl && \overline{ Z(x_1,t|y_1,t') Z(x_2,t|y_2,t') \ldots Z(x_n,t|y_n,t') } =\langle x_1,x_2 , \ldots, x_n  | e^{- (t-t') H_{n}}
| y_1, \ldots , y_{n}  \rangle
\eea

\begin{figure}[!ht]
\centering
\includegraphics[scale=0.5]{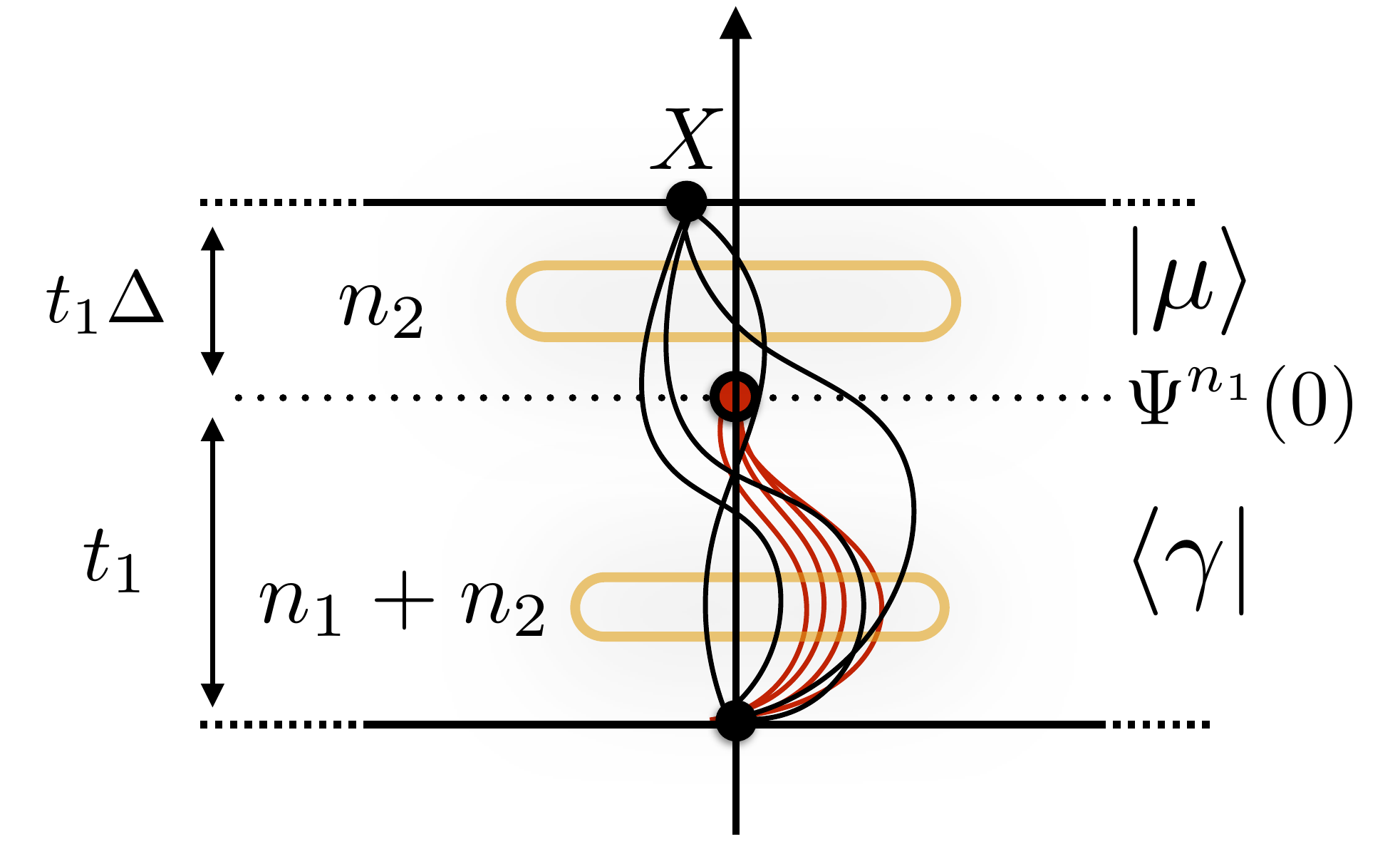}
\caption{Graphical representation of the moments of the partition function from the delta-Bose gas replica picture
displayed in equations \eqref{eq:partitionfunctionbosons}
and \eqref{momgen1} (with $X_1=0$ and $X_2=X$): At time $t=0$ and point $0$, $n_1 + n_2$ interacting bosons (replica) are created. From these $n_1$ propagate to the point $0$ at time $t_1$ where they are annihilated. The other $n_2$ propagate to point $X$ at time $t_2=(1+ \Delta) t_1$ (and their positions at time $t_1$ are denoted $y_\alpha$). The replica partition function is evaluated by summing over all the eigenstates $| \gamma \rangle $ of the Lieb-Liniger Hamiltonian  in the time interval $t \leq t_1$ with $n_1 + n_2$ bosons, and all the eigenstates $|\mu \rangle $ with $n_2$ bosons for $t_1 < t \leq t_2$.    }
\label{fig:replicas}
\end{figure}

Using these relations, and the statistical independence of
the partition sum (i.e. of the noise) on the time intervals $[0,t_1]$ and $[t_1,t_2]$,
we can rewrite the joint two-time moments as
\begin{align}\label{eq:partitionfunctionbosons}
Z_{n_1,n_2}(t_1,t_2) & =  \left(\prod_{j=1}^{n_2}\int_{-\infty}^{\infty} dy_j e^{-w |y_j-X_1|} \right) \underbrace{ \langle X_2 , \ldots, X_2}_{n_2}   | e^{- \Delta t_1 H_{n_2}} | y_1, \ldots , y_{n_2}  \rangle \nn \\& \times \langle  \underbrace{y_1, \ldots , y_{n_2}}_{n_2}, \underbrace{X_1  , \ldots , X_1 }_{n_1} | e^{- t_1 H_{n_2+n_1} } | \underbrace{0  , \ldots , 0  }_{n_1 + n_2}\rangle
\end{align}
with $t_2-t_1=\Delta t_1$ (see Fig. \ref{fig:replicas}). Here
we have assumed $n_1 \geq 1$ and $n_2 \geq 1$. Clearly
$Z_{n_1,0}(t_1,t_2)= \overline{Z_1^{n_1}} = Z_{n_1}(t_1)$ and
$Z_{0,n_2}(t_1,t_2)= \overline{Z_2^{n_2}} = Z_{n_2}(t_2)$ reduce to
a one-time problem $Z_n(t)$, which was studied in \cite{we} (for $w=0$) and which
we do not recall in details here.

Let us now use the decomposition of the evolution operator into the eigenstates of $H_n$, namely
$e^{- t H_n} = \sum_{\mu \in \Lambda_n} |\mu \rangle e^{- t E_\mu} \langle \mu |$. Here
$\Lambda_n$ denotes a complete orthogonal basis of eigenstates $| \mu \rangle$ of $H_n$,
$E_\mu$ their eigenenergy and $||\mu||$ their norms. The wave-functions in the coordinate basis
are denoted as $\psi_\mu(x_1,..x_n) \equiv  \langle x_1,..x_n |\mu\rangle$. We obtain
\bea \label{momgen1}
\fl && Z_{n_1,n_2}(t_1,t_2) =
\sum_{\mu \in \Lambda_{n_2},\gamma \in \Lambda_{n_1+n_2}}
\frac{ \psi_\gamma^*(0,..,0) \psi_\mu(X_2,..X_2)}{||\gamma||^2 ||\mu ||^2 } e^{- \Delta t_1 E_\mu - t_1 E_\gamma } F^{n_2;n_1+n_2}_{\mu ; \gamma}
\eea
where $F^{n_2;n_1+n_2}_{\mu ;\gamma}$ is a ``form factor" defined as
\bea \label{eq:ff}
\fl && F^{n_2;n_1+n_2}_{\mu ; \gamma} \equiv  \prod_{\alpha=1}^{n_2} \int dy_\alpha e^{-w |y_j-X_1|}
 \psi_\mu^*(y_{1},..,y_{n_2})
\psi_\gamma(\underbrace{X_1  , \ldots , X_1 }_{n_1} ,y_{1},..,y_{n_2})
\eea
For $w=0$, it is proportional to the form factor of the operator $[\Psi(X)]^{n_1}$ in
the second-quantized formulation (which we will not use in our work).
Until now the formula are very general, they hold for arbitrary $L$, and
even apply to a broader class of time uncorrelated noise $\eta(x,t)$
upon a suitable generalization of $H_n$.

\subsection{Bethe ansatz in large size limit and strings}

We now use the known properties of the LL model \cite{ll}, which allows to treat the
continuum KPZ equation when the noise $\eta(x,t)$ is a Gaussian white noise. The eigenstates are parametrized by a set of distinct (in general complex) quasi-momenta or rapidities
$\mu \equiv \{ \lambda_1,..\lambda_n\}$. These are determined by the
coupled Bethe equations, which are the quantization rules for the rapidities and that enforce the periodicity of the eigenstate $|  \{ \lambda_1,..\lambda_n\} \rangle $. For $|\mu \rangle \in \Lambda_n$, the (un-normalized) eigenfunctions are totally symmetric in the $x_\alpha$, and are such that, on the sector
$x_1 \leq x_2 \leq .. \leq x_n$, they are equal to (for $\bar c=1$)
\bea \label{def1}\fl &&
\psi_\mu(x_1,..x_n) =  \sum_P A_P \ e^{i \sum_{\alpha=1}^n \l_{P_\alpha} x_\alpha} \, , \quad
A_P=\prod_{n \geq \beta > \alpha \geq 1} \Big(1 + \frac{i  
}{\lambda_{P_\beta} - \lambda_{P_\alpha}}\Big)\,
\eea
where the sum runs over all $n!$ permutations $P$ of the rapidities $\l_j$. The eigenenergies associated to \eqref{def1} are given by a sum over all the single particle energies
$E_\mu=\sum_{\alpha=1}^n \lambda_\alpha^2$. Note that the eigenfunction \eqref{def1} computed with all the particles sitting at the same point is given by $\psi_\mu(x,..x) = n! e^{  i x \sum_\alpha \lambda_\alpha}$.
Focusing on the limit $L \to +\infty$, the general eigenstate is built by partitioning the $n$ particles into a set of $n_s \leq n$ bound states called {\it strings} \cite{m-65}
formed by $m_j \geq 1$ particles with $n=\sum_{j=1}^{n_s} m_j$.
Their associated rapidities are
\be\label{stringsol}
\l_{j, a_j}=k_j +\frac{i}2(m_j+1-2a_j)
\ee
where $m_j k_j \in \mathbb{R}$ is the total momentum of string $j$.
Here, $a_j = 1,...,m_j$ labels the rapidities within the string $j=1,\dots n_s$.
From now on a string state is denoted as $|\mu \rangle = |{\bf k}, {\bf m} \rangle$
and labeled by the set of $(k_j,m_j)_{j=1,..n_s}$. Inserting
(\ref{stringsol}) in (\ref{def1}) leads to the Bethe eigenstates of the infinite system with corresponding eigenvalues:
\be \label{en}
E_\mu= E({\bf k},{\bf m}) \equiv  \sum_{j=1}^{n_s} m_j k_j^2-\frac{1}{12} m_j^3
\ee
Note that these are the eigenvalues of the shifted Hamiltonian $H_n - \frac{1}{12} n$, which is the proper Hamiltonian to consider, in accordance
with the redefinition (\ref{redef}).

The (inverse squared) norm of these eigenstates are given by \cite{cc-07}
\bea
&& \fl  \frac{1}{||\mu ||^2} = \frac{1}{n!  L^{n_s} } \Phi[{\bf k},{\bf m}] \prod_{j=1}^{n_s} \frac1{m_j^{2}}
\quad , \quad \Phi[{\bf k},{\bf m}] = \prod_{1\leq i<j\leq n_s}
\frac{4(k_i-k_j)^2 +(m_i-m_j)^2}{4(k_i-k_j)^2 +(m_i+m_j)^2}
 \nn \\
&& \fl \label{norm}
\eea
In the large size limit $ L \to \infty$ the string momenta $k_j$ follow the same quantization rules as the ones for free particles $k_j = \frac{2 \pi I_j}{L}$ (where $I_j$ are integer numbers),
therefore the sums over the eigenstates in \eqref{momgen1} can be recast into integrals over the whole real line, one for each string momenta. In particular we can write
\begin{equation} \label{sumstates}
\lim_{L \to +\infty} \sum_{\mu \in \Lambda_n} \frac{1}{||\mu ||^2} = \frac{1}{n!} \sum_{n^{\mu}_s=1}^{\infty}
\frac{1}{n^{\mu}_s!}
\prod_{j=1}^{n_s^\mu} \left( \sum_{m^\mu_j=1}^{n_s^\mu} \int \frac{dk_j}{2 \pi m_j } \right)
 \delta_{n,\sum_{j=1}^{n_s^\mu} m_j} \Phi[{\bf k},{\bf m}]
 \end{equation}

We are now finally in position to write the final formula for the joint moments \eqref{momgen1} in the large $L$ limit.
The sums in (\ref{momgen1}) are now involving string states $|\gamma \rangle=|{\bf q},{\bf m}^\gamma \rangle$, for the time interval $[0,t_1]$, and $|\mu \rangle=|{\bf p},{\bf m}^\mu \rangle$ for the time
interval $[t_1,t_2]$. Using (\ref{sumstates}) for each sum, it leads to
\bea
\fl && Z_{n_1,n_2}(t_1,t_2)=\nn \\
\fl &&    \sum_{n^{\mu}_s=1}^{n_2}  \sum_{n^{\gamma}_s=1}^{n_1+n_2}
\frac{1}{n^{\mu}_s! n^{\gamma}_s!}  \prod_{j=1}^{n_s^\mu} \left(\sum_{m^\mu_j =1}^\infty \int \frac{dp_j}{2 \pi m_j^\mu }  \right)  \prod_{j=1}^{n_s^\gamma} \left(\sum_{m^\gamma_j =1}^\infty  \int \frac{dq_j}{2 \pi m_j^\gamma }
 \right)e^{- \Delta t_1 E({\bf p}, {\bf m}^\mu) - t_1 E({\bf q}, {\bf m}^\gamma) }
 \nn  \\
\fl &&
  \times  F^{n_2;n_1+n_2}_{{\bf p} , {\bf m}^\mu;{\bf q} , {\bf m}^\gamma }  \Phi[\boldsymbol{p}, \boldsymbol{m}^\mu]  \Phi[\boldsymbol{q}, \boldsymbol{m}^\gamma]   \delta_{n_2, \sum_{j=1}^{n_s^\mu} m^\mu_j}
\delta_{n_1+n_2,\sum_{j=1}^{n_s^\gamma} m_j^\gamma}
e^{ + i (X_2-X_1) \sum_{j=1}^{n_s^\mu} p_j m_j^\mu + i X_1 \sum_{j=1}^{n_s^\gamma} q_j m_j^\gamma} \nn \\
\fl && \label{momform1}
\eea
where $F^{n_2;n_1+n_2}_{{\bf p} , {\bf m}^\mu;{\bf q} , {\bf m}^\gamma }$ is the
form factor given in \eqref{eq:ff} specialized to the case $X_1=0$ and
evaluated on generic strings states. The explicit dependence
on $X_1,X_2$ has been extracted, using that for any Bethe state $\psi_\mu(x_1+X,..,x_n+X)
= e^{i  X \sum_{\alpha=1}^n \lambda_\alpha}$. Apart from the form factors all the other factors are explicit functions.

\subsection{Two-time generating function: definitions}

In order to obtain the probability distribution from the moments, we now
introduce a two-time generating function. To avoid proliferation of factors
$2^{1/3}$ in the calculation,
we remind that we parametrize the time $t_1$ as
\begin{equation}
\lambda = \left( t_1/4  \right)^{1/3}
\end{equation}
and, from now on, we use interchangeably the two variables $\sigma$ and $s$ (or $\tilde s$, in the case
$X_2=X \neq 0$, $X_1=0$), related via
\be
\sigma_1 = 2^{-2/3} s_1 \quad , \quad \sigma = 2^{-2/3} \tilde s \quad , \quad \tilde s = s + \frac{X^2}{4 \Delta t_1}
\ee
so that $\lambda s_1=t_1^{1/3} \sigma_1$ and $\lambda \tilde s=t_1^{1/3} \sigma$.

We define the generating function of the two-time problem as
\begin{equation} \label{gdef1}
g_{\Delta, \lambda} (s_1, s)= \overline{ e^{- Z_1 e^{ - \lambda s_1}}  e^{- Z_2 e^{ - \lambda s_1-  \lambda \Delta^{1/3}  s}}  }
\end{equation}
It can be rewritten in terms of the (minus) the scaled DP free energies, i.e. the scaled KPZ heights
$h_1$ and $h_{21} \equiv h$ defined in (\ref{defH12}),(\ref{defH}),(\ref{defh1}), and when
$X \neq 0$
in \eqref{defhX}, leading to
\begin{equation} \label{gdef2}
g_{\Delta, \lambda} (s_1, s) = \overline{ e^{-  e^{ \lambda (\tilde h_1 - s_1)}}  e^{- e^{\lambda (  \tilde h_1 - s_1  +    \Delta^{1/3}  (\tilde h  -s)) }} }
\end{equation}
where we have defined $\tilde h_1=2^{2/3} h_1$ and $\tilde h=2^{2/3} (h - \hat X^2)$.
If this function is known, then one can retrieve in principle the JPDF of
$h_1$ and $h$, following e.g. \cite{we}. For now let us indicate how to do
it only in the limit of both times being large, i.e. $\lambda$ large, at fixed $\Delta$,
which is our main focus here.

In that case the two double exponentials in (\ref{gdef2})
converge to the Heaviside step function $\theta(x)$ in the limit $\lambda \to \infty$
leading to
\bea
\! \! \! \! \!  \! \!  g_{\Delta} (s_1, s)  = \lim_{\lambda \to \infty} g_{\Delta, \lambda}  (s_1, s) &=& \overline{ \theta(\tilde h_1 - s_1) \theta ( \tilde h_1 - s_1 + \Delta^{1/3} (\tilde h  -s) )   } \nn \\
\! \! \! \! \! \! &=& \overline{ \theta(h_1 - \sigma_1) \theta ( h_1 - \sigma_1 + \Delta^{1/3} (h  -\sigma) )   }
\eea
It is then easy to see that the JPDF of $h_1$ and $h$, i.e. $P_{\Delta}(h_1,h)$ defined in (\ref{defP})-(\ref{defPinfty}), is obtained by taking derivatives respect to $\sigma_1,\sigma$ as follows
\begin{equation} \label{deriv1}
P_{\Delta}(\sigma_1,\sigma)  = \left( \partial_{\sigma_1} \partial_{\sigma} -  {\Delta^{- 1/3}} \partial^{2}_{\sigma} \right) g_{\Delta}(s_1,s)
\end{equation}

\subsection{Generating function: expression as a series in the number of strings}

We can now expand the exponential in (\ref{gdef1}) and write the following expansion of the generating function in terms of the two-time joint moments:
\bea
&& g_{\Delta, \lambda}(s_1, s)  = \sum_{n_1,n_2 \geq 0}
\frac{(-1)^{n_1+n_2}}{n_1! n_2!} e^{- \Delta^{1/3} \lambda s n_2 -  \lambda s_1  (n_1 +n_2) }
\overline{Z_1^{n_1} Z_2^{n_2}} \nn \\
&& = -1 + g_{\lambda}(s_1;X_1) + g_{\lambda \Delta^{1/3}}(s_1  +  \Delta^{1/3} s;X_2)
+ \hat g_{\Delta, \lambda}(s_1, s)  \label{gsum}
\eea
where we have separated the pieces which can be expressed
in terms of $g_\lambda(s;X)$, the generating function of the one-time problem
(i.e. point to point DP problem, droplet IC for KPZ), namely
\begin{equation}
g_{\lambda}(s;X) \equiv  \overline{ e^{- Z_\eta (X,t=4\lambda^3|0,0) e^{- \lambda s}} }
\end{equation}
for $n_1=0$ or $n_2=0$, and the piece which genuinely depends on the two
times
\bea
\hat g_{\Delta, \lambda}(s_1, s) = \sum_{n_1, n_2 \geq 1} \frac{(-1)^{n_1+n_2}}{n_1! n_2!} e^{- \Delta^{1/3} \lambda s n_2 -  \lambda s_1  (n_1 +n_2) }
Z_{n_1,n_2}(t_1,t_2)
\eea
Note that we have written the second term in (\ref{gsum}) directly in the limit $w=0$. The analysis
of this term for $w>0$, and a generalization thereof, is slighlty more involved and is performed in the Appendix.

Note that in the limit of both times being large one thus obtains (setting $X_1=X_2=0$ in that equation for
simplicity)
\bea
g_{\Delta}(s_1, s) \equiv \lim_{\lambda \to \infty }   g_{\Delta, \lambda}(s_1, s) = -1 + F_2(\sigma_1) + F_2(\sigma_1  +  \Delta^{1/3} \sigma)
+ \hat g_{\Delta}(s_1, s) \nonumber  \\
\eea
where, we recall $F_2(s)$ is the CDF of the GUE-TW distribution and $\hat g_{\Delta}(s_1, s)  = \lim_{\lambda \to \infty}  \hat g_{\Delta, \lambda}(s_1, s)$.
We note that the one-time terms,{ namely the ones that depend only on one single rescaled
height at either $t_1$, i.e. only on $\sigma_1$ (from $H_1$), or at $t_2$, i.e. only on $\sigma_1  +  \Delta^{1/3} \sigma$ (from $H_2$),} do not produce any contribution to
the JPDF, hence (\ref{deriv1}) leads to
\begin{equation} \label{deriv2}
P_{\Delta}(\sigma_1,\sigma)  = \left( \partial_{\sigma_1} \partial_{\sigma} -  {\Delta^{- 1/3}} \partial^{2}_{\sigma} \right) \hat g_{\Delta}(s_1,s)
\end{equation}

We can now use (\ref{momform1}) to express the generating function for arbitrary finite times as a series of terms
with different values of the numbers of strings $n_s^\mu,n_s^\gamma$ as
\bea \label{hatg}
\hat g_{\Delta, \lambda} (s_1, s)  =    \sum_{n_s^\mu \geq 1, n_s^\gamma \geq 1} \frac{1}{n_s^\gamma!} \frac{1}{n_s^\mu !}  \hat Z_{n_s^\gamma, n_s^\mu}(s_1,s)
\eea
where

\bea \label{Znsns}
\fl && \hat Z_{n_s^\gamma, n_s^\mu}(s_1,s)  =   \sum_{n_1=1}^\infty \sum_{n_2=1}^\infty \frac{1}{n_1! n_2!}
 \prod_{j=1}^{n_s^\mu} \left(\sum_{m^\mu_j =1}^\infty \int \frac{dp_j}{2 \pi m_j^\mu }  \right)  \prod_{j=1}^{n_s^\gamma} \left(\sum_{m^\gamma_j =1}^\infty (-1)^{m_j^\gamma}  \int \frac{dq_j}{2 \pi m_j^\gamma }  \right)\nn \\
 \fl && ~~~~~~~~~~~~~~~~ \times e^{ i (X_2-X_1) \sum_{j=1}^{n_s^\mu} p_j m_j^\mu + i X_1 \sum_{j=1}^{n_s^\gamma} q_j m_j^\gamma} \Phi[\boldsymbol{p}, \boldsymbol{m}^\mu]  \Phi[\boldsymbol{q}, \boldsymbol{m}^\gamma]
 F^{n_2;n_1+n_2}_{{\bf p} , {\bf m}^\mu;{\bf q} , {\bf m}^\gamma }
\nn \\
\fl &&
    ~~~~~~~~~~~~~~~~  \times \delta_{n_2, \sum_{j=1}^{n_s^\mu} m_j}
\delta_{n_1+n_2,\sum_{j=1}^{n_s^\gamma} m_j^\gamma}
e^{- \Delta t_1 E({\bf p}, {\bf m}^\mu) - t_1 E({\bf q}, {\bf m}^\gamma) - \Delta^{1/3} \lambda s n_2 -  \lambda s_1  (n_1 +n_2) } \nn \\
\fl &&
\eea

\subsection{Generating function: tail approximation at large \texorpdfstring{$h_1$}{h(1)} and single string contribution}
\label{sec:gener2}

In principle the formula (\ref{hatg}), (\ref{Znsns}), if properly evaluated, should lead to
the full result for the two-time JPDF via (\ref{deriv2}). This requires however to
obtain closed expressions for all the form factors. While such a program is in progress,
it is not yet available. In particular, the recent attempt in \cite{dotsenko2times1}, is, as we will show,
incorrect.

\begin{figure}[!ht]
\centering
\includegraphics[scale=0.5]{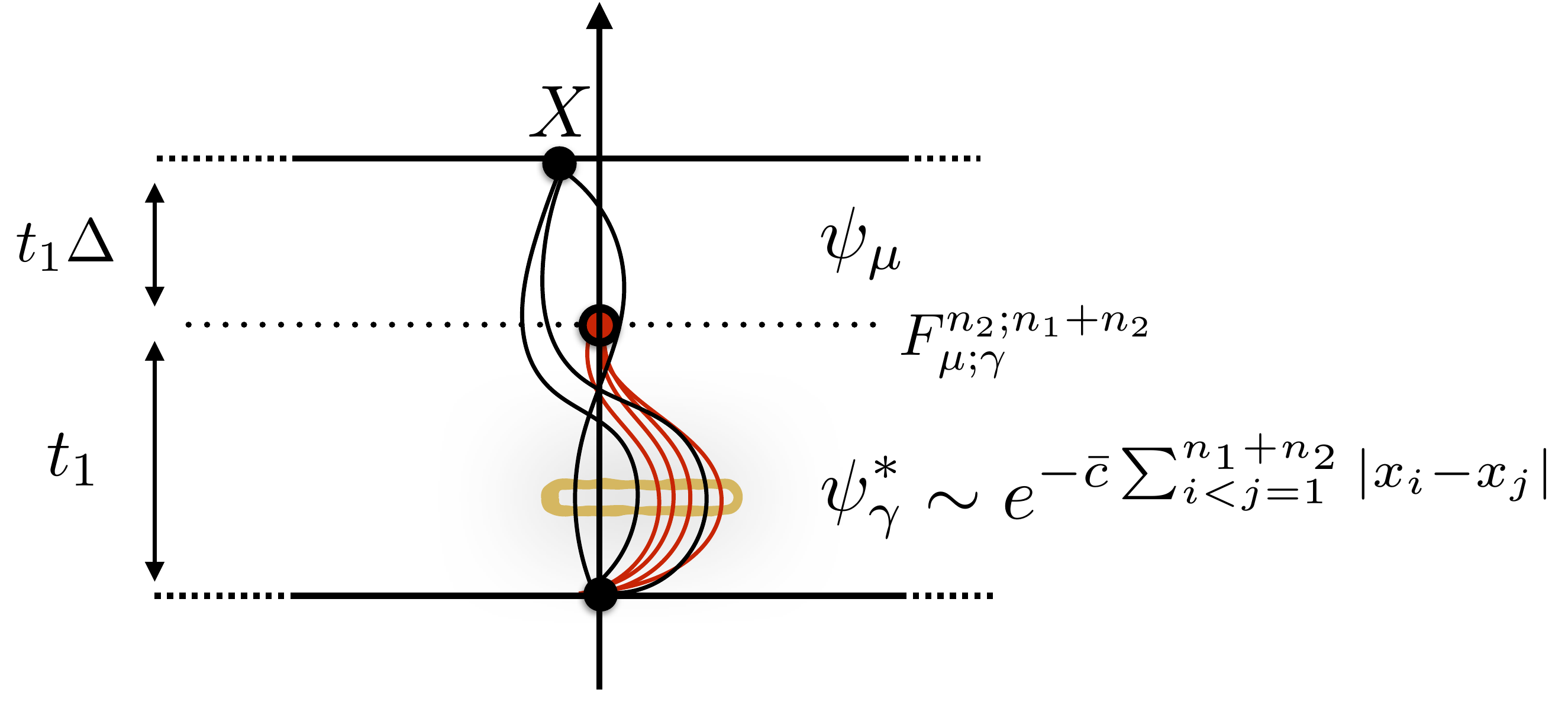}
\caption{Graphical representation of the one-string approximation. We restrict the sum over all the eigenstates $|\gamma \rangle $ in the sector $t \leq t_1$ to a reduced sum over all the states such that all the $n_1+n_2$ bosons are in the same bound state (string). On the other hand we give no restrictions for the eigenstates in the time interval $t> t_1$, where we sum over all the possible configurations with $n_2$ bosons for the state $| \mu \rangle$. The one-string to multiple-string form factor $F^{n_2; n_1 + n_2}_{\mu; \gamma}$ \eqref{ffformula} connects the two states at $t = t_1$ in a non-trivial manner. The summations can now
be performed {\it exactly}.}
\label{fig:one_string_approx}
\end{figure}

Our aim will thus be more modest. We will focus on {\it the tail approximation of the JPDF} for large $h_1$, but arbitrary $h$. To this aim, instead of calculating the complete sum (\ref{hatg}), we will obtain an exact expression for the partial sum
\bea \label{hatg1}
\hat g_{\Delta, \lambda}^{(1)} (s_1, s)  \equiv     \sum_{n_s^\mu \geq 1}  \frac{1}{n_s^\mu !}  \hat Z_{n_s^\gamma=1, n_s^\mu}(s_1,s)
\eea
i.e. restrict to states such that the initial string state (in the interval $[0,t_1]$) is
made of only one string, $n^\gamma_s=1$, of length $m_1^\gamma=n_1+n_2$, as shown in Fig. \ref{fig:one_string_approx}. The main reason
of this restriction is that this is a case where the form factor can be explicitly calculated, as
shown in the next Section.

We will then be able to compute explicitly in the limit of both times large the
partial sum $\hat g_{\Delta, \lambda}^{(1)} (s_1, s)$ and we
will argue that it contains the complete leading large $s_1$ correction, with
\bea
\hat g_{\Delta} (s_1, s) = \hat g_{\Delta}^{(1)}(s_1, s) + O(e^{- \frac{4}{3} s_1^{3/2}})
\eea
and $\hat g_{\Delta, \lambda}(s_1, s) = O(e^{- \frac{2}{3} s_1^{3/2}})$ uniformly in $s$. Hence,
via (\ref{deriv2}), we will obtain the leading tail of the JPDF, $P^{(1)}_{\Delta, \lambda}(s_1, s)$,
as described in (\ref{defP1}) , with
\begin{equation} \label{deriv3}
P^{(1)}_{\Delta}(\sigma_1,\sigma)  = \left( \partial_{\sigma_1} \partial_{\sigma} -  {\Delta^{- 1/3}} \partial^{2}_{\sigma} \right) \hat g^{(1)}_{\Delta}(s_1,s)
\end{equation}
It is interesting to note that there is also a {\it double tail} approximation where
{\it both $h_1$ and $h$ are large} (i.e. both $\sigma$ and $\sigma_1$ are large):
\bea
&& \hat g_{\Delta}^{(1)} (s_1, s) = \hat g^{(1,1)}_{\Delta}(s_1, s) + O(e^{- \frac{4}{3} s^{3/2}}) \\
&& \hat g_{\Delta, \lambda}^{(1,1)} (s_1, s)  \equiv   \hat Z_{n_s^\gamma=1, n_s^\mu=1}(s_1,s)
\eea
such that
\begin{equation} \label{deriv4}
P^{(1,1)}_{\Delta }(\sigma_1,\sigma)  = \left( \partial_{\sigma_1} \partial_{\sigma} -  {\Delta^{- 1/3}} \partial^{2}_{\sigma} \right) \hat g^{(1,1)}_{\Delta}(s_1,s) \sim O(e^{- \frac{4}{3} \sigma_1^{3/2} - \frac{4}{3} \sigma^{3/2} })
\end{equation}
is the leading term in the tail of the JPDF in both variables. The calculation of this
particular term will be performed independently in Appendix \ref{sec:Z11}.
\section{A one-string to multiple-string form factor in the \texorpdfstring{$\delta-$}{delta-}Bose Gas}

\label{ff}

Consider now the form factor $F^{n_2;n_1+n_2}_{\mu;\gamma } = F^{n_2;n_1+n_2}_{{\bf p} , {\bf m}^\mu;{\bf q} , {\bf m}^\gamma }$ introduced in (\ref{momform1}), and evaluated on two string states such that $|\gamma \rangle$ contains a single string,
i.e $n_s^\gamma=1$ and $|\gamma \rangle = |q,n_1+n_2 \rangle$ while the second string state $| \mu  \rangle = |{\bf p}, {\bf m} \rangle$
is an arbitrary string state with $n_s^\mu \geq 1$. Clearly the strings in the state  $| \mu  \rangle $ must satisfy the constraint $\sum_{j=1}^{n_s^\mu} m_j=n_2$.

It is shown in Appendix \ref{app:ff} that this form factor admits a relatively simple expression, namely
\begin{align} \label{ffformula}
& F^{n_2;n_1+n_2}_{{\bf p} , {\bf m}^\mu;q , n_1+n_2 }
\nn \\& =  n_2! (n_1+n_2)! (2 w + n_1)_{n_2}
\prod_{j=1}^{n_s} \frac{1}{(w + i (q - p_j) + \frac{n_1+n_2-m_j}{2})_{m_j}
(w - i (q - p_j) + \frac{n_1+n_2-m_j}{2})_{m_j}}
\end{align}
Its calculation turns out to be very similar to the one of the overlap in the case of KPZ with a
stationary initial condition for KPZ. Interestingly enough, as discussed later,
the stationary KPZ will also play an important role in the physics of the present problem.

\section{Calculation of the tail of the two-time JPDF}
\label{sec:calc}

\subsection{Aim and starting formula}

Here our aim is to calculate the function $\hat g^{(1)}_{\Delta, \lambda}(s_1, s)$
defined in (\ref{hatg1}). As discussed in Section \ref{sec:gener2} it is a partial sum over states, which contains
only contribution of single-string states $|\gamma \rangle=|q,n_1+n_2 \rangle$, i.e. $n_s^\gamma=1$,
with arbitrary string states $|\mu \rangle$ and we will denote here for simplicity
$n_s^\mu \equiv n_s$ and $m_j^\mu \equiv m_j$.
We first obtain expressions valid for arbitrary $\lambda$, then we
study large $\lambda$, i.e. calculate $\hat g^{(1)}_{\Delta}(s_1, s)$, which
contains the desired result about the tail of the JPDF via (\ref{deriv3}) . We recall that it is given as a series in
increasing number of strings $n_s$ as
\bea \label{hatg1new}
\hat g_{\Delta, \lambda}^{(1)} (s_1, s)  \equiv     \sum_{n_s \geq 1}  \frac{1}{n_s !}  \hat Z_{1, n_s}(s_1,s)
\eea
where from now on we use the
shorthand notation $\hat Z_{n_s^\gamma=1, n_s^\mu}(s_1,s) \equiv \hat Z_{1, n_s}(s_1,s)$.
Using (\ref{Znsns}) and the expression for the form factor (\ref{ffformula}) we obtain our starting formula
for the partition sums at fixed $n_s$ as
\begin{align}
& \hat Z_{1,n_s}(s_1,s)
= \sum_{n_1 \geq 1,n_2 \geq 1}\frac{(-1)^{n_1 + n_2}}{n_1!  n_2!}  \prod_{j=1}^{n_s} \sum_{m_j \geq 1}
\int \frac{d p_j}{4\pi \Delta^{1/3} \lambda m_j} \delta_{\sum_{j=1}^{n_s} m_j,n_2} \Phi({\bf p}/(2 \Delta^{1/3} \lambda),{\bf m})
\nn \\&
\int \frac{d q}{4\pi \lambda (n_1+n_2)}   e^{-\lambda s_1 (n_1 + n_2) - \Delta^{1/3} \lambda s n_2}
e^{  \frac{1}{3} \Delta \lambda^3 \sum_{j=1}^{n_s} m_j^3   -  \Delta^{1/3} \lambda  m_j p_j^2  + \frac{1}{3} \lambda^3
(n_1+n_2)^3  - \lambda (n_1+n_2) q^2 } \nn \\
& \times e^{ i \lambda \Delta^{1/3} \tilde X \sum_{j=1}^{n_s} p_j m_j} n_2! (n_1+n_2)! (2 w + n_1)_{n_2}
\left|\prod_{j=1}^{n_s} \frac{1}{(w + i (\frac{q}{2 \lambda} - \frac{p_j}{2 \Delta^{1/3} \lambda}) + \frac{n_1+n_2-m_j}{2})_{m_j} }  \right|^2 \label{start2}
\end{align}
where we have set $X_1=0$ and $X_2=X$ with no loss of generality. We have defined the scaled position
\bea
\tilde X = X/(2 \Delta^{2/3} \lambda^2)
\eea
replaced $t_1= 4 \lambda^3$ everywhere, and then performed the
rescalings $p_j \to p_j/(2 \Delta^{1/3} \lambda)$, $q \to q/(2 \lambda)$, convenient for later use.

The challenge is now to simplify $\hat Z_{1,n_s}(s_1,s)$ so as to be able to perform the sum.
This will be done below in the limit of large time $\lambda \to +\infty$.
Before doing so let us point out that the single string term $\hat Z_{1,1}(s_1,s)$ is somewhat
easier to evaluate, and its detailed calculation is given in \ref{sec:Z11}. It will be
checked that it can be recovered from the full calculation in the limit of large and positive $s$. In addition,
as discussed above and below, it contains information about the double tail of the JPDF.

\subsection{Calculation of \texorpdfstring{$\hat Z_{1n_s}(s_1,s)$}{FIXME} }

One of the main difficulty in performing the sums in (\ref{start2}) is that, due to the
Kronecker delta, $\delta_{n_2,\sum_{j=1}^{n_s} m_j}$, the sum over the $m_j$ is constrained
and not independent from the sum over $n_2$. Alternatively replacing $n_2$ by
$\sum_{j=1}^{n_s} m_j$ in the expression for the form factor lead
to a complicated coupling of the different strings, making then impossible to sum over all the possible string lengths. To perform the decoupling we use the identity
\begin{equation}
\delta_{\sum_{j=1}^{n_s} m_j,n_2}  = \oint   \frac{dz}{2 \pi i z} \:  z^{  n_2  -  \sum_{j=1}^{n_s} m_j}
\end{equation}
encircling the zero in the positive sense.  We can now invoke the expression for the Cauchy determinant and the consequent identities
used in previous works \cite{we} to decouple the strings in the inverse norm factor
\begin{align}
& \left( \prod_{j=1}^{n_s} \frac{1}{2 \Delta^{1/3} \lambda m_j} \right) \Phi({\bf p}/(2 \Delta^{1/3} \lambda),{\bf m})  = \det_{i,j=1}^{n_s} \frac{1}{i (p_i - p_j) + \Delta^{1/3} \lambda (m_i + m_j)}\nn \\& = \prod_{j=1}^{n_s}\int_0^\infty du_j  \det_{i,j=1}^{n_s} e^{i p_j (u_i - u_j) - \Delta^{1/3} \lambda(u_i + u_j) m_j }
\end{align}
where a set of $n_s$ real (positive) auxiliary variables has been introduced. Using properties of
determinants upon line or column multiplication all summations over $m_j$ and integrations over $p_j$ can be brought inside the determinant a, using the identity
\bea
&& \prod_{j=1}^{n_s} \sum_{m_j \geq 1} \int \frac{d p_j ~ z^{-m_j} }{2\pi}
\det_{i,j=1}^{n_s} e^{i p_j (u_i - u_j) - \Delta^{1/3} \lambda(u_i + u_j) m_j }\nn \\
&&
=
\det_{i,j=1}^{n_s} \sum_{m \geq 1} \int \frac{d p ~ z^{-m} }{2\pi}
e^{i p (u_i - u_j) - \Delta^{1/3} \lambda(u_i + u_j) m }
\eea

This leads to
\begin{align}
 \hat Z_{1n_s}(s_1,s)
=&  \sum_{n_1 \geq 1,n_2 \geq 1} \frac{(-1)^{n_1 }  }{n_1!n_2!} \oint   \frac{dz}{2 \pi i z^{1-n_2}}
\int \frac{d q}{4\pi \lambda (n_1+n_2)} \nn   \\
& \times  e^{- \lambda s_1 (n_1 + n_2)  }
e^{ \frac{1}{3} \lambda^3
(n_1+n_2)^3  - \lambda (n_1+n_2) q^2 }  n_2! (n_1+n_2)! (2 w + n_1)_{n_2}
\nn \\& \times  \prod_{j=1}^{n_s}
\int_{0}^\infty  du_j    \det K_{n_1,n_2,q,z}(u_i,u_j) \label{hatZres}
\end{align}
where the Kernel is given by
\begin{align}
 & K_{n_1,n_2,q,z}(u_i,u_j) =  { \theta(u_i)  \theta(u_j) }{}  \sum_{m \geq 1}   (-1)^m
\int \frac{d p}{2\pi   } e^{ \frac{1}{3} \Delta \lambda^3 m^3 - \Delta^{1/3} \lambda m p^2
- \Delta^{1/3} \lambda (u_i + u_j)  m } \nn \\&\times  e^{i p(u_i - u_j)}         z^{-m} e^{ - \Delta^{1/3} \lambda s  m }
e^{ i \lambda \Delta^{1/3} \tilde X  p m}  \:\:  \nn  \\& \times
\frac{\Gamma \left(-\frac{m}{2}+w+\frac{n_1}{2}+\frac{n_2}{2}+\frac{i q}{2 \lambda}-\frac{i p}{2 \Delta^{1/3} \lambda}\right) \Gamma \left(-\frac{m}{2}+w+\frac{n_1}{2}+\frac{n_2}{2}-\frac{i q}{2 \lambda}+\frac{i p}{2 \Delta^{1/3} \lambda}\right)}{\Gamma \left(\frac{m}{2}+w+\frac{n_1}{2}+\frac{n_2}{2}+\frac{i q}{2 \lambda}-\frac{i p}{2 \Delta^{1/3} \lambda}\right) \Gamma \left(\frac{m}{2}+w+\frac{n_1}{2}+\frac{n_2}{2}-\frac{i q}{2 \lambda}+\frac{i p}{2 \Delta^{1/3} \lambda}\right)} \label{defKnnqz}
\end{align}
where we have used the representation of the Pochhammer symbols $(x)_n$ in terms of
Gamma functions, i.e. $(x)_n = \Gamma(x+n)/\Gamma(x)$. Note that this
representation allows for an analytic continuation to complex $n$, which is
used later.

It turns out that this kernel has a very similar structure as the kernel obtained in the
study of stationary KPZ in \cite{SasamotoStationary}. We now make this more
apparent by providing an equivalent representation involving the deformed Airy function
$\Ai_{\Gamma}^{\Gamma}(a,b,c,d)$ introduced in \cite{SasamotoStationary}.
This function is defined as \footnote{For convenience, we use an integration variable $z=-i z'$ where $z'$ denotes the one used in \cite{SasamotoStationary}.}
\begin{align}
\Ai_{\Gamma\Gamma}^{\Gamma\Gamma}\left(a,b,c,d\right)
=
\frac{1}{2\pi i }
\int_{\Gamma_{\frac{c,d}{b}}}dz e^{-az+z^3/3}
\frac{\Gamma(-bz+c)\Gamma(- bz+d)}{\Gamma(bz+c)\Gamma(bz+d)}
\label{AiGG}
\end{align}
with $\Gamma_{\frac{c,d}{b}}$ goes from $-i\infty$ to $+ i \infty$ passing on the left of the poles of the Gamma functions, e.g. a possible choice being a straight line from
$-i\infty + \alpha$ to $+ i \infty + \alpha$ with $0 \leq \alpha< Re [c/b]$ and $0 \leq \alpha < Re [d/b]$. Note that
the condition $\alpha \geq 0$ is necessary in order for the integral to be convergent
\footnote{Note also that $b>0$ has been assumed (so that all poles of the Gamma function are then to the right of the contour), a condition which will always be satisfied here.}.

Our kernel can be equivalently written in terms of this deformed Airy function using the generalization of
the Airy trick introduced in \cite{SasamotoStationary}, which is
valid for $p \in \mathbb{R}$, $m, \Delta^{1/3} \lambda \geq 0$  and $a,b > n/2$, namely
\begin{align}
& \frac
{\Gamma\left(-ip+a-\frac{m}{2}\right)\Gamma\left(ip+b-\frac{m}{2}\right)}
{\Gamma\left(-ip+a+\frac{m}{2}\right)\Gamma\left(ip+b+\frac{m}{2}\right)}
e^{\frac{1}{3} \Delta \lambda^3 m^3}
  =\int_{-\infty}^{\infty}dy
\Ai_{\Gamma\Gamma}^{\Gamma\Gamma}
\left(y,\frac{1}{2 \Delta^{1/3} \lambda},-ip+a,ip+b\right) e^{m y \Delta^{1/3} \lambda},
\end{align}
Inserting in (\ref{defKnnqz}), our kernel becomes
\begin{align}
 K_{n_1,n_2,q,z}(u_i,u_j) &  =   \theta(u_i)  \theta(u_j)
\int_{-\infty}^{\infty}dy  \int \frac{d p}{2\pi   }  e^{i p(u_i - u_j)}
\nn \\ & \Ai_{\Gamma\Gamma}^{\Gamma\Gamma}
\left(y,\frac{1}{2 \Delta^{1/3} \lambda},-\frac{ip}{2 \Delta^{1/3} \lambda}+c_+,\frac{ip}{2 \Delta^{1/3} \lambda}+c_-\right) \nn \\&   \times  \sum_{m \geq 1}   (-1)^m    e^{ m y \Delta^{1/3} \lambda
- \Delta^{1/3} \lambda m p^2  -  \Delta^{1/3} \lambda (u_i + u_j) m   -  \Delta^{1/3} \lambda m s  +
i \lambda \Delta^{1/3} \tilde X  p m}
z^{-m}
\label{defKnnqz2}
\end{align}
where we have defined the following coefficients
\begin{equation} \label{cpm}
c_\pm =  w+\frac{n_1}{2}+\frac{n_2}{2} \pm \frac{i q}{2 \lambda}
\end{equation}

This expression allows to sum over the string length $m \geq 1 $, after performing
the shift $y \to y + p^2 + u_i  + u_j + s - i \tilde X p$ \footnote{for $\tilde X \neq 0$ one can
check that the complex shift is legitimate, as in Eq. (68) of \cite{PLDCrossoverDropFlat}}
, leading to
\begin{align}
  K_{n_1,n_2,q,z}(u_i,u_j) &  =  -  \theta(u_i)  \theta(u_j)
\int_{-\infty}^{\infty}dy  \int \frac{d p}{2\pi   }    \frac{e^{i p(u_i - u_j)}   }{1 + e^{- y \Delta^{1/3} \lambda }z}
\nn \\& \Ai_{\Gamma\Gamma}^{\Gamma\Gamma}
\left(y + p^2 + u_i  + u_j + s  - i \tilde X p ,\frac{1}{2 \Delta^{1/3} \lambda},-\frac{ip}{2 \Delta^{1/3} \lambda}+c_+,\frac{ip}{2 \Delta^{1/3} \lambda}+c_-\right)    \label{defKnnqz3}
\end{align}

An equivalent form for the kernel can be given, which involves now two (simpler) deformed Airy functions.
These are defined as follows
\begin{equation}
\Ai_\Gamma^\Gamma (a,b,c,d) = \frac{1}{2 \pi i } \int_{\Gamma_{ d/b}} e^{-z a +  z^3/2} \frac{\Gamma(-b z + d)}{\Gamma( b z + c)}
\end{equation}
with similar contours $\Gamma_{\frac{d}{b}}$ as in \eqref{AiGG} that goes from $-i\infty$ to $+ i \infty$ passing on the left of the
poles of the Gamma functions.
We can now use the doubling formula given in \cite{SasamotoStationary}, which generalizes a similar standard
identity for Airy functions. It states that, for $a,b,x\in \mathbf{R}$ and $w > 0$, one has
\begin{align}\label{decomposition_gamma}
&~\int_{-\infty}^{\infty} \frac{dp}{2 \pi}
\Ai_{\Gamma\Gamma}^{\Gamma\Gamma}
\left(p^2+v- i \tilde X p,w,-iwp+a,iwp+b\right)e^{ipx}\notag\\
&=
\frac{1}{2^{\frac{1}{3}}}
\Ai_\Gamma^\Gamma\left(2^{-\frac{2}{3}}(\tilde v +x),2^{\frac{1}{3}}w,\tilde a,\tilde b
-\frac{w \tilde X}{2}\right)
\Ai_\Gamma^\Gamma\left(2^{-\frac{2}{3}}(\tilde v -x),2^{\frac{1}{3}}w,\tilde b,\tilde a\right)
e^{- \frac{1}{2} x \tilde X}, \\
& \tilde v=v+\frac{\tilde X^2}{4}  \quad , \quad \tilde a=a+\frac{w \tilde X}{2} \quad , \quad \tilde b =
b-\frac{w \tilde X}{2}
\end{align}
where $\Ai_\Gamma^\Gamma$ is defined above. We have slightly generalized this
identity using a change of variable $p \to p+ i \tilde X/2$ followed by a shift of contour
back to the real axis, thereby generalizing the Eq.(83) of \cite{PLDCrossoverDropFlat}
for standard Airy functions.
We can now use the relation \eqref{decomposition_gamma} to integrate over $p$
in (\ref{defKnnqz3}) and obtain
\begin{align}
 & K_{n_1,n_2,q,z}(u_i,u_j) =  -  2^{-1/3 }\theta(u_i)  \theta(u_j)
\int_{-\infty}^{\infty}dy
\Ai_{\Gamma}^{\Gamma}
\left(2^{-2/3}(y  + 2 u_i+ \tilde s) ,\frac{2^{1/3}}{2 \Delta^{1/3} \lambda}, \tilde c_+, \tilde c_-\right)   \nn\\& \times   \Ai_{\Gamma}^{\Gamma}
\left(2^{-2/3}(y  + 2 u_j +\tilde s ),\frac{2^{1/3}}{2 \Delta^{1/3} \lambda}, \tilde c_-, \tilde c_+\right)
\frac{1}{1 + e^{- y \Delta^{1/3} \lambda }z}  \label{defKnnqz4}
\end{align}
with, using (\ref{cpm}),
\bea
&& \tilde c_\pm = w + \frac{n_1}{2} + \frac{n_2}{2} \pm \frac{i q}{2 \lambda} \pm \frac{\tilde X}{4 \Delta^{1/3} \lambda}
\quad , \quad \tilde s = s + \frac{\tilde X^2}{4}
\eea
Note that the kernel in (\ref{defKnnqz4}) should be multiplied
by the factor $e^{- \frac{1}{2} (u_i-u_j) \tilde X}$. However it is easy to check that this
factor is immaterial when calculating the Fredholm determinant, hence we omit it and we reintroduce it only at the very end of the calculation.

This kernel is thus very similar to the one for the stationary case, e.g. $\bar K$ in (4.22) in
\cite{SasamotoStationary}. The parameter $\gamma_t \sim t^{1/3}$ there, is replaced by $2^{2/3} \Delta^{1/3} \lambda=(t_2-t_1)^{1/3}$ here. The drifts
$v,w$ there are replaced by $\tilde c_+,\tilde c_-$. There is however
an important difference. In the present case, there are additional degrees of freedom, $i q $
and $n_1 + n_2$ (in $\tilde c_+,\tilde c_-$) that need to be summed over.


\subsection{Calculation of \texorpdfstring{$\hat g_{\Delta, \lambda}^{(1)} (s_1, s)$}{FIXME}: Fredholm determinant and
summation over the number of strings}

To recapitulate, we have obtained an expression
for $\hat Z_{1n_s}(s_1,s)$, Eq. (\ref{hatZres}), in terms of a $n_s$-uple integral over auxiliary variables
$u_j$, of the determinant of a kernel, $K_{n_1,n_2,q,z}$, and also involving ``external" integration (or summation) variables
$n_1,n_2,q,z$. We obtained three equivalent expressions for this kernel. It is thus now possible to formally sum over $n_s$ to obtain a representation for $\hat g_{\Delta, \lambda}^{(1)} (s_1, s)$ involving a
Fredholm determinant
\bea \label{startDet}
\fl && \hat g_{\Delta, \lambda}^{(1)} (s_1, s) = \sum_{n_s \geq 1} \frac{1}{n_s!} \hat Z_{1n_s}(s_1,s)
= \sum_{n_1 \geq 0,n_2 \geq 1} \frac{(-1)^{n_1 }  }{n_1!n_2!} \oint   \frac{dz}{2 \pi i z^{1-n_2}}
\int \frac{d q}{4\pi \lambda (n_1+n_2)} \nn   \\
\fl && \times  e^{- \lambda s_1 (n_1 + n_2)  }
e^{ \frac{1}{3} \lambda^3
(n_1+n_2)^3  - \lambda (n_1+n_2) q^2 }  n_2! (n_1+n_2)! (2 w + n_1)_{n_2} {\rm Det}[ I - K_{n_1,n_2,q,z}]
\eea

It is now possible to choose (or move) the integration contour of $z$ to encircle the negative real axis to
obtain an equivalent formula. The details of the procedure are given in the Appendix
\ref{app:sums}, where more general considerations about similar constrained sums
are also discussed. As a result we obtain
\begin{align}
& \hat g_{\Delta, \lambda}^{(1)} (s_1, s)
= \sum_{n_1 \geq 1,n_2 \geq 1}  {(-1)^{n_1+  n_2 }  }{}
\int \frac{d q}{4\pi \lambda (n_1+n_2)} \nn  \\
& \times  e^{- \lambda s_1 (n_1 + n_2)  }
e^{ \frac{1}{3}  \lambda^3
(n_1+n_2)^3  - \lambda (n_1+n_2) q^2 }   \frac{(n_1+n_2)!}{n_1!}  \frac{\Gamma(2 w + n_1 + n_2)}{\Gamma(2 w +n_1)}
   \nn \\&  \times \int_{-\infty}^{\infty} d\kappa e^{- \Delta^{1/3} \lambda \kappa n_2} \left( {\rm Det}[ I + B_{n_1,n_2,q,\kappa} + P_{n_1,n_2,q,\kappa} ] - {\rm Det}[I +  B_{n_1,n_2,q,\kappa}] \right)
   \label{Det2}
\end{align}

with
\begin{align}
 & B_{n_1,n_2,q,\kappa}(u_i,u_j) =  2^{-1/3 }\theta(u_i)  \theta(u_j)
\nn \\& PV \int_{-\infty}^{\infty}dy
\frac{\Ai_{\Gamma}^{\Gamma}
\left(2^{-2/3}(y    + 2 u_i+\tilde s) ,\frac{2^{1/3}}{2 \Delta^{1/3} \lambda}, \tilde c_+, \tilde c_-\right)    \Ai_{\Gamma}^{\Gamma}
\left(2^{-2/3}(y  + 2 u_j +\tilde s),\frac{2^{1/3}}{2 \Delta^{1/3} \lambda}, \tilde c_-, \tilde c_+\right) } {e^{- (y + \kappa) \Delta^{1/3} \lambda  }-1}
\end{align}

\begin{align}
  P_{n_1,n_2,q,\kappa}(u_i,u_j)  & =    2^{-1/3 }\theta(u_i)  \theta(u_j)
  \int_{-\infty}^{\infty}dy
\Ai_{\Gamma}^{\Gamma}
\left(2^{-2/3}(y + 2 u_i+\tilde s) ,\frac{2^{1/3}}{2 \Delta^{1/3} \lambda}, \tilde c_+, \tilde c_-\right) \nn \\&    \times  \Ai_{\Gamma}^{\Gamma}
\left(2^{-2/3}(y  + 2 u_j +\tilde s),\frac{2^{1/3}}{2 \Delta^{1/3} \lambda}, \tilde c_-, \tilde c_+\right)     \delta  (y + \kappa)
\end{align}
where we recall 
$\tilde c_\pm = w + \frac{n_1+n_2}{2} \pm \frac{i q}{2 \lambda} \pm \frac{\tilde X}{4 \Delta^{1/3} \lambda}$
and $\tilde s = s + \frac{\tilde X^2}{4}$.
These expressions are valid for arbitrary times $t_1,t_2$, i.e. arbitrary $\lambda,\Delta$.
The remaining summations over $n_1,n_2$ still involves a cubic exponential which we
can now eliminate using the standard Airy trick
\bea
e^{ \frac{1}{3}  \lambda^3 (n_1+n_2)^3 } = \int dy Ai(y_1) e^{\lambda (n_1+n_2) y_1}
\eea
For convenience we also eliminate one denominator using the identity $1/(\lambda (n_1+n_2))=\int_0^{+\infty} dv
e^{- v \lambda (n_1+n_2)}$. Upon shifting $y_1 \to y_1 +s_1+q^2+ v$ we obtain
\bea
&& \hat g_{\Delta, \lambda}^{(1)} (s_1, s)
=  \int \frac{d q}{4\pi}  \int dy_1 \int_0^{+\infty} dv \Ai(y_1 + s_1+q^2+ v) \nn \\
&& \times
\sum_{n_1 \geq 1,n_2 \geq 1}  (-1)^{n_1+  n_2 }
e^{- \lambda y_1 (n_1 + n_2)  }  \frac{(n_1+n_2)!}{n_1!}  \frac{\Gamma(2 w + n_1 + n_2)}{\Gamma(2 w +n_1)} \\
&&
\times  \int_{-\infty}^{\infty} d\kappa e^{- \Delta^{1/3} \lambda \kappa n_2} \left( {\rm Det}[ I + B_{n_1,n_2,q,\kappa} + P_{n_1,n_2,q,\kappa} ] - {\rm Det}[I +  B_{n_1,n_2,q,\kappa}] \right) \nn
   \label{Det3}
\eea

This is our final result for arbitrary times, i.e. arbitrary $\lambda$ and $\Delta$. In the next
Section we will give an equivalent expression where summations over $n_1,n_2$ are
replaced by contour integrals. Further progress on these formula will then be achieved, but
only in the limit of large times, i.e. $\lambda \to \infty$.

\subsection{Large times limit of \texorpdfstring{$\hat g_{\Delta, \lambda}^{(1)} (s_1, s)$}{FIXME}  }

We now use the so-called Mellin-Barnes transformation of the sums over $n_1$ and $n_2$. The
basic identity is
\begin{equation}
\sum_{n_1 \geq 1,n_2 \geq 1}  (-1)^{n_1+n_2} f(n_1,n_2) = \int_{C_1} \frac{dz_1}{2 i } \int_{C_2} \frac{dz_2}{2
i } \frac{1}{\sin(\pi z_1)} \frac{1}{\sin(\pi z_2)} f(z_1,z_2)
\end{equation}
where each contour encircles all the solutions of $\sin(\pi z) = 0$ with $Re(z) >0$, i.e.
all strictly positive integers. As such it is strictly true for a function $f(z_1,z_2)$ which is analytic near each of these
points. The usual use of this formula is then to move these contours parallel
to the imaginary axis as $C_j = a + i \mathbb{R}$ with $0<a<1$, and we
will assume in the following that this can be done.

It is useful for the following to rescale the new variables $z_1,z_2$ by $1/\lambda$, and to define a rescaled
slope $\tilde w$ as
\begin{equation}
w = \frac{\tilde{w}}{\lambda \Delta^{1/3}}
\end{equation}
We then obtain

\bea
\fl && \hat g_{\Delta, \lambda}^{(1)} (s_1, s)
=  \int \frac{d q}{4\pi}  \int dy_1 \int_0^{+\infty} dv \Ai(y_1 + s_1+q^2+ v) \nn \\
\fl && \times \int_{C_1} \frac{dz_1}{2 i } \int_{C_2} \frac{dz_2}{2 i } \frac{1}{\lambda \sin(\pi z_1/\lambda)} \frac{e^{- y_1 (z_1+z_2)} }{\lambda \sin(\pi z_2/\lambda)}  \frac{\Gamma( \frac{z_1}{\lambda} + \frac{z_2}{\lambda} + 1)}{\Gamma (\frac{z_1}{\lambda }+ 1)} \frac{\Gamma\left( \frac{2  \tilde{w} }{ \lambda \Delta^{1/3} }  + \frac{z_1 + z_2}{\lambda }\right)}{\Gamma\left( \frac{2  \tilde{w} }{ \lambda \Delta^{1/3} }  + \frac{z_1 }{\lambda }\right) }
 \nn \\
 \fl &&
 \times  \int_{-\infty}^\infty d\kappa e^{-  \kappa z_2 \Delta^{1/3}}  \left( {\rm Det} [ I + B_{q,z_1/\lambda , z_2/\lambda, \kappa}  + P_{q,z_1/\lambda,z_2/\lambda,\kappa} ]  -  {\rm Det}  [ I +   B_{q,z_1/\lambda,z_2/\lambda,\kappa}] \right)
\eea

\begin{align}
 & B_{q,z_1/\lambda , z_2/\lambda, \kappa}(u_i,u_j) =  2^{-1/3 }\theta(u_i)  \theta(u_j)
PV \int_{-\infty}^{\infty}dy   \frac{1}{e^{- (y + \kappa) \Delta^{1/3} \lambda  }-1}
\nn \\& \Ai_{\Gamma}^{\Gamma}
\left(2^{-2/3}(y + \tilde s  + 2 u_i) ,\frac{2^{1/3}}{2 \lambda \Delta^{1/3}}, \frac{\tilde{w}+ \frac{\tilde X}{4}}{\lambda \Delta^{1/3}}+\frac{z_1}{2 \lambda}+\frac{z_2}{2 \lambda}+\frac{i q}{2 \lambda}, \frac{\tilde{w} -  \frac{\tilde X}{4}}{\lambda \Delta^{1/3}}+\frac{z_1}{2 \lambda}+\frac{z_2}{2 \lambda}-\frac{i q}{2 \lambda}\right)  \nn  \\&
 \Ai_{\Gamma}^{\Gamma}
\left(2^{-2/3}(y + \tilde s  + 2 u_j) ,\frac{2^{1/3}}{2 \lambda \Delta^{1/3}}, \frac{\tilde{w} - \frac{\tilde X}{4}}{\lambda \Delta^{1/3}}+\frac{z_1}{2 \lambda}+\frac{z_2}{2 \lambda}-\frac{i q}{2 \lambda}, \frac{\tilde{w} +\frac{\tilde X}{4}}{\lambda \Delta^{1/3}}+\frac{z_1}{2 \lambda}+\frac{z_2}{2 \lambda}+\frac{i q}{2 \lambda}\right)
\end{align}

\begin{align}
 & P_{q,z_1/\lambda , z_2/\lambda, \kappa}(u_i,u_j) =  2^{-1/3 }\theta(u_i)  \theta(u_j)
PV \int_{-\infty}^{\infty}dy  \delta(y + \kappa)
\nn \\&
\Ai_{\Gamma}^{\Gamma}
\left(2^{-2/3}(y + \tilde s  + 2 u_i) ,\frac{2^{1/3}}{2 \lambda \Delta^{1/3}}, \frac{\tilde{w} + \frac{\tilde X}{4}}{\lambda \Delta^{1/3}}+\frac{z_1}{2 \lambda}+\frac{z_2}{2 \lambda}+\frac{i q}{2 \lambda}, \frac{\tilde{w} - \frac{\tilde X}{4}}{\lambda \Delta^{1/3}}+\frac{z_1}{2 \lambda}+\frac{z_2}{2 \lambda}-\frac{i q}{2 \lambda}\right)    \nn \\&
 \Ai_{\Gamma}^{\Gamma}
\left(2^{-2/3}(y + \tilde s  + 2 u_j) ,\frac{2^{1/3}}{2 \lambda \Delta^{1/3}}, \frac{\tilde{w}- \frac{\tilde X}{4}}{\lambda \Delta^{1/3}}+\frac{z_1}{2 \lambda}+\frac{z_2}{2 \lambda}-\frac{i q}{2 \lambda}, \frac{\tilde{w}+ \frac{\tilde X}{4}}{\lambda \Delta^{1/3}}+\frac{z_1}{2 \lambda}+\frac{z_2}{2 \lambda}+\frac{i q}{2 \lambda}\right)
\end{align}

So until now, the formula are still valid for arbitrary times, i.e. arbitrary $\lambda$ and $\Delta$.

We will now take the limit of large times $t_1,t_2 \to \infty$, at fixed $\Delta$, i.e. the
limit of large $\lambda$ in the above formula.
A first simplification which occurs in that limit is that:

\bea
&&\!\!\!\!\!\! \lim_{\lambda \to \infty} \Ai_\Gamma^\Gamma (a,b/\lambda,c/\lambda,d/\lambda) =  \lim_{\lambda \to \infty} \frac{1}{2 \pi i }\int_{\Gamma_{ d/b}} e^{- a z +  z^3/3} \frac{\Gamma(- b z/\lambda + d/\lambda)}{\Gamma(  b z/\lambda + c/\lambda)}\nn \\
&& = \Ai^{+}_{-} \left(a ,c/b ,d/b \right) \equiv
\frac{1}{2\pi i }
\int_{\Gamma_{d/b}}dz e^{- a z+z^3/3}
\frac{z+ \frac{c}{b}  }{-z+ \frac{d}{b}}
\eea
where we recall that a choice for the contour $\Gamma_{d/b}$ is a straight line from
$-i\infty + \alpha$ to $+ i \infty + \alpha$ with $0 \leq \alpha < Re [d/b]$.

In addition in the denominators $\lambda \sin(\pi z_i/ \lambda) \to \pi z_i$ and the
contours of integration become $C_{1,2} \to C= \epsilon + i \mathbb{R}$ with
$\epsilon>0$ infinitesimal, i.e. they pass to the right of zero.
We thus obtain (using again the notation $\sigma=2^{-2/3} \tilde s$ and $\sigma_1=2^{-2/3} s_1$)
\begin{align}
&\hat g_{\Delta}^{(1)} (s_1, s) = \lim_{\lambda \to \infty}  \hat g_{\Delta, \lambda}^{(1)} (s_1, s)
=  \int \frac{d q}{4\pi   } \int_0^\infty dv  \int  {dy_1}{}   \Ai \Big( y_1 + q^2 + 2^{2/3} \sigma_1   + v  \Big)  \nn \\& \times  \int_{-i\infty +  \epsilon }^{i\infty +  \epsilon}  \frac{dz_1}{2 \pi i  } \int_{-i\infty +  \epsilon }^{i\infty +  \epsilon} \frac{dz_2}{2 \pi i }
   e^{y_1 (z_1 + z_2)}    \frac{1}{  z_2  }  \frac{z_1 + 2 \frac{\tilde{w}}{\Delta^{1/3}}}{z_1 (z_1 + z_2 + 2 \frac{\tilde{w}}{\Delta^{1/3}})}
\nn \\&\times  \int_{-\infty}^\infty d\kappa e^{  \kappa z_2 (4\Delta)^{1/3}}   \partial_{\kappa} {\rm  Det} [I - P_{ \kappa+\sigma}\mathcal{F}^{\tilde{w}}_{ z_1 + z_2 , q} P_{  \kappa+ \sigma} ]
\end{align}
Here we have performed manipulations as in the Appendix, going there from (\ref{appres1})
to (\ref{appres2}) where we have changed $\kappa \to -\kappa$ and then
the integration variable inside the kernel $y \to y - \sigma$. For convenience
we define
\bea
&& \hat w = 2^{2/3} \tilde w \quad , \quad \hat X =  2^{2/3} \frac{\tilde X}{4} = \frac{X}{2 (\Delta t)^{2/3}} \\
&& \hat w_\pm =\hat w \pm \hat X
\eea
The new kernel is given by (see the analogous formula (\ref{kinfty}) of the
Appendix)
 \begin{align} \label{Fw}
 &  \mathcal{F}^{\tilde{w}}_{ z,q}(u_i,u_j)    \\
   & =
  \int_{0}^{\infty}dy
\Ai^{+}_{-}
\left( y +   u_i  ,  \hat{w}_+ +  (4\Delta)^{1/3} \left(\frac{z}{2} + i \frac{q}{2  }\right), \hat{w}_- +  (4\Delta)^{1/3} \left(\frac{z}{2} - i \frac{q}{2  }\right) \right)  \nn \\&  \times \Ai^{+}_{-}
\left( y +   u_j ,   \hat{w}_- + (4\Delta)^{1/3}\left(\frac{z}{2} - i \frac{q}{2  }\right),\  \hat{w}_+ +  (4\Delta)^{1/3}\left(\frac{z}{2} + i \frac{q}{2  }\right) \right) \nn
 \end{align}
 Note that we have performed a change of variable $y \to 2^{2/3} y$ and a similarity transformation
 in the Fredholm determinant $u_i \to 2^{-1/3} u_i$ (which absorbs also a factor $2^{1/3}$ in
 front of the kernel from the integration measure).

We can now change variables to $z_1 + z_2 \to z$ and $z_2 \to z_2$ in the complex
integral, as well as $\kappa \to \kappa - \sigma$,
which allows to rewrite the expression as
\begin{align}\label{eq:final_largetime}
& \hat g_{\Delta}^{(1)} (s_1, s)  \nn \\ &
=   \int \frac{d q}{4\pi   } \int_0^\infty dv  \int  {dy_1}{} \int_{-\infty}^\infty d\kappa  \Ai \Big( y_1 + q^2 + 2^{2/3} \sigma_1   + v  \Big)
 \nn \\&  \times \int_{-i\infty   + 2\epsilon}^{i\infty   + 2\epsilon}
\frac{dz}{2 \pi i }   \int_{-i\infty  + \epsilon }^{i\infty  + \epsilon } \frac{dz_2}{2 \pi i }
     \frac{   e^{y_1 z + (\kappa - \sigma)z _2 (4\Delta)^{1/3} }   }{z_2 (z - z_2) }
         \left(  \frac{z - z_2 + 2 \tilde{w}\Delta^{-1/3}}{z + 2 \tilde{w}\Delta^{-1/3}} \right)  \partial_{ \kappa}  {\rm Det}[I - P_{ \kappa}\mathcal{F}^{\tilde{w}}_{ z,q} P_{ \kappa} ]
\end{align}

We now take the limit $\tilde{w} \to 0^+$. The case of a finite $\tilde w$ is
studied in the Appendix \ref{app:w}. Here we obtain
\begin{align}
\hat g_{\Delta}^{(1)} (s_1, s) &
=    \int \frac{d q}{4\pi   } \int_0^\infty dv  \int  {dy}{} \int_{-\infty}^\infty d\kappa  \Ai \Big( y + q^2 + 2^{2/3} \sigma_1   + v  \Big)    \nn \\&  \int_{-i\infty   + 2 \epsilon}^{i\infty    + 2  \epsilon}
\frac{dz}{2 \pi i }   \int_{-i\infty  + \epsilon }^{i\infty  + \epsilon } \frac{dz_2}{2 \pi i }
     \frac{   e^{y z + (\kappa - \sigma)z _2 (4\Delta)^{1/3} }   }{z_2 z}
         \partial_{ \kappa}    {\rm Det} [I - P_{ \kappa}\mathcal{F}_{ z,q} P_{ \kappa} ]   \label{new11}
\end{align}
where now
\begin{align}
   \mathcal{F}_{ z,q}(u_i,u_j)  = &
  \int_{0}^{\infty}dy
\Ai^{+}_{-}
\left( y +   u_i , \hat X+ (4\Delta)^{1/3} \left(\frac{z}{2} + i \frac{q}{2  }\right),
- \hat X + (4\Delta)^{1/3} \left(\frac{z}{2} - i \frac{q}{2  }\right) \right)  \nn \\&  \Ai^{+}_{-}
\left( y +   u_j ,   - \hat X + (4\Delta)^{1/3}\left(\frac{z}{2} - i \frac{q}{2  }\right),
\hat X + (4\Delta)^{1/3}\left(\frac{z}{2} + i \frac{q}{2  }\right) \right) \label{Fzq2}
 \end{align}
We note that the last arguments of the kernel are given by $b=\hat X+ (4\Delta)^{1/3} \left(\frac{z}{2} + i \frac{q}{2  }\right)$
and $c=- \hat X + (4\Delta)^{1/3} \left(\frac{z}{2} - i \frac{q}{2  }\right)$, therefore is possible to choose the contours $\Gamma_b,\Gamma_c$
as $ X + \epsilon + i \mathbb{R}$ and $ -X  + i \mathbb{R}$ such that we can integrate over $y$ using that $ \Re [ z_1 +  z_2] >0$
\begin{align}
& \int_0^\infty dy \Ai^{+}_{-} \left(y + u_i ,b ,c \right) \Ai^{+}_{-} \left(y + u_j ,c ,b \right) \nn \\& =
\left( \frac{1}{2\pi i } \right)^2
\int_{-\hat{X}-i\infty}^{-\hat{X}+ i \infty}d\tilde{z}_1 \int_{\hat{X} + \epsilon-i\infty}^{\hat{X} + \epsilon+ i \infty} d\tilde{z}_2  e^{- u_i \tilde{z}_1+\tilde{z}_1^3/3} e^{- u_j \tilde{z}_2+\tilde{z}_2^3/3} \frac{1}{\tilde{z}_1  + \tilde{z}_2}
\frac{( \tilde{z}_1 +b)  }{  (\tilde{z}_1 -c)}  \frac{( \tilde{z}_2 +c)  }{  (\tilde{z}_2 -b)}
\end{align}
and by rearranging
\begin{equation}
\frac{1}{\tilde{z}_1 +  \tilde{z}_2}
\frac{( \tilde{z}_1 +b)  }{  (\tilde{z}_1 -c)}  \frac{( \tilde{z}_2 +c)  }{  (\tilde{z}_2 -b)}  = \frac{1}{\tilde{z}_1 +  \tilde{z}_2}  +  (b +c )  \frac{1}{\tilde{z}_1  - c} \frac{1}{\tilde{z}_2  - b}
\end{equation}
we can rewrite the kernel as
\bea \label{id3}
\fl &&  \int_0^\infty dy \Ai^{+}_{-} \left(y + u_i ,b ,c \right) \Ai^{+}_{-} \left(y + u_j ,c ,b \right) =
K_\Ai(u_i,u_j) + (b+c) {\cal B}_{b}(u_i) {\cal B}_{c}(u_j)
\eea
in terms of the Airy kernel $K_{\rm Ai}$ defined in (\ref{airyK2}).
We have also used the integral representation of the Airy function (with $\alpha \geq 0$)
\bea \label{Airydef}
Ai(x) = \int_{-i\infty  + \alpha}^{+ i \infty + \alpha}  \frac{dz}{2 i \pi} e^{- x z + z^3/3}
\eea
and defined the functions $\mathcal{B}_{a}(x)$ as follows
\bea
&& \mathcal{B}_{a}(x) \equiv  \int_{-i\infty}^{+ i \infty} \frac{dz}{2 i \pi} \frac{1}{a-z}
 e^{- x z + z^3/3} \nn \\
 && =
e^{a^3/3 - x a} - \int_0^\infty dy e^{a y } \Ai(x  + y) \equiv \int_{-\infty}^0 d y e^{y a } \Ai(y  + x )
\eea
The second expression is valid for $Re(a)>0$, which is the domain of interest here, however
the first expression is valid for all $a$. This function can also be determined as the
solution of $(a+\partial_x)  \mathcal{B}_{a}(x)= \Ai(x)$ which vanishes at $x=+\infty$.
Note that (\ref{newexpression}) is formula (6.22)
in  \cite{SasamotoStationary} but here it is extended to the needed complex values of the argument.

Using (\ref{id3}) we can rewrite the above kernel (\ref{Fzq2}) as follows
 \begin{align} \label{newexpression}
  \mathcal{F}_{ z,q}(u_i,u_j)     = &
   K_{\Ai} (u_i,u_j)   + (4\Delta)^{1/3} z  \  \mathcal{B}_{   \hat X+ (4\Delta)^{1/3} \left(\frac{z}{2} + i \frac{q}{2  }\right)}(u_i)  \mathcal{B}_{ - \hat X  +  (4\Delta)^{1/3}\left(\frac{z}{2} - i \frac{q}{2  }\right)} (u_j)
 \end{align}

Since (\ref{newexpression}) involves a rank one projector, one can
rewrite the Fredholm determinant as
\begin{align}
& {\rm Det} [I - P_{\kappa}\mathcal{F}_{ z,q} P_{\kappa} ]   = {\rm Det} [I -  P_{\kappa}K_{\Ai} P_{\kappa}]   \bigg(1 -   (4\Delta)^{1/3} z   \nn \\& \times \int_{\kappa}^\infty du dv \mathcal{B}_{ \hat X +  (4\Delta)^{1/3} \left(\frac{z}{2} + i \frac{q}{2  }\right)}(u) \mathcal{B}_{ - \hat X +  (4\Delta)^{1/3}  \left(\frac{z}{2} - i \frac{q}{2  }\right)}(v) [I- P_\kappa K_{\Ai}  P_\kappa ]^{-1}(u,v)  \bigg)
\end{align}
This expansion generates then two terms which, once inserted in equation \eqref{new11}, can be integrated over $z$ and $z_2$. For the first term we have simply
\begin{align}
&         \int_{-\infty}^\infty  {d\kappa}{} \int_{-i\infty   + 2\epsilon}^{i\infty   + 2\epsilon}
\frac{dz}{2 \pi i }   \int_{-i\infty  + \epsilon }^{i\infty  + \epsilon } \frac{dz_2}{2 \pi i }
     \frac{   e^{y z +  (\kappa - \sigma)z_2 (4\Delta)^{1/3} }   }{z_2 z }
         \partial_\kappa    {\rm Det} [I - P_{\kappa} K_{\Ai} P_{\kappa}] \nn \\&
        =  \int_{-\infty}^\infty  {d\kappa}{ }  \ \theta(y) \theta(\kappa - \sigma) \partial_{\kappa} {\rm Det} [I - P_{\kappa} K_{\Ai} P_{\kappa}]
\end{align}
We now use that
\begin{equation}
\mathcal{B}_{\pm \hat X + (4\Delta)^{1/3}\left(\frac{z}{2} \pm i \frac{q}{2  }\right)}(x)   = \int_{-\infty}^0 d y_1 e^{y_1 (\pm \hat X + (4\Delta)^{1/3} \left(\frac{z}{2} \pm i \frac{q}{2  }\right)) } \Ai(y_1  + x )
\end{equation}
to integrate the second piece as follows
\begin{align}
&  -  \int_{-\infty}^\infty d \kappa    \int_{-i\infty   + 2\epsilon}^{i\infty   + 2\epsilon}
\frac{dz}{2 \pi i }   \int_{-i\infty  + \epsilon }^{i\infty  + \epsilon } \frac{dz_2}{2 \pi i }
     \frac{   e^{y z +  (\kappa - \sigma)z_2 (4 \Delta)^{1/3} }   }{z_2  z } (4\Delta)^{1/3} z
     \partial_\kappa \bigg(  {\rm Det} [I - P_{\kappa}K_{\Ai} P_{\kappa}]
\nn  \\& \times
\int_\kappa^\infty du dv \mathcal{B}_{ \hat X+ (4\Delta)^{1/3} \left(\frac{z}{2} + i \frac{q}{2  }\right)}(u)
\mathcal{B}_{ - \hat X +  (4\Delta)^{1/3}  \left(\frac{z}{2} - i \frac{q}{2  }\right)}(v) [I- P_\kappa K_{\Ai}  P_\kappa ]^{-1}(u,v)
\bigg) \nn\\ =
  &  - (4 \Delta)^{1/3}  \int_{-\infty}^\infty d \kappa   \int_{-i\infty   + 2\epsilon}^{i\infty   + 2\epsilon}
\frac{dz}{2 \pi i }   \int_{-i\infty  + \epsilon }^{i\infty  + \epsilon } \frac{dz_2}{2 \pi i }
     \frac{   e^{y z +  (\kappa - \sigma)z_2 (4 \Delta)^{1/3} }   }{z_2  }
     \nn    \\&  \partial_\kappa  \bigg( {\rm Det} [I - P_{\kappa}K_{\Ai} P_{\kappa}]    \int_\kappa^\infty du dv \int_{-\infty}^0 dy_1 dy_2   e^{y_1 (4 \Delta)^{1/3} \left(\frac{z}{2} - i \frac{q}{2  }\right) }   e^{y_2 (4 \Delta)^{1/3} \left(\frac{z}{2} + i \frac{q}{2  }\right) } e^{\hat X (y_2-y_1)}
        \nn \\&
     \times [1- P_\kappa K_{\Ai}  P_\kappa ]^{-1}(u,v)\Ai(u  + y_1)  \Ai(v  + y_2)  \bigg)
\end{align}
the integration over $z$ gives a $ \delta\left(y + \frac{(4 \Delta)^{1/3}}{2} (y_1 + y_2) \right)$ and the one over $z_2$ produces a $\theta (\kappa - \sigma)$, leading to
\begin{align}
  &     = -   (4 \Delta)^{1/3}  \int_{-\infty}^\infty d \kappa     \theta(\kappa - \sigma)   \int_{-\infty}^0 dy_1 dy_2  \delta\left(y + \frac{(4 \Delta)^{1/3}}{2} (y_1 + y_2) \right)
  e^{( \hat X + i \frac{q}{2} (4 \Delta)^{1/3} ) (y_2 - y_1)}
       \nn  \\&  \partial_\kappa \left( {\rm Det} [I - P_{\kappa}K_{\Ai} P_{\kappa}]   \int_\kappa^\infty du dv   [I- P_\kappa K_{\Ai}  P_\kappa ]^{-1}(u,v)  \Ai(u + y_1)  \Ai(v  + y_2)  ) \right)
\end{align}
 Plugging this second piece in \eqref{new1} we can integrate over $y$  and move the derivative on the $\theta$ function (by integration by parts)
\begin{align}
&- (4 \Delta)^{1/3}  \int \frac{d q}{4\pi   } \int_0^\infty dv  \int  {dy}{} \int_{-\infty}^\infty d\kappa  \Ai \Big( y + q^2 + 2^{2/3} \sigma_1   + v  \Big)  \nn \\&  \times \int_{-\infty}^\infty d \kappa     \theta(\kappa - \sigma)   \int_{-\infty}^0 dy_1 dy_2  \delta\left(y + \frac{(4 \Delta)^{1/3}}{2} (y_1 + y_2) \right)
e^{( \hat X + i \frac{q}{2} (4 \Delta)^{1/3} ) (y_2 - y_1)}
       \nn  \\&  \times \partial_\kappa \left( {\rm Det} [I - P_{\kappa}K_{\Ai} P_{\kappa}]   \int_\kappa^\infty du dv   [1- P_\kappa K_{\Ai}  P_\kappa ]^{-1}(u,v)  \Ai(u + y_1)  \Ai(v  + y_2)  ) \right)\nn \\&
       = +(4 \Delta)^{1/3}
       \int_0^\infty dv  \int_{-\infty}^\infty d \kappa    \ \delta(\kappa - \sigma)   \nn \\& \times   \int_{-\infty}^0 dy_1 dy_2    \int \frac{d q}{4\pi   }  \Ai \Big( - \frac{(4 \Delta)^{1/3}}{2} (y_1 + y_2) + q^2 + 2^{2/3} \sigma_1   + v  \Big)
       e^{( \hat X + i \frac{q}{2} (4 \Delta)^{1/3} ) (y_2 - y_1)}
       \nn  \\&  \times   \left( {\rm Det} [I - P_{\kappa}K_{\Ai} P_{\kappa}]   \int_\kappa^\infty du dv   [I - P_\kappa K_{\Ai}  P_\kappa ]^{-1}(u,v)  \Ai(u + y_1)  \Ai(v  + y_2)  ) \right)
\end{align}

We can then finally write an expression for the two-time generating function in the large time limit
\begin{align}\label{finalgenerating_notes}
&    \hat{g}^{(1)}_{\Delta }(s_1  , s)    \nn \\
& =    \left(  \int_0^\infty du K_{\Ai}(u + \sigma_1, u + \sigma_1) \right) \left(1 -  {\rm Det} [I - P_{\sigma}K_{\Ai} P_{\sigma} ] \right) \nn \\&
+ { \Delta^{1/3}}  {\rm Det} [I - P_{\sigma}K_{\Ai} P_{\sigma} ]   \int_{-\infty}^0 dy_1 dy_2 K_{\Ai} (   -y_1 \Delta^{1/3} + \sigma_1  ,- y_2 \Delta^{1/3}+ \sigma_1) e^{\hat X (y_2 - y_1)} \nn \\& \int_\sigma^\infty du dv \ \Ai(  y_1 + u)
\Ai( y_2  + v ) [I-P_{\sigma}K_{\Ai}P_{\sigma} ]^{-1}(u,v)
\end{align}
  where have used the identity \cite{Vallee}
\begin{equation} \label{idAiry}
\int_{-\infty}^{\infty} \frac{dq}{2 \pi} \Ai(a + q^2) e^{i q c}= 2^{-1/3} \Ai(2^{-2/3} (a + c)) \Ai(2^{-2/3} (a - c))
\end{equation}
which implies
\begin{equation} \label{id2}
 \int \frac{d q}{4\pi   } \int_0^\infty dv  \int_0^\infty  {dy}{}   \Ai \Big( y + q^2 + 2^{2/3} \sigma_1   + v  \Big) = \int_0^\infty du  K_{\Ai}(\sigma_1 + u , \sigma_1 + u) = \text{Tr}[ P_{\sigma_1} K_\Ai ]
\end{equation}
We have used everywhere the first definition of the Airy kernel in (\ref{airyK2}).

Now, recalling the definition (\ref{TWtail}) of the tail of the TW-GUE cumulative distribution
\begin{equation}
F_2^{(1)}(\sigma_1)- 1  =  -  \int_0^\infty du K_{\Ai}(u + \sigma_1, u + \sigma_1)
\end{equation}
we can simplify the formula, to obtain our final result for the generating function {
\begin{align} \label{final0}
  \hat{g}^{(1)}_{\Delta }(s_1,s)   & = 1  -  F^{(1)}_2(\sigma_1)  - F_2(\sigma)
  \text{Tr}[ P_{\sigma_1} K_\Ai ]
   + \Delta^{1/3}    {F_2(\sigma)}  \text{\text{Tr}} \left[ P_\sigma K^{\Delta}_{\sigma_1}   P_\sigma   (I-P_{\sigma} K_{\Ai}P_{\sigma} )^{-1}   \right]
\end{align}}
where we have introduced a new kernel $K^{\Delta}_{\sigma_1}$ defined as
\begin{align}
& K^{\Delta}_{\sigma_1}(u,v)  \nn \\
&  =  \int_{0}^{+\infty} dy_1 dy_2 \Ai(-{y_1}{ }   +  u) K_{\Ai}(y_1 \Delta^{1/3}+ \sigma_1, y_2 \Delta^{1/3} + \sigma_1)  \Ai( -{y_2}{ }  +  v) e^{\hat X (y_2 - y_1)}
\label{K4def}
\end{align}
{\blue Using our starting formula \eqref{deriv3} we then arrive to an expression for the JPDF
\begin{equation}
P^{(1)}_{\Delta}(\sigma_1,\sigma)  = \left( \partial_{\sigma_1} \partial_{\sigma} -  {\Delta^{- 1/3}} \partial^{2}_{\sigma} \right) \left(  {F_2(\sigma)}  \text{\text{Tr}} \left[   \Delta^{1/3}  P_\sigma K^{\Delta}_{\sigma_1}   P_\sigma   (I-P_{\sigma} K_{\Ai}P_{\sigma} )^{-1}   - P_{\sigma_1} K_\Ai  \right] \right).
\end{equation}
}

Note that to the precision which we use here, i.e. up to higher order corrections in $\sigma_1$ of order $O(e^{- \frac{8}{3} \sigma_1^{3/2}})$, the last term in the
expression \eqref{final0} can be rewritten as a difference between two Fredholm determinants
 \begin{align}\label{finalfinal}
  \hat{g}^{(1)}_{\Delta }(s_1,s)   & = 1  -  F^{(1)}_2(\sigma_1)  -  F_2(\sigma) \text{Tr}[ P_{\sigma_1} K_\Ai ]
 \nn \\ \!\!\! \!\!\!\!  - & \Delta^{1/3} F_2(\sigma)  \left(  \text{Det} \left[ I - P_\sigma( K_{\Ai} +  K^{\Delta}_{\sigma_1}  )P_\sigma \right] -   \text{Det} \left[ I - P_\sigma K_{\Ai}  P_\sigma \right]  \right) + O(e^{- \frac{8}{3} \sigma_1^{3/2}})
\end{align}
suitable for numerical evaluations.

The kernel $K^{\Delta}_{\sigma_1}$ is trace-class (i.e its trace is finite) for $\Delta>0$ (see below), and it is interesting to note
that it can also be written as a square $K^{\Delta}_{\sigma_1} = M_{\sigma_1,-\hat X}  M_{\sigma_1,\hat X}^T$ where
$M_{\sigma_1,\hat X}(u,v)=\int_0^\infty dy \Ai(-y + u) \Ai(\Delta^{1/3} y + v) \theta(v-\sigma_1) e^{\hat X y}$.
Another observation, used below, is that one can write formally
\bea
&& K^{\Delta}_{\sigma_1}(u,v) = \mathcal{B}_{\hat X - \Delta^{1/3} \partial_\rho}(u)
\mathcal{B}_{- \hat X - \Delta^{1/3} \partial_{\rho'}}(v) K_\Ai(\rho,\rho') |_{\rho=\rho'=\sigma_1}
\label{K4operator}
\eea
where the function $\mathcal{B}_a(u)$ has been defined in (\ref{defBB})
and (\ref{K4operator}) is taken in the sense of operators.


We will now study this result in various limiting cases, namely large and small $\Delta$
and large $\sigma_1$.

\subsection{Small \texorpdfstring{$\Delta$}{Delta} limit}
\label{sec:smallDelta}

In this Section we study the expansion for small $\Delta$ of our final expression \eqref{final0}. For simplicity we
first set $X=0$, the general case is sketched at the end.

If one performs the limit $\Delta \to 0$ directly on the expression (\ref{K4def}) of the
new kernel $K^{\Delta}_{\sigma_1}$ one finds, keeping also the leading correction
\begin{equation} \label{limK4}
K_{\sigma_1}^{\Delta}(u,v)
= K_{\Ai}(\sigma_1, \sigma_1)  \mathcal{B}_0(u) \mathcal{B}_0(v)
+ \frac{1}{2} \Delta^{1/3} \Ai(\sigma_1)^2 \partial_a|_{a=0} [ \mathcal{B}_a(u) \mathcal{B}_a(v) ]
+ (\Delta^{2/3})
\end{equation}
where we recall $\mathcal{B}_a(u) = \int_{-\infty}^0 dy e^{a y} \Ai(y+u)
=e^{a^3/3- u a} - \int_{0}^{+\infty} dy e^{a y} \Ai(y+u)$. In particular we use later
$\partial_a \mathcal{B}_a(u) |_{a=0} = - u \mathcal{B}_0(u)
+ \Ai'(u)$. The expression \eqref{limK4} is correct at fixed $u,v$, but formal as an
operator since the resulting limit is not trace class (the integral of $\mathcal{B}_0(u)$
diverges at large $u$).
To make the manipulations more rigorous in the limit we now
put the kernel $K^{\Delta}_{\sigma_1}$ in an alternative form, more suitable to take the
$\Delta \to 0$ in formula \eqref{final0}. First we rewrite
\begin{align}
K^{\Delta}_{\sigma_1}(u,v)   & =  \int_0^\infty df \int_{-\infty}^{\infty} dy_1 dy_2 \Ai({y_1}{}   +  u) \Ai( f- \Delta^{1/3} y_1 + \sigma_1 ) (1-\theta(y_1))  \nn \\&  \times  \Ai( f- \Delta^{1/3} y_2 + \sigma_1 )  \Ai(  {y_2}{}  +  v) (1-\theta(y_2))
\end{align}
We then use the following integration formula for the Airy functions obtained
using the integral representation (\ref{Airydef})
\begin{equation}
\int_{-\infty}^{\infty} dy \Ai(-\Delta^{1/3} y + a) \Ai(  y + b)  = \frac{1}{(1  + \Delta)^{1/3}} \Ai\left( \frac{a + \Delta^{1/3} b}{(1 + \Delta)^{1/3}} \right)
\end{equation}
Expanding the factors, this allows to rewrite the kernel as follows
\begin{align}
 & K_{\sigma_1}^{\Delta}(u,v) 
  =  \frac{1}{ (1 + \Delta)^{1/3}} K_{\Ai} \left( \frac{ \Delta^{1/3} u  +   \sigma_1}{(1+ \Delta)^{1/3}} , \frac{\Delta^{1/3} v  +   \sigma_1}{(1+ \Delta)^{1/3}} \right) \nn \\& -  \frac{1}{(1  + \Delta)^{1/3}}  \int_0^\infty dy df \Ai\left( \frac{\Delta^{1/3}u +  (f + \sigma_1)}{(1 + \Delta)^{1/3}} \right) \Ai( - y \Delta^{1/3}+ f   + \sigma_1)  \Ai(y + v ) + (u \leftrightarrow v) \nn \\&
  + \int_0^\infty dy_1 dy_2 K_{\Ai}(- y_1 \Delta^{1/3}  + \sigma_1, - y_2 \Delta^{1/3}  + \sigma_1)  \Ai(y_1 + v ) \Ai(y_2 + u ) \label{K4alt}
\end{align}
Using this formula one can easily recovers (\ref{limK4}) at fixed $u,v$. Next, using this new
representation we can explicity calculate the building blocks in \eqref{final0} in an expansion
at small $\Delta$ and we find
\bea
\fl &&  \Delta^{1/3} \text{Tr}[ P_\sigma K^{\Delta}_{\sigma_1} ] = \text{Tr}[ P_{\sigma_1} K_\Ai ] - \Delta^{1/3}
(1 + {\cal L}(\sigma)) K_\Ai(\sigma_1,\sigma_1)  + \Delta^{2/3} \Ai(\sigma_1)^2 {\cal M}(\sigma)
+ O(\Delta) \nn  \\ \fl &&
\eea
with
\begin{align}
  \mathcal{L}(\sigma)&  =
- 1 +\sigma +  2 \int_\sigma^\infty du \int_{0}^\infty dy \Ai(u + y) - \int_\sigma^\infty du \int_{0}^\infty  dy_1 dy_2  \Ai(u + y_1)  \Ai(u + y_2) \label{funL}
\end{align}
where the second, third and fourth terms come from the
first, second and third line, respectively in (\ref{K4alt}). The next order
involves the function
\bea \label{funM}
 {\cal M}(\sigma)& =& \frac{\sigma^2}{2} + \int_{\sigma}^{+\infty} du \int_{0}^{+\infty} dy
(u - y) \Ai(u+y) \\
& +  &
 \int_{\sigma}^\infty du \int_0^{+\infty} dy_1 dy_2 y_1 \Ai(y_1 + u ) \Ai(y_2 + u )  \nn
\eea

We also find, for any integer $q \geq 1$,
\bea
\fl && ~~~~~ \Delta^{1/3} {\rm \text{Tr}}[ P_\sigma K^{\Delta}_{\sigma_1} ( P_\sigma K_\Ai P_\sigma)^q ]
= \Delta^{1/3} {\rm \text{Tr}}[  ( P_\sigma K_\Ai P_\sigma)^q {\cal B}_0 {\cal B}_0^T]
K_\Ai(\sigma_1,\sigma_1) \nn \\
\fl && ~~~~~~~~~~~~ ~~~~~~~~~~~~ + \frac{1}{2} \Delta^{2/3} \Ai(\sigma_1)^2
 {\rm \text{Tr}}[  ( P_\sigma K_\Ai P_\sigma)^q \partial_a|_{a=0} [ \mathcal{B}_a \mathcal{B}_a^T ]  + O(\Delta)
\eea
Here we can equivalently use (\ref{limK4}), since the convergence of the integrals involved in the
trace is guaranteed by the properties of the Airy kernel. Here and below we denote interchangeably
the rank one projector ${\cal B}_0 {\cal B}_0^T = | {\cal B}_0 \rangle  \langle {\cal B}_0 |$, i.e.
\bea \label{b0}
| {\cal B}_0 \rangle  \langle {\cal B}_0 | (u,v) = {\cal B}_0 {\cal B}_0^T (u,v)={\cal B}_0(u) {\cal B}_0(v)
\eea

We can now substitute in \eqref{finalfinal} where we represent $(I-P_\sigma K_\Ai P_\sigma)^{-1}=I + \sum_{q \geq 1} (P_\sigma K_\Ai P_\sigma)^q$, and we obtain the expansion
 \bea
\fl &&  \hat{g}^{(1)}_{\Delta }(s_1,s)    = 1  -  F^{(1)}_2(\sigma_1)  - \Delta^{1/3}
       K_\Ai(\sigma_1,\sigma_1) \label{gsmall1}  \\
\fl   && \times F_2(\sigma)
    \left( 1 + {\cal L}(\sigma)  -  \text{Tr}[  P_\sigma K_\Ai P_\sigma (I- P_\sigma K_\Ai P_\sigma)^{-1} {\cal B}_0 {\cal B}_0^T] \right)  \nn \\
\fl &&    +
    \Delta^{2/3} F_2(\sigma) \Ai(\sigma_1)^2 \big( {\cal M}(\sigma)
    + \frac{1}{2} \text{Tr}[  P_\sigma K_\Ai P_\sigma (I- P_\sigma K_\Ai P_\sigma)^{-1}
    \partial_a|_{a=0} [ \mathcal{B}_a \mathcal{B}_a^T ] \big) + O(\Delta)
      \nonumber
 \eea
where a cancellation of two leading terms $O(\Delta^0)$ has occured. The trace in the second line can
be rewritten using determinants as follows (supressing temporarily the $O(\Delta^{2/3})$ term)
\bea
\! \! \! \! \! \! \! \!  &&  \hat{g}^{(1)}_{\Delta }(s_1,s)    = 1  -  F^{(1)}_2(\sigma_1)  - \Delta^{1/3}
       K_\Ai(\sigma_1,\sigma_1) \\
\! \! \! \! \! \! \! \!    && \times
    \left( F_2(\sigma) {\cal L}(\sigma)  + {\rm Det}[ 1 - P_\sigma K_{\Ai} P_\sigma -
      P_\sigma K_{\Ai}  P_\sigma \mathcal{B}_0 \mathcal{B}_0^T ] \right)  +
    O(\Delta^{2/3}) \nonumber
 \eea
 recalling that $F_2(\sigma)={\rm Det}[ 1 - P_\sigma K_{\Ai} P_\sigma]$. We have used
 that $P_\sigma K_{\Ai}  P_\sigma \mathcal{B}_0 \mathcal{B}_0^T=
 | P_\sigma K_{\Ai}  P_\sigma  \mathcal{B}_0 \rangle \langle \mathcal{B}_0 |$ is also a
 rank one projector, hence one can apply formula (\ref{detproj2}).


We can now calculate the JPDF for small $\Delta$ using the formula (\ref{deriv3})
\bea
P^{(1)}_{\Delta}(\sigma_1,\sigma) =  (\partial_{\sigma_1} \partial_{\sigma} - \Delta^{-1/3} \partial_{\sigma}^2 )  \hat{g}^{(1)}_{\Delta }(s_1,s) \label{derivnew}
\eea
Let us first discuss $\Delta=0$.
Clearly in the limit $\Delta \to 0$ the first term vanishes and the only non-zero term is given by the second term.
It leads to a probability distribution factorized in a term dependent on $\sigma_1$ and a one (the non-trivial one)
depending on $\sigma$, which takes the form of a second derivative, and we finally find
\begin{align} \label{ressmalldelta}
P^{(1)}_{\Delta \to 0}(\sigma_1  , \sigma)  & =    F_2^{(1)}{}'(\sigma_1)  F_0'(\sigma)
\end{align}
where $F_2^{(1)}{}'(\sigma_1) = K_{\Ai}(\sigma_1, \sigma_1)$ is the tail of the PDF of the
TW-GUE distribution, and
$F'_0( \sigma)$ is precisely the PDF of {\it the Baik-Rains distribution}, of associated CDF
\begin{align} \label{BR1}
F_0(  \sigma)  & =  \partial_{\sigma} \Big[  F_2(\sigma) \mathcal{L}(\sigma) +  \text{Det} \left( 1 - P_\sigma K_{\Ai} P_\sigma-      P_\sigma K_{\Ai}  P_\sigma \mathcal{B}_0  \mathcal{B}_0^T \right)   \Big]
\end{align}
in a form similar to the one given in \cite{SasamotoStationary} in the study of the KPZ equation with
Brownian initial conditions (see Appendix \ref{app:airy}, \ref{app:w} and \ref{app:BR}
for an alternative derivation of this result, a more detailed comparison with \cite{SasamotoStationary}
(the reader should be aware of the presence of misprints in Eqs (2.37-2.38) in \cite{SasamotoStationary}). The above result (\ref{ressmalldelta}) corresponds to a product of a GUE random variable and a Baik-Rains random variable.
Thus, we find here that in the limit where $1 \ll t_2-t_1 \ll t_1$, the height increment $h(t_2,0)-h(t_1,0)$ becomes statistically independent of the height at the earlier time, and furthermore, that it is distributed as in
large time stationary KPZ. This result will be confirmed by a different approach in Appendix \ref{app:airy}.

The above results about the limit $\Delta=0$ extend for a finite $\hat X$, and even in presence of a slope $\hat w$ as
is shown in Appendix \ref{app:w} and \ref{app:BR} . The BR distribution $F'_0(\sigma)$ in (\ref{ressmalldelta})
is then replaced by the extended BR distribution, $\partial_\sigma F_0(\sigma-\hat X^2;\hat X, \hat w)$.

We now obtain the first correction to the GUE-TW times BR distribution at small but finite $\Delta$.
Applying the differentiation operator \eqref{derivnew} to
\eqref{gsmall1} and keeping the next order correction we find
\bea \label{ressmalldelta2}
\fl && P^{(1)}_{\Delta}(\sigma_1  , \sigma)   =    F_2^{(1)}{}'(\sigma_1)  F_0'(\sigma) + \Delta^{1/3} \Ai(\sigma_1)^2 \Big( F_0(\sigma)
\\
\fl && ~~~~  - \partial_\sigma^2
[ F_2(\sigma) ({\cal M}(\sigma)+1)  -
\text{Det} \left( 1 - P_\sigma K_{\Ai} P_\sigma-      P_\sigma K_{\Ai}  P_\sigma \mathcal{B}_0
 \partial_a|_{a=0} \mathcal{B}_a^T  \right)  ] \Big) + O(\Delta^{2/3}) \nn
\eea
Quite remarkably, a comparison with (\ref{derBR}) shows that
\bea
\fl && ~~~~ P^{(1)}_{\Delta}(\sigma_1  , \sigma)   =    F_2^{(1)}{}'(\sigma_1)  F_0'(\sigma)
- \frac{1}{2} \Delta^{1/3} F_2^{(1)}{}''(\sigma_1)  \partial_\sigma \partial_{\hat w} F_0(\sigma;0;\hat w)]|_{\hat w=0}
+ O(\Delta^{2/3})
\label{p1br}
\eea
where we recall that $F_2^{(1)}{}'(\sigma_1) = K_\Ai(\sigma_1,\sigma_1)$ and
$F_2^{(1)}{}''(\sigma_1) = - \Ai(\sigma_1)^2$. This means that the term in the first order in $\Delta^{1/3}$ is related to a shift in a finite
slope $\hat w$ of the BR distribution. This is perfectly consistent with the
formula \eqref{K4operator}. Furthermore, since
$F_2^{(1)}{}''(\sigma_1) \simeq_{\sigma_1 \to +\infty} - 2  \sigma_1^{1/2} F_2^{(1)}{}'(\sigma_1)$, in the
double limit of large $\sigma_1$ and small $\Delta$ such that $\hat w_{\sigma_1,\Delta}= \Delta^{1/3} \sigma_1^{1/2}$ is constant, the JPDF in (\ref{ressmalldelta2})-(\ref{p1br}) remains
decoupled into the product of (the tail of) the GUE-TW distribution and the extended BR distribution with a
finite slope given by $\hat w_{\sigma_1,\Delta}$. While
this property is shown here for a fixed but small value of  $\hat w_{\sigma_1,\Delta} \ll 1$ it
is shown in Section \ref{sec:crossover} that it holds for
arbitrary value of $\hat w_{\sigma_1,\Delta}$.

From the result \eqref{p1br} one can calculate conditional cumulants of $\sigma$ at
fixed $\sigma_1$. One finds
\bea \label{small_delta_expansion_final}
\fl && \langle \sigma^p \rangle_{\sigma_1} =
\langle \sigma^p \rangle_{\mbox{BR}}  + \Delta^{1/3} \frac{\Ai(\sigma_1)^2}{K_\Ai(\sigma_1,\sigma_1) } A_p
+ O(\Delta^{2/3}) \quad , \quad A_p = \int d\sigma \sigma^p G(\sigma)
\eea
with
\bea
\fl &&
G(\sigma) = F_0(\sigma) - \partial_\sigma^2
[ F_2(\sigma) ({\cal M}(\sigma)+1)  -
\text{Det} \left( 1 - P_\sigma K_{\Ai} P_\sigma-      P_\sigma K_{\Ai}  P_\sigma \mathcal{B}_0
 \partial_a|_{a=0} \mathcal{B}_a^T  \right)  ]  \nn \\ \fl &&
\eea

\subsection{Large \texorpdfstring{$\Delta$}{Delta} limit }
\label{sec:largeDelta}

Let consider again the generating function in \eqref{final0}. To perform the large $\Delta$ limit
it is convenient to make a change of variable $y_j \to y_j \Delta^{-1/3}$ in the definition of
$K^{\Delta}_{\sigma_1}$ and expand as
\bea
\fl && K^{(4),{\rm sym}}_{\sigma_1}(u,v) =    \int_{0}^{+\infty} \frac{dy_1 dy_2}{\Delta^{2/3}} \Ai(-{y_1} \Delta^{-1/3}   +  u) \nn  \\
\fl && ~~~~~~~~~~~~~~~ \times K_{\Ai}(y_1 + \sigma_1, y_2  + \sigma_1)  \Ai( -{y_2}{ }  \Delta^{-1/3}  +  v)
\cosh (\bar X (y_2 - y_1) )
\nn \\
\fl &&  =  \frac{1}{\Delta^{2/3}} \Ai(  u) \Ai(  v) \phi_{00}(\sigma_1)   - \frac{1}{\Delta} [ \Ai'(  u) \Ai(  v) + \Ai(u) \Ai'(v) ]
\phi_{10}(\sigma_1) \nn \\
\fl &&  +  \frac{1}{\Delta^{4/3}}  \Ai'(  u) \Ai'(  v) \phi_{11}(\sigma_1)  + \frac{1}{2\Delta^{4/3}}
[ u \Ai(  u)  \Ai(  v) + \Ai(u) v \Ai(v) ] \phi_{21}(\sigma_1) + O(\Delta^{-1}) \label{explarge}
\eea
where we denote
\bea
\!\!\!\! \! \!\! \phi_{pq}(\sigma_1) \equiv  \int_{0}^{+\infty} dy_1 dy_2 y_1^p y_2^q  K_{\Ai}(y_1 + \sigma_1, y_2  + \sigma_1)
\cosh (\bar X (y_2 - y_1) )
\eea
with $\phi_{pq}(\sigma_1)=\phi_{qp}(\sigma_1)$, and we have defined
\bea \label{scales}
\bar X = \Delta^{-1/3} \hat X = \frac{X}{2 \Delta t_1^{2/3}} = \frac{X}{2 (t_2-t_1) t_1^{-1/3}}
\eea
what appears to be the relevant scale along $x$ in the large $\Delta$ limit. Note that
we are using here the symmetrized version $K^{(4),{\rm sym}}_{\sigma_1}$ of the operator
$K^{\Delta}_{\sigma_1}$, where $e^{\hat X (y_2 - y_1)}$ has been replaced by $\cosh( \hat X (y_2 - y_1))$.
This is allowed since one can use in \eqref{final0}
the symmetry of the operator $(I-P_{\sigma} K_{\Ai}P_{\sigma} )^{-1}$
to symmetrize in $u,v$, hence in $y_1,y_2$ and one can use there either form (symmetrized
or unsymmetrized) of $K^{\Delta}_{\sigma_1}$.

Each of the terms in (\ref{explarge}) is a rank one projector, hence using (\ref{detproj2}) we
can expand (using that the Airy kernel is symmetric)
\begin{align}
& \Delta^{1/3}    {F_2(\sigma)}  \text{\text{Tr}} \left[ P_\sigma K^{\Delta}_{\sigma_1}   P_\sigma   (1-P_{\sigma} K_{\Ai}P_{\sigma} )^{-1}   \right]
\nn \\& =
 \frac{1}{\Delta^{1/3}} \phi_{00}(\sigma_1)
( {\rm  Det }[ I- P_\sigma (K_{\Ai} - \Ai \Ai^T) P_\sigma ]-F_2(\sigma))
  \nn\\& - \frac{2}{\Delta^{2/3}} \phi_{10}(\sigma_1)
  \left( {\rm  Det } [ I - P_\sigma( K_{\Ai}  - \Ai' \Ai^T) P_\sigma] - F_2(\sigma)\right)\nn \\&
+ \frac{1}{\Delta}   \phi_{11}(\sigma_1)  \left( {\rm  Det } [ I- P_\sigma (K_{\Ai} - \Ai' {\Ai'}^T) P_\sigma ] - F_2(\sigma)\right)  \nn \\ &
+ \frac{1}{\Delta }   \phi_{20}(\sigma_1) \left( {\rm  Det } [ I - P_\sigma (K_{\Ai} - \Ai'' \Ai^T) P_\sigma ] - F_2(\sigma)\right)
+ O(\Delta^{-4/3})
\end{align}
where we have used similar notation as in (\ref{b0}) to denote rank one projectors.

We can now expand the generating function including all terms up to $O(1/\Delta)$ and obtain
\bea
 \fl && \hat{g}^{(1)}_{\Delta }(s_1  , s)   = 1 - F_2^{(1)}(\sigma_1)
  - F_2(\sigma) \int_{\sigma_1}^{+\infty} dy K_\Ai(y,y) +
 \frac{ \phi_{00}(\sigma_1) }{\Delta^{1/3}} F_2'(\sigma)  - \frac{\phi_{10}(\sigma_1)}{\Delta^{2/3}}  F_2''(\sigma) \nn \\
  \fl &&
+ \frac{1}{\Delta}   \phi_{11}(\sigma_1)  \left( {\rm  Det } [ I- P_\sigma (K_{\Ai} - \Ai' {\Ai'}^T) P_\sigma ] - F_2(\sigma)\right)  \nn \\
\fl &&
+ \frac{1}{\Delta }   \phi_{20}(\sigma_1) \left( {\rm  Det } [ I - P_\sigma (K_{\Ai} - \Ai'' \Ai^T) P_\sigma ] - F_2(\sigma)\right)
+ O(\Delta^{-4/3})
\eea
where we have used formula (\ref{derGUE}) and (\ref{der2GUE}) to simplify the first two subleading orders in terms
of the derivatives of the TW-GUE distribution.

The first consequence is that in the infinite time separation limit the joint PDF is simply the product of two PDF of GUE distributions
\begin{align}
 \lim_{\Delta \to \infty} P_{\Delta}(\sigma_1,\sigma) =
 \lim_{\Delta \to \infty}  [\partial_{\sigma_1} \partial_{\sigma} - \Delta^{-1/3} \partial_{\sigma}^2] & {g}^{(1)}_{\Delta, \infty}(s_1  , s)    =  F_2^{(1)}{}'(\sigma_1)   F_2'(\sigma)
\end{align}
as expected. Note that it is obtained here in the tail region for $\sigma_1$ but it is
expected to hold for any $\sigma_1$.

In addition we obtain the corrections to the JPDF as
\bea
\fl && P_{\Delta}(\sigma_1,\sigma) =  [\partial_{\sigma_1} \partial_{\sigma}  - \Delta^{-1/3} \partial_{\sigma}^2]  {g}^{(1)}_{\Delta, \infty}(s_1  , s)    =  F_2^{(1)}{}'(\sigma_1)   F_2'(\sigma)  - \frac{1}{\Delta^{1/3}}  F_2''(\sigma) \tilde R_{1/3}(\sigma_1) \nn \\
\fl && + \frac{1}{\Delta^{2/3}}  F_2'''(\sigma) \tilde R_{2/3}(\sigma_1)   + \frac{1}{\Delta} C_1(\sigma_1,\sigma) + \frac{1}{\Delta^{4/3}} C_{4/3}(\sigma_1,\sigma)
+ O(\frac{1}{\Delta^{5/3}})  \label{Pexp}
\eea
where we have defined
\bea
&& \tilde R_{1/3}(\sigma_1) = - \phi'_{00}(\sigma_1)  - \int_{\sigma_1}^\infty dy K_{\Ai}(y,y)
 \\
&& \tilde R_{2/3}(\sigma_1) = - \phi'_{10}(\sigma_1)  -   \phi_{00}(\sigma_1)
\eea
and the functions $C_1(\sigma_1,\sigma)$ and
$C_{4/3}(\sigma_1,\sigma)$ have more complicated expressions,
given in the Appendix \ref{app:higher}.

Dividing by $F_2^{(1)}{}'(\sigma_1)= K_\Ai(\sigma_1,\sigma_1)$, this leads to a rather simple exact formula for the three leading orders in the large $\Delta$ expansion of the {\it conditional moments}
$\langle h^p \rangle_{h_1}$
defined in Section. \ref{sec:condmom}. For convenience here we use the notation
$\langle \sigma^p \rangle_{\sigma_1} \equiv \langle h^p \rangle_{h_1}$.
Using the above, and
integration by parts we obtain for any integer $p \geq 1$
\bea
\fl &&  \langle \sigma^p \rangle_{\sigma_1}
= \langle \sigma^p \rangle_{\mbox{\tiny GUE}} +  \frac{p}{\Delta^{1/3}}
 \langle \sigma^{p-1} \rangle_{\mbox{\tiny GUE}} ~ R_{1/3}(\sigma_1)
+  \frac{p(p-1)}{\Delta^{2/3}}  \langle \sigma^{p-2} \rangle_{\mbox{\tiny GUE}} ~ R_{2/3}(\sigma_1)
  \nn \\
 \fl  && -  \frac{1}{\Delta} \left( p(p-1)(p-2) \langle \sigma^{p-3} \rangle_{\mbox{\tiny GUE}} ~ R_{11}(\sigma_1)
 + a_p R_{12}(\sigma_1) + b_p R_{13}(\sigma_1) \right) + O(\Delta^{-4/3}) \nn \\
\fl && ~~~~~~~ + O(e^{- \frac{4}{3} \sigma_1^{3/2}}) \label{condmomexp}
\eea
where we recall that the values of the cumulants of the GUE-TW and BR distribution
are given in (\ref{cumGUE}) and (\ref{cumBR}). We have defined
\bea
R_{1/3}(\sigma_1) = \frac{\tilde R_{1/3}(\sigma_1)}{K_\Ai(\sigma_1,\sigma_1)}
\quad , \quad R_{2/3}(\sigma_1) = \frac{\tilde R_{2/3}(\sigma_1)}{K_\Ai(\sigma_1,\sigma_1)}
\eea
the functions $R_{11},R_{12},R_{13}$ and the coefficients $a_p,b_p$ being
given in the Appendix \ref{app:higher}. The functions $R_\alpha$ contain all the dependence
in $\bar X$. Note that (\ref{condmomexp}) is exact
at large $\sigma_1$, with corrections expected to be only of the order $O(e^{- \frac{4}{3} \sigma_1^{3/2}})$.

Let us now discuss some properties of the functions $R_{1/3},R_{2/3}$ and their consequences.
Let us focus on the case $\bar X=0$. Then one has
\bea \label{R13}
\fl && R_{1/3}(\sigma_1)= \frac{  \left[\int_{\sigma_1}^\infty dy \Ai(y )\right]^2 - \int_{\sigma_1}^\infty dy K_{\Ai}(y,y) }{K_\Ai(\sigma_1,\sigma_1)}  \\
\fl && R_{2/3}(\sigma_1)= \frac{
\left[\int_{0}^\infty dy y \Ai(y+\sigma_1 )\right]  \left[\int_{\sigma_1}^\infty dy \Ai(y )\right]
- \int_{\sigma_1}^{+\infty}  dy_1 dy_2 K_{\Ai}(y_1, y_2)  }{K_\Ai(\sigma_1,\sigma_1)}
\eea
and one finds that at large $\sigma_1 \to +\infty$
\bea
R_{1/3}(\sigma_1) \simeq  \frac{3}{2 \sigma_1^{1/2}} -\frac{13}{8
   \sigma_1^2}+\frac{327}{64}
   \frac{1}{\sigma_1^{7/2}} +O\left(\frac{1}{\sigma_1^5}\right)
\eea
and that keeping this three term approximation gives 1 percent accuracy around $\sigma_1=4.5$
and 10 percent around $\sigma_1=2.5$. Specific values are
$R_{1/3}(1)=0.984861$, $R_{1/3}(0)=1.20144$, $R_{1/3}(-1)=1.552$ and
$R(\langle \sigma \rangle_{\mbox{\tiny GUE}} )=1.87372$.
One also finds that at large $\sigma_1 \to +\infty$
\bea
R_{2/3}(\sigma_1) \simeq  \frac{1}{\sigma_1} - \frac{5}{2 \sigma_1^{5/2}} + ..
\eea
and that keeping this two term approximation gives
1 percent accuracy around $\sigma_1=10.$ and
10 percent around $\sigma_1=4.5$. Specific values are
$R_{2/3}(1)=0.40778$, $R_{2/3}(0)=0.59372$ and
$R_{2/3}(-1)=0.971346$.

From the above formula for the moments, we obtain the expansion for the
conditional cumulants
\bea \label{mean1}
&&  \langle \sigma \rangle_{\sigma_1}
= \langle \sigma \rangle_{\mbox{\tiny GUE}} +  \frac{1}{\Delta^{1/3}}  R_{1/3}(\sigma_1) + O(\frac{1}{\Delta},e^{- \frac{4}{3} \sigma_1^{3/2}})
\\
\label{var1}
&& \langle \sigma^2 \rangle^c_{\sigma_1} =
\langle \sigma^2 \rangle^c_{\mbox{\tiny GUE}}
- \frac{1}{\Delta^{2/3}}  ( R_{1/3}(\sigma_1)^2 - 2 R_{2/3}(\sigma_1) ) + O(\frac{1}{\Delta},e^{- \frac{4}{3} \sigma_1^{3/2}})
\eea
Note that the term $\Delta^{-1/3}$ cancels in the variance. The correction to the variance
is thus $O(\Delta^{-2/3})$ with a {\it negative} coefficient
(equal to $-0.466009$, $-0.256029$ and $-0.154392$
for $\sigma_1=-1,0,1$ respectively and $\simeq - \frac{1}{4 \sigma_1}$ for large $\sigma_1$).
Since the variance of the BR distribution is larger than the GUE one,
this implies {\it non-monotonicity} in $\Delta$ for the variance. For the higher cumulants we find
\bea
 \langle \sigma^3 \rangle^c_{\sigma_1} =  \langle \sigma^3 \rangle^c_{\mbox{\tiny GUE}} + O(\frac{1}{\Delta})
\quad , \quad  \langle \sigma^4 \rangle^c_{\sigma_1} =  \langle \sigma^4 \rangle^c_{\mbox{\tiny GUE}} + O(\frac{1}{\Delta})
\eea
and the $O(\frac{1}{\Delta})$ term is calculated explicitly in the Appendix \ref{app:higher}.
Since the variance is already corrected to order $O(\Delta^{-2/3})$ we easily
obtain the leading contribution to the skewness and kurtosis
\bea
&& {\rm Sk} =  {\rm Sk}_{\mbox{\tiny GUE}}  \left( 1 + \frac{3}{2 \Delta^{2/3}}
\frac{1}{\langle \sigma^2 \rangle^c_{\mbox{\tiny GUE}}} ( R_{1/3}(\sigma_1)^2 - 2 R_{2/3}(\sigma_1)) \right)
+ O(\frac{1}{\Delta})  \\
&&  {\rm Ku} =  {\rm Ku}_{\mbox{\tiny GUE}}  \left( 1 + \frac{2}{\Delta^{2/3}}
\frac{1}{\langle \sigma^2 \rangle^c_{\mbox{\tiny GUE}}} ( R_{1/3}(\sigma_1)^2 - 2 R_{2/3}(\sigma_1)) \right)
+ O(\frac{1}{\Delta})  \label{SkKu}
\eea

\begin{figure}
\centering
\includegraphics[scale=0.8]{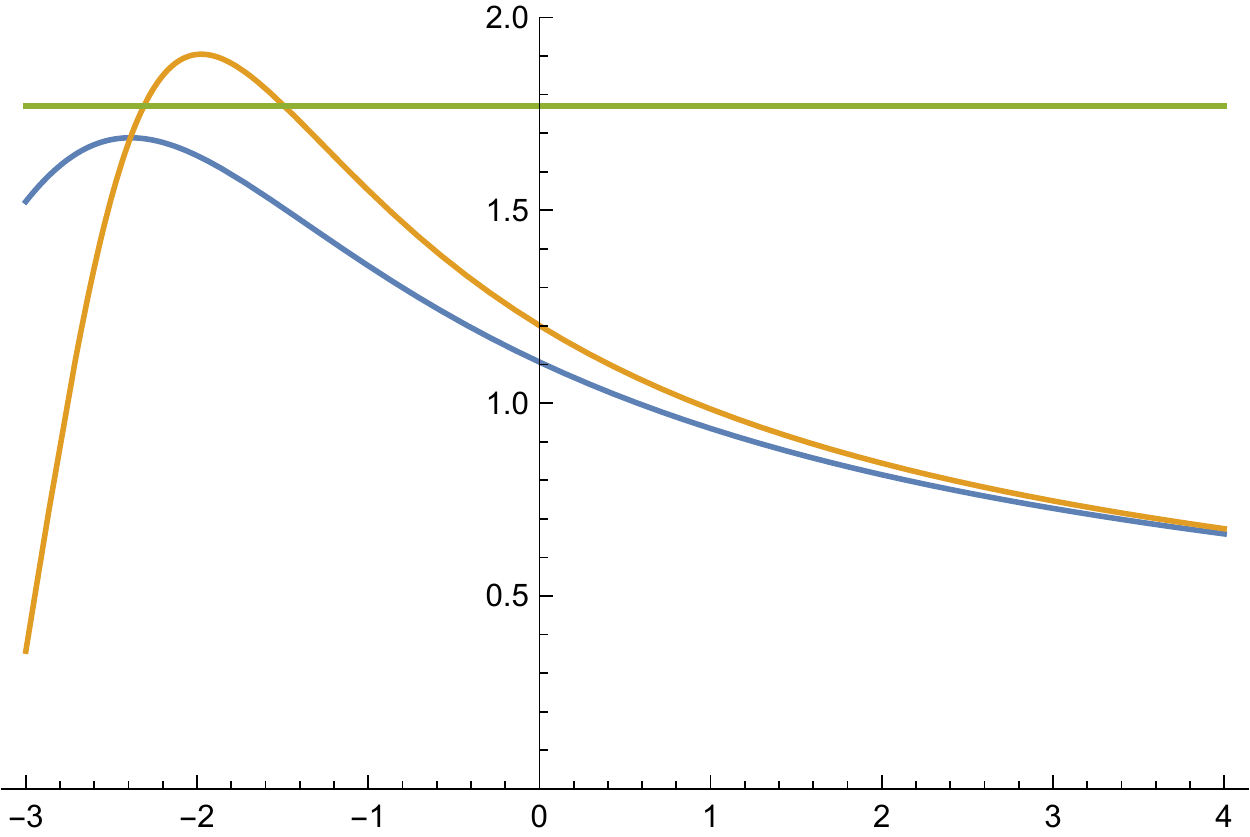}
\caption{Mean of the scaled height difference $h=\sigma$ in the large $\Delta$ limit: Plot of the
coefficient of the correction $O(\Delta^{-1/3})$ to the mean, defined as
(i) orange curve: conditional mean $\langle \sigma \rangle_{\sigma_1}$, i.e.
the function $R_{1/3}(\sigma_1)$ plotted as function of $\sigma_1$,
(ii) blue curve: integrated conditional mean
$\langle \sigma \rangle_{\sigma_1>\sigma_c}$, i.e. the function $R^{>}_{1/3}(\sigma_c)$
plotted as function of $\sigma_c$. (iii) horizontal line at $1.77109$: exact result
for unconditionned mean. The (unknown) exact correction to the integrated conditional mean
should asymptote $R^{>}_{1/3}(\sigma_c)$ at large positive $\sigma_c$ and the exact value
 $1.77109$ at large negative $\sigma_c$, see \eqref{exactmean}.}
\label{fig:coeff}
\end{figure}

The case $X \neq 0$ can be analyzed similarly. One finds that as long as $\bar X < 1$, i.e.
$\hat X < \Delta^{1/3}$, there is essentially no change in the JPDF. When $\bar X$ increases
beyond unity, the function $R_\alpha$ keeps increasing, which presumably means
that the present tail approximation is valid only for larger values of $\sigma_1$
as $\bar X \gg 1$. \\

Finally, we can study the integrated conditional moments $\langle h^p \rangle_{h_1 > \sigma_c}$
as defined in Section. \ref{sec:condmom}. The large $\Delta$ expansion formula for the moments
and cumulants, Eqs
(\ref{condmomexp}),(\ref{mean1}),(\ref{var1}),(\ref{SkKu}), are still valid
but with the functions $R_\alpha$ replaced as follows
\bea  \label{condintegr}
R_\alpha(\sigma_1) \to  R^{>}_{\alpha}(\sigma_c) = \frac{\int_{\sigma_c}^{+\infty} d\sigma_1 \tilde R_\alpha(\sigma_1)}{\int_{\sigma_c}
d \sigma_1 K_\Ai(\sigma_1,\sigma_1)}
\eea
On obtains, for instance, for $X=0$, $R^{>}_{1/3}(\sigma_c)=
\{0.934421, 1.10668, 1.3563, 1.58566\}$ for $\sigma_c = \{1,0,-1,\langle \sigma \rangle_{\mbox{\tiny GUE}} \}$.

{ In the case of the mean value, it is possible to compare the above result for the
conditional mean (\ref{mean1}) and the integrated conditional mean (\ref{condintegr}),
with the exact result (\ref{mean1tot0}) and its large $\Delta$ expansion in Eq. (\ref{asympt1}).
The comparison of the $O(\Delta^{-1/3})$ coefficient is plotted in Fig.
\ref{fig:coeff}. At eyesight it gives a rough estimate of the value of
$\sigma_c$ where our tail approximation in the RBA calculation must break down
(around $\sigma_c \approx -1.5$). Interestingly, the conditioned and unconditioned
mean share the absence of a $O(\Delta^{-2/3})$ correction term.}

\begin{figure}
\centering
\includegraphics[scale=1.8]{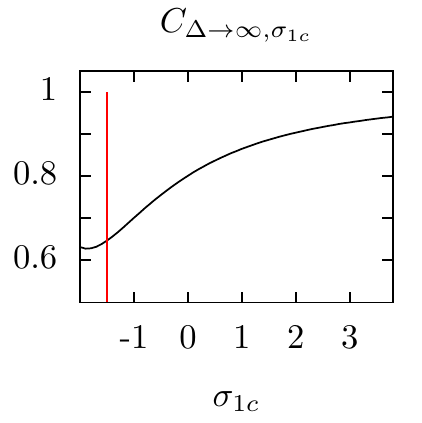}
\caption{
{  Two-time conditional connected correlation functions, $c_{\sigma_c} = \lim_{\Delta \to +\infty} c_{\Delta=+\infty,\sigma_c}$
defined in \eqref{condcorrfun}, as a function of $\sigma_c$, in the infinite
time separation limit $t_2/t_1 \gg 1$.
The formula  \eqref{condcorrfun} gives an excellent approximation for all
$\sigma_c > -1.5$ (blue vertical line). The curve approaches $1$ for large positive $\sigma_c$
as expected, and a naive extrapolation { (see text)
yields $c \approx 0.58 \pm 0.05$} (red horizontal line) for $\sigma_c \to - \infty$, close to the observation
in Fig. 12 of \cite{Takeuchi}.}}
\label{fig:coeff4}
\end{figure}

{

\subsection{Two-time conditional correlations at large \texorpdfstring{$\Delta$}{FIXME}: persistence of correlations}
\label{sec:cond}

Here we introduce and study the correlation functions for the two height profiles, conditioned to realizations for which $H_1$ is larger than a certain value. To be more precise we recall the meaning of the
conditional averages as:
\bea
\langle O(H_1,H_2) \rangle^c_{h_1 > \sigma_c} := \frac{\int_{H_1>\sigma_c t_1^{1/3},H_2} dH_1 dH_2 {\cal P}(H_1,H_2) O(H_1,H_2)}{
\int_{H_1>\sigma_c t_1^{1/3},H_2} dH_1 dH_2 {\cal P}(H_1,H_2)}
\eea
where ${\cal P}$ is the JPDF of $H_1$ and $H_2$, and similarly for the rescaled quantities. Hence
it is a correlation function, but calculated only over the events where $H_1 > \sigma_c t_1^{1/3}$.

We define the following universal two-time conditional connected correlations
 \begin{equation}\label{condcorrfun}
c_{\Delta, \sigma_c} = \frac{\langle H_1 H_2 \rangle^c_{h_1 > \sigma_c}}{\langle H_1^2 \rangle^c_{h_1 > \sigma_c}} = 1 + \frac{\langle H H_1 \rangle^c_{h_1 > \sigma_c}}{\langle H_1^2 \rangle^c_{h_1 > \sigma_c}}  = 1 + \Delta^{1/3} \frac{\langle h h_1 \rangle^c_{h_1 > \sigma_c}}{\langle h_1^2 \rangle^c_{h_1 > \sigma_c}}
 \end{equation}
Clearly in the limit $\sigma_c \to - \infty$ we recover the dimensionless standard
(i.e. complete) two-time connected correlation functions
\begin{equation}
\lim_{\sigma_c \to - \infty} c_{\Delta, \sigma_c}  = \frac{\langle H_1 H_2 \rangle^c }{\langle H_1^2 \rangle^c }
 \end{equation}
which has been studied in experiments (see Fig. 12 in \cite{Takeuchi})
and found there to reach a non-zero limit,
$c = \lim_{\Delta \to +\infty,\sigma_c \to - \infty}$ at large $\Delta$ as
also argued from theory (see Eq. \eqref{persist} in Appendix \ref{app:airy} here,
and \cite{FerrariSpohn2times,GueudreUnpub}). This phenomenon of
breaking of ergodicity can be probed in a much finer way here, by conditioning
it to the value of $h_1>\sigma_c$. Indeed, the larger $\sigma_{c}$ the
more the two polymer paths are expected to overlap, and the higher the ratio should
be. We now calculate this ratio as a function of $\sigma_c$:  our result is exact
for large positive $\sigma_c$ and extrapolates nicely to negative values of $\sigma_c$, leading to a
quite reasonable estimate for $c$ (see discussion below).

Using the result of the previous Section we can rewrite the factors in (\ref{condcorrfun}) respectively as
\begin{equation}\label{num}
{\langle h h_1 \rangle^c_{h_1 > \sigma_c}}{} =  \mathcal{N}_{\sigma_c}^{-1}\left( \int_{\sigma_c}^{+\infty} d\sigma_1 \sigma_1 F_2^{(1)}{}'(\sigma_1) \langle h \rangle_{\sigma_1}   -   {\langle  h_1 \rangle_{h_1 > \sigma_c}}{} \int_{\sigma_c}^{+\infty} d\sigma_1  \langle h \rangle_{\sigma_1} F_2^{(1)}{}'(\sigma_1) \right)
\end{equation}
and
\begin{equation}
  {\langle  h_1^2 \rangle^c_{h_1 > \sigma_c}}{} =  \mathcal{N}_{\sigma_c}^{-1} \int_{\sigma_c}^{+\infty} d\sigma_1 \sigma_1^2 F_2^{(1)}{}'(\sigma_1)  - \mathcal{N}_{\sigma_c}^{-2} \left( \int_{\sigma_c}^{+\infty} d\sigma_1 \sigma_1 F_2^{(1)}{}'(\sigma_1)   \right)^2
\end{equation}
\begin{equation}
  {\langle  h_1 \rangle_{h_1 > \sigma_c}}{} =  \mathcal{N}_{\sigma_c}^{-1} \int_{\sigma_c}^{+\infty} d\sigma_1 \sigma_1 F_2^{(1)}{}'(\sigma_1)
\end{equation}
where $\mathcal{N}_{\sigma_c} = 1- F^{(1)}_2(\sigma_c)=\int_{\sigma_1}^\infty dy K_{\Ai}(y,y)$.
We can now substitute
$\langle h \rangle_{\sigma_1} \equiv \langle \sigma \rangle_{\sigma_1}$ from Eq. (\ref{mean1}).
The leading term $\langle \sigma \rangle_{\mbox{\tiny GUE}}$ cancels in the
numerator (\ref{num}) and we obtain the limit
\be \label{explicit}
c_{\sigma_c} := \lim_{\Delta\to+\infty,\sigma_c} c_{\Delta, \sigma_c} =  1 +  \frac{\int_{\sigma_c}^{+\infty} d\sigma_1 \sigma_1 \tilde R_{1/3}(\sigma_1)
-    {\langle  h_1 \rangle_{h_1 > \sigma_c}}{}  \int_{\sigma_c}^{+\infty} d\sigma_1 \tilde R_{1/3}(\sigma_1)}{
 {   {\langle  h_1^2 \rangle^c_{h_1 > \sigma_c}}{} \ \mathcal{N}_{\sigma_c}}{}}
 \ee
where we recall that
$\tilde R_{1/3}(\sigma_1)= \left[ \int_{\sigma_1}^\infty dy \Ai(y )\right]^2 - \int_{\sigma_1}^\infty dy K_{\Ai}(y,y)$.
The first order corrections in $1/\Delta$ are given by
\begin{equation}
c_{\sigma_c, \Delta} = c_{\sigma_c} +\Delta^{-2/3} \left( \frac{ -\int_{\sigma_c}^\infty d\sigma_1 \sigma_1 F_2'{}^{(1)}(\sigma_1) \langle h \rangle_{\sigma_1}^{[1]} +   \langle  h_1 \rangle_{h_1 > \sigma_c} \int_{\sigma_c}^\infty d\sigma_1 F_2'{}^{(1)}(\sigma_1)  \langle h \rangle_{\sigma_1}^{[1]}  } {  {\langle  h_1^2 \rangle^c_{h_1 > \sigma_c}}{} \ \mathcal{N}_{\sigma_c}} \right) + O(\Delta^{-1})
\end{equation}
where
\begin{equation}
\langle h \rangle_{\sigma_1}^{[1]} =( a_1 R_{11}(\sigma_1) + b_1 R_{12}(\sigma_1))
\end{equation}
and the coefficients $ a_1, R_{11}, b_1, R_{12}$ defined in Appendix \ref{app:higher}.
Note that here the existence of this finite limit, and breaking of ergodicity, is demonstrated from
an explicit calculation based on the RBA method.
We also note that it exhibits no dependence in $\hat X$, hence the ratio $c_{\sigma_c}$ is
{\it independent of the endpoint position} in the scaling region $\hat X = O(1)$.

The ratio $c_{\sigma_c}$ is plotted as a function of $\sigma_c$ in Fig. \ref{fig:coeff4} and it approaches $1$ with corrections of order $\sim \sigma_c^{-1}$ for large and positive $\sigma_c$.
{ As $\sigma_c$ decreases below $\approx -1.5$ the tail approximation becomes unreliable and
indeed the curve $c_{\sigma_c}$ in Fig. \ref{fig:coeff4} exhibits a minimum at $\sigma_c \approx -2$
while the exact minimum is expected at $\sigma_c \approx - \infty$. Considering the value at the
minimum, and in view of Fig. \ref{fig:coeff} (where the maximum of the curve was
found not far from the exact value) we can estimate the value $c=c_{\sigma_c=-\infty}
= 0.58 \pm 0.05$ for the unconditionned ratio.}
}

\subsection{Limit of \texorpdfstring{$\sigma_1 \sim \Delta^{-2/3}$}{FIXME} large, and extended Baik-Rains distribution with a slope}
\label{sec:crossover}

Here we compute the leading tail of JPDF when $\sigma_1$ is large and positive,
at fixed $\sigma$. We use the asymptotics of the Airy kernel
for $\sigma_1 \to +\infty$ at fixed $y_{1,2} \Delta^{1/3}$ as follows
\begin{equation}
K_{\Ai}(- y_1 \Delta^{1/3} + \sigma_1, - y_2 \Delta^{1/3} + \sigma_1)    \simeq \frac{e^{- \frac{4}{3} \sigma_1^{3/2}}}{8 \pi \sigma_1} e^{(y_1 + y_2) \Delta^{1/3} \sigma_1^{1/2}} (1 - \frac{\Delta^{2/3} (y_1^2 + y_2^2)}{4 \sigma_1^{1/2}} + ..)
\end{equation}

Inserting in the definition (\ref{K4def}) this leads to
\bea
K^{\Delta}_{\sigma_1}(u,v)   \simeq_{\sigma_1 \to \infty}  \frac{e^{- \frac{4}{3} \sigma_1^{3/2}}}{8 \pi \sigma_1}
{\cal B}_{\Delta^{1/3}\sigma_1^{1/2}+ \hat X}(u) {\cal B}_{\Delta^{1/3}\sigma_1^{1/2}-\hat X}(v)
\eea
where we recall that for $a>0$
\bea
{\cal B}_{a}(u) = \int_{-\infty}^0 dy \ e^{ a y  } \Ai(  {y}{ } + u)
\eea
Thus it is clear that the most interesting scaling regime is in the {\it double limit}
where $\sigma_1 \to +\infty$ while $\Delta \to 0$ so that the combination
\bea
\hat{w}_{\sigma_1,\Delta} =\sigma_1^{1/2} \Delta^{1/3}
\eea
remains fixed, as well as fixed $\hat X$. If instead one fixes $\Delta$ and takes
$\sigma_1 \to +\infty$ the expression degenerates to the one studied previously
in the large $\Delta$ expansion (see below). We thus now consider only
this scaling limit. Let us set $\hat X=0$ for notational simplicity, and restore it at the end.
We obtain
%
\bea
\fl && \hat{g}^{(1)}_{\Delta }(s_1  , s)   - 1 + F_2^{(1)}(\sigma_1) \\
\fl &&  ~~~~ =
\frac{e^{-\frac{4}{3} \sigma_1^{3/2}} }{16 \pi \sigma_1^{3/2}} \left(
- F_2(\sigma) + 2 \Delta^{1/3} \sigma_1^{1/2} F_2(\sigma) \text{Tr} [ (I- P_\sigma K_\Ai P_\sigma)^{-1} {\cal B}_{\Delta^{1/3}\sigma_1^{1/2}} {\cal B}_{\Delta^{1/3}\sigma_1^{1/2}}^T ] \right) \nn
\eea

We are now in position to compute the JPDF in this scaling limit, that we denote as
\begin{equation}
\hat{P}^{\hat{w}_{\sigma_1,\Delta}}_{\Delta}(\sigma_1,\sigma) = \lim_{\sigma_1 \to \infty}\lim_{\Delta \to 0} \Big|_{\sigma_1^{1/2} \Delta^{1/3} =\hat{w}_{\sigma_1,\Delta}} {P}_{\Delta}(\sigma_1,\sigma)
\end{equation}
we obtain
\bea
\fl &&\hat{P}^{\hat{w}_{\sigma_1,\Delta}}_{\Delta}(\sigma_1,\sigma) =  [\partial_{\sigma_1} \partial_{\sigma}  - \Delta^{-1/3} \partial_{\sigma}^2]  {g}^{(1)}_{\Delta, \infty}(s_1  , s)  \\
\fl && = \frac{e^{-\frac{4}{3} \sigma_1^{3/2}} }{16 \pi \sigma_1^{3/2}} \bigg(
2 \sigma_1^{1/2} \partial_\sigma ( F_2(\sigma) - 2 \Delta^{1/3} \sigma_1^{1/2} F_2(\sigma) \text{Tr} [ (I- P_\sigma K_\Ai P_\sigma)^{-1} {\cal B}_{\Delta^{1/3}\sigma_1^{1/2}} {\cal B}_{\Delta^{1/3}\sigma_1^{1/2}}^T ] ) \nn \\
\fl &&  + \frac{1}{\Delta^{1/3}} \partial^2_\sigma (F_2(\sigma) - 2 \Delta^{1/3} \sigma_1^{1/2} F_2(\sigma) \text{Tr} [ (I- P_\sigma K_\Ai P_\sigma)^{-1} {\cal B}_{\Delta^{1/3}\sigma_1^{1/2}} {\cal B}_{\Delta^{1/3}\sigma_1^{1/2}}^T ] )
\bigg) \nn
\eea
which can be rewritten as (we reintroduce now the dependence on $\hat X$)
\bea\label{largesigma1_BR}
&&\hat{P}^{\hat{w}_{\sigma_1,\Delta}}_{\Delta}(\sigma_1,\sigma)  =  \frac{e^{-\frac{4}{3} \sigma_1^{3/2}} }{8 \pi \sigma_1} \bigg(
(\partial_\sigma + \frac{1}{2 \hat{w}_{\sigma_1,\Delta} } \partial^2_\sigma )
   {\rm Det}[I - P_\sigma B_{\hat{w}_{\sigma_1,\Delta}, \hat X} P_\sigma]
\bigg)
\eea
in terms of the kernel
\bea
B_{\hat{w}+ \hat X, \hat{w}- \hat X}(u_i,u_j) = K_\Ai(u_i,u_j) + 2 \hat w ~ \mathcal{B}_{ \hat{w}+\hat{X}}(u_i)
\mathcal{B}_{ \hat{w}-\hat{X}}(u_j)
\eea
Hence we find that in this scaling regime the (leading order in the tail) JPDF ``simplifies" into the
product of the of the GUE in the variable $\sigma_1$,
and, in the variable $\sigma$ the generalized Baik-Rains probability distribution
(with a fixed slope $\hat{w}_{\sigma_1,\Delta} =\sigma_1^{1/2} \Delta^{1/3}$),
$\partial_\sigma F_0(\sigma-\hat X^2;\hat X, \hat w_{\sigma_1,\Delta})$
as recalled in formula \eqref{lim1} (and studied there). In that regime the
effect of a large $\sigma_1$ is thus to bias the longer polymer path
to remain closer to $x=0$ at time $t=t_1$. Note also that balancing
$\hat X$ and $\hat{w}_{\sigma_1,\Delta}$ produces a scale,
$X \sim \Delta t_1^{1/3} \sigma_1^{1/2}$ which seems to cross
over, for $\sigma_1 = O(1)$, to the one identified in (\ref{scales})
in the large $\Delta$ analysis.
\\

Note that when $ \hat{w}_{\sigma_1,\Delta} \gg 1$ (e.g. large $\sigma_1$ at fixed $\Delta$
or large $\Delta$ at fixed $\sigma_1$)
we obtain simply a product of two GUE-TW probability distributions
\label{finalfinalexpinsigma1_2}
\begin{equation}
\hat{P}^{\hat{w}_{\sigma_1,\Delta}}_{\Delta}(\sigma_1,\sigma)  \simeq F'_2(\sigma_1) F'_2(\sigma)
\end{equation}
which can be seen by noticing that the leading behavior of the tail of a GUE-TW distribution in $\sigma_1$ is given by
\begin{equation}
F'_{2}(\sigma_1)   \sim \frac{e^{-\frac{4}{3} \sigma_1^{3/2}} }{8 \pi \sigma_1}  \:\:\:\:\: \sigma_1 \gg 1
\end{equation}

%
%
%
%
%
%
%
%

\section{Arguments using the Airy processes}

\label{app:airy}

In this Section we analyze the two-time problem in terms of
Airy processes \cite{spohn2000,QuastelVariational,ferrariAiry-a}. In the two-time problem this approach was pioneered by
Corwin and Hammond \cite{privateC}, and recently in \cite{FerrariSpohn2times}.
There the focus was on the second moment of the two-time height correlations.
Here we push the approach a bit forward, and use this to discuss some properties of the full JPDF,
as well as to confirm and explain some of the results obtained in this paper. We also derive new results for the large $\Delta$ correction to the mean,
and for the dependence on the endpoint position $\hat{X}$. {\blue Everywhere in this Section we consider the KPZ height function in the large time limit. }

\medskip

\subsection{\texorpdfstring{Airy$_2$}{Airy2} process and droplet initial condition}

Let us recall a few facts about Airy processes. At large time the droplet
solution of the KPZ equation takes the form
\bea \label{hdrop}
h(x,t|y,0)  = t^{1/3} ( {\cal A}_2(\hat x - \hat y) - (\hat x - \hat y)^2 )  \quad , \quad \hat x=\frac{x}{2 t^{2/3}}
\eea
Here
${\cal A}_2(\hat x)$ is the Airy process, which is stationary, parity invariant,
with power law correlations over a scale of order unity.
Its marginal PDF at one point is the GUE-TW distribution, and explicit Fredholm determinant formula exist
for its multi-point correlations (in terms of the extended Airy kernel). Equation \eqref{hdrop}
is true as a process, either in $x$ at fixed $y$, or $y$ at fixed $x$. This means that it allows to obtain
multi-point correlations of $h(x,t|y,0)$ in $x$, for a fixed value of $y$, or the reverse. It is not true if
 $h(x,t|y,0)$ is seen as a process in both $x,y$ \footnote{This implies that if information about varying $y$ is used, e.g. as in (\ref{hstat}), only the one-point PDF for one value of $x$ can be obtained} (that process is called the Airy sheet and is
much less is known about it \cite{KPZFixedPoint,Pimentel}).

\subsection{\texorpdfstring{Airy$_{\rm stat}$}{Airy stat} process and stationary initial condition}

Consider now the solution $h^{w_L,w_R}_{stat}(x,t)$ of the KPZ equation with
Brownian initial condition plus drifts, $h(y,t=0)=B_0(y) + y (w_L \theta(-y) - w_R \theta(y))$, where $B_0(y)$ is
a two-sided unit Brownian with $B_0(0)=0$ and $w_{\pm}$ are the drifts. At large time
\bea \label{hstat2}
h^{w_L,w_R}_{stat}(x,t) =  t^{1/3} {\cal A}^{w_L,w_R}_{\rm stat}(\hat x)
\eea
where, denoting $\hat w_{L,R}=w_{L,R} t^{1/3}$, using \eqref{hdrop}
\bea
\fl && ~~~~~ {\cal A}^{w_L,w_R}_{\rm stat}(\hat x)  =
\max_{\hat y} ( {\cal A}_2(\hat x -\hat y) - (\hat x - \hat y)^2 + \sqrt{2} B(\hat y)
+ 2 \hat y (\hat w_L \theta(-\hat y)  - \hat w_R \theta(\hat y))  )
\eea
Its one point CDF is given by the (extended) Baik-Rains distribution
\bea \label{aga}
&& Prob( {\cal A}^{w_L,w_R}_{\rm stat}(\hat x) < \zeta) = H(\zeta + \hat x^2 ; \frac{\hat w_L  + \hat x}{2}, \frac{\hat w_R - \hat x}{2}) \nn \\
&& = F_0(\zeta;\hat x, \frac{\hat w_L+\hat w_R}{2})
\eea
where the function $H(x;w_+,w_-)$ is defined in definition 3 in \cite{png}
and is symmetric in $w_+,w_-$. Here we
have slightly extended the Theorem 1.5 in \cite{BaikLiuAiry}, using the fact that the
STS symmetry fixes some of the dependence in the variables $(\sigma,\hat w_{L,R},\hat x)$.
We have also introduced an alternative notation for the extended BR distribution.
The RBA solution of the KPZ equation with these initial conditions, obtained in
\cite{SasamotoStationary}, shows that
\bea
&&  \partial_\sigma H(\sigma  ; \frac{\hat w_L  + \hat x}{2}, \frac{\hat w_R - \hat x}{2})
 = \partial_\sigma F_0(\sigma - \hat x^2; \hat x, \frac{\hat w_L+\hat w_R}{2}) \\
 && \equiv   \partial_\sigma \left( 1 + \frac{\partial_{\sigma}}{\hat{w}_L + \hat{w}_R}\right)  {\rm Det} [I - P_{\sigma }B_{\hat{w}_L + \hat x,\hat{w}_R - \hat x}   P_{\sigma} ]   \label{BR1BR2}
\eea
remembering the shift $\zeta = \sigma - \hat x^2$ in the definition (2.5) in \cite{SasamotoStationary}.

Let us now consider now the true stationary limit $w_{L,R}=0$. At large time
\bea \label{hstat}
h^{0,0}_{stat}(x,t) =  t^{1/3} {\cal A}_{\rm stat}(\hat x)
\eea
as a process in $x$. Here ${\cal A}_{\rm stat}(\hat x)={\cal A}^{0,0}_{\rm stat}(\hat x) $ is the so-called
Airy process with stationary initial data \cite{BaikFerrariPeche2010,QuastelVariational}, which
has been much studied. Since it arises from the Brownian initial condition with no drifts,
one has \cite{QuastelVariational}
\bea \label{Astatx}
 {\cal A}_{\rm stat}(\hat x) &=& \max_{\hat y} ( {\cal A}_2(\hat x - \hat y) - (\hat x - \hat y)^2 + \sqrt{2} B(\hat y) ) \\
& =& \max_{\hat y} ( {\cal A}_2(\hat y) - \hat y^2 + \sqrt{2} B(\hat x - \hat y) ) \label{Astatx2} \\
&=& \sqrt{2} B(\hat x) + \max_{\hat y} ( {\cal A}_2(\hat y) - \hat y^2 + \sqrt{2} \hat B(\hat y) ) \label{last}
\eea
where the second equality is obtained by performing the change $\hat y \to \hat x - \hat y$ in
the first one. In the third one we have defined $\hat B(\hat y)\equiv B(\hat x - \hat y)-B(\hat x)$
which is also a two-sided Brownian motion (as can be checked by computing its two point correlator).
Thus one can write in law that ${\cal A}_{\rm stat}(\hat x) = \sqrt{2} B(\hat x) + {\cal A}_{\rm stat}(0)$,
i.e. a sum of a two-sided unit Brownian motion and a uniform global shift,
 ${\cal A}_{\rm stat}(0)$, distributed according to the Baik-Rains distribution \cite{png}.
Since $B(\hat x)$ and $\hat B(\hat x)$ are correlated, $B(\hat x)$ and ${\cal A}_{\rm stat}(0)$ are also correlated. Indeed ${\cal A}_{\rm stat}(\hat x)$ is a non-trivial process whose
multi-point correlations are given by an explicit Fredholm determinant formula
(see \cite{BaikFerrariPeche2010}). The equalities (\ref{Astatx})-(\ref{last})
are valid a priori only for the one point PDF of ${\cal A}_{\rm stat}(\hat x)$, not as
a process in $x$ (the same is true for the process ${\cal A}^{w_L,w_R}_{\rm stat}(\hat x)$ above,
less studied). The one point PDF
of ${\cal A}_{\rm stat}(\hat x)$ (see e.g. Section 2.4 and formula (63) in \cite{CorwinLiuWang})
is the extended BR distribution (with no drifts), a specialization of \eqref{aga}
\bea
\text{Prob}( {\cal A}_{\rm stat}(\hat x) < \zeta) = H(\zeta + {\hat x^2} ; \frac{\hat x}{2}, - \frac{\hat x}{2})
= F_0(\zeta;\hat x)
\label{defH}
\eea
Note also the notation $F_{\hat x}(\zeta)$ introduced in \cite{FerrariSpohnStationary2006}
(see Appendix A and formula (1.20) there), to which we prefer the
notation $F_0(\zeta;\hat x)$. This distribution has zero mean for all $\hat x$ \cite{png},
and is even in $\hat x$,
a consequence of the statistical parity invariance of the process ${\cal A}_{\rm stat}(\hat x)$.


\subsection{Two-time problem}

From (\ref{hdrop}) we have, for the earlier time
\bea
&& h(0,t_1|0,0)  = t_1^{1/3}  {\cal A}_2(0)
\eea
For the later time $t_2$, we write
\bea
h(x,t_2|0,0) = \int dy ~ h(x,t_2|y,t_1) h(y,t_1|0,0)
\eea
where the two terms are statistically independent. In the large time limit it becomes
\bea
\fl && h(x,t_2|0,0) =  t_1^{1/3} \max_{\hat y} [  {\cal A}_2(\hat y) - \hat y^2
+  \Delta^{1/3}  \tilde {\cal A}_2(\Delta^{-2/3} (\hat x - \hat y)) - \Delta^{-1}  (\hat x - \hat y)^2 ]
\eea
where $\tilde {\cal A}_2$ is a second Airy process, independent of ${\cal A}_2$.
Here, and from now on we define
\bea
\! \! \! \! \! \! \! \!  \hat x=\frac{x}{2 t_1^{2/3}}  \quad , \quad \hat y=\frac{y}{2 t_1^{2/3}}  \quad , \quad  \hat X=\frac{x}{2 ( \Delta t_1)^{2/3}}  \quad , \quad  \hat Y=\frac{y}{2 ( \Delta t_1)^{2/3}} \label{defXY}
\eea

In other words, the scaled height variables introduced in (\ref{defh1}) and \eqref{defhX}
are expressed in terms of Airy processes
\bea
\fl ~~~~~~~~~~~~~~~~~h_1 &=&  {\cal A}_2(0) \nn \\
\fl  ~~h_1+ \Delta^{1/3} (h - \hat X^2)
&=& \max_{\hat y} [  {\cal A}_2(\hat y) - \hat y^2
+  \Delta^{1/3}  \tilde {\cal A}_2(\Delta^{-2/3} (\hat x - \hat y)) - \Delta^{-1}  (\hat x - \hat y)^2 ] \label{Airy1}
\eea

This correspondence allows to make some statements about the
large and small $\Delta$ limit, following, and then extending,
the analysis of \cite{FerrariSpohn2times}.
In both cases one uses the property
that the Airy process is locally a Brownian, i.e. one has, for $a \ll 1$
and as a process in $\hat y$
\bea \label{Brown}
{\cal A}_2(\hat z + a \hat y) = {\cal A}_2(\hat z) + \sqrt{2 a} B(\hat y) + a ~ {\cal C}(\hat y;\hat z) + o(a)
\eea
The precise mathematical statements can be found in
\cite{Haag,QuastelAiry1Brownian}. A crucial point, whose consequences
were not emphasized before, is that
$B(\hat y)$ and ${\cal A}_2(0)$ are uncorrelated
\footnote{a simple qualitative argument for that was pointed to us by I. Corwin. The
Airy process is the limit of the top curve of independent Brownians conditioned
not to intersect. Since at any typical point there is a finite gap between the
two top curves, the process is locally Brownian and uncorrelated to
its height.}. Note however that the regime of scales $a \hat y$ where the (uncorrelated) Brownian
approximation holds presumably becomes more narrow
when conditioning to a larger positive value of ${\cal A}_2(0)$ (as studied
below). Much less is known about the next order process
${\cal C}(\hat y;\hat z)$.

\subsection{ Large \texorpdfstring{$\Delta$}{Delta} limit}

We will consider the limit of large $\Delta$. In a first stage we will keep $\hat X$ fixed,
i.e. $X \sim (\Delta t_1)^{2/3}$ which is the standard KPZ scaling.
In that limit we can thus replace in (\ref{Airy1}), as a process in $\hat y$
\bea
\Delta^{1/3}  \tilde {\cal A}_2(\Delta^{-2/3} (\hat x - \hat y)) =
\Delta^{1/3} \tilde {\cal A}_2(\hat X) + \sqrt{2} \tilde B(\hat y) + O(\Delta^{-1/3}) \label{approxB}
\eea
where $\tilde B$ is a unit Brownian, uncorrelated with both $\tilde {\cal A}_2(\hat X)$ and with
with ${\cal A}_2(\hat y)$. Denoting for convenience and from now on $\sigma_1=h_1$ and
$\sigma=h$ the two random variables whose JPDF is studied in this work, we arrive at
the equalities in law
\bea
\fl  \sigma_1 &=&  {\cal A}_2(0) \\
\fl   \sigma &=& \tilde {\cal A}_2(\hat X) \label{sigmaexp}    \\
\fl  &+&  \Delta^{-1/3} \big( \max_{\hat y} [  {\cal A}_2(\hat y) - \hat y^2 + \sqrt{2} \tilde B(\hat y)
+ 2 \Delta^{-1/3} \hat X \hat y - \Delta^{-1} \hat y^2 + O(\Delta^{-1/3})]
- {\cal A}_2(0) \big) \nn
\eea
Thus we see that to retain a non-trivial dependence in the space variable of the leading correction in the
large $\Delta$ limit we must indeed introduce $\bar X = \Delta^{-1/3} \hat X$
as was found in the RBA calculation in Section \ref{sec:largeDelta}, see Eq. \eqref{scales}.
From \eqref{sigmaexp} we obtain
the following expansion in powers of $\Delta^{-1/3}$
\bea
&& \sigma_1 =  {\cal A}_2(0) \quad , \quad
\sigma =  \tilde {\cal A}_2(\hat X)  + \Delta^{-1/3} {\cal C}_1 + \Delta^{-2/3} {\cal C}_2 + ..  \label{AiryExp0} \\
&&
{\cal C}_1= \max_{\hat y} [  {\cal A}_2(\hat y) - \hat y^2 + \sqrt{2} \tilde B(\hat y) + 2 \bar X \hat y ] - {\cal A}_2(0) \label{AiryExp}
\eea
Since the point $\hat y^*$ where the maximum is attained is $O(1)$,
the quadratic term $\Delta^{-1} \hat y^2$ in (\ref{sigmaexp}) can be neglected, being subdominant.
One word of caution is needed here. Since we have little control on the
neglected term $O(\Delta^{-1/3})$ in \eqref{sigmaexp} (originating from the
neglected term in \eqref{approxB}) the dependence
on $\bar X$ given here and below is tentative. The only certain
result is that the correction ${\cal C}_1$ is independent of $\hat X$ at
fixed $\hat X$ (i.e. the considerations below are safest at the point $\bar X=0$).

The first and important conclusion of (\ref{AiryExp0}) is that
for $\Delta \to +\infty$ the JPDF of $\sigma,\sigma_1$ becomes the product of two independent
GUE-TW distributions.
Let us examine the corrections. First let us compare (\ref{AiryExp}) and \eqref{Astatx}.
One can rewrite, as an equality in law at fixed $\bar X$
\bea
&& \max_{\hat y} [  {\cal A}_2(\hat y) - \hat y^2 + \sqrt{2} \tilde B(\hat y) + 2 \bar X \hat y ] \nn \\
&& = \max_{\hat y} [  {\cal A}_2(\hat y) - (\hat y-\bar X)^2 + \sqrt{2} \tilde B(\hat y)  ] + \bar X^2
= {\cal A}_{\rm stat}(\bar X)  + \bar X^2
\eea
where we have used the stationarity of the Airy process, i.e. that
${\cal A}_2(\hat y)$ has the same statistics as ${\cal A}_2(\hat X - \hat y)$,
and \eqref{Astatx}. Hence we see that the first term in (\ref{AiryExp}),
equal to ${\cal C}_1 + {\cal A}_2(0)$, is distributed, up to a shift,
as ${\cal A}_{\rm stat}(\bar X)$, i.e.
according to the (extended) Baik-Rains distribution.
It is however correlated with ${\cal A}_2(0)$ in a
non-trivial way. Our RBA result for the conditional first moment (\ref{mean1}) gives some information about
this correlation: it predicts that at large $\sigma_1$
\bea
\fl && E( \max_{\hat y} [  {\cal A}_2(\hat y) - \hat y^2 + \sqrt{2} \tilde B(\hat y) + 2 \bar X \hat y ] |  {\cal A}_2(0) = \sigma_1) = \sigma_1 + R_{1/3}(\sigma_1) + O(e^{- \frac{4}{3} \sigma_1^{3/2}}) \nn \\
\fl &&  \label{res1}
\eea
In fact one can also check the reverse, i.e. calculate from (\ref{AiryExp0}) the
leading $O(\Delta^{-1/3})$ corrections to $P(\sigma,\sigma_1)$.
One precisely finds the decoupled form $-\Delta^{-1/3} F_2''(\sigma) \tilde R_{1/3}(\sigma_1)$
obtained in \eqref{Pexp},
if one assumes \eqref{res1} as well as
statistical independence of ${\cal C}_1$ and $\tilde {\cal A}_2(\hat X)$.
Thus, to that order, there is perfect agreement with our RBA result.
Note that in the limit $\sigma_1 \to +\infty$ one can argue that the position of the maximum
in (\ref{res1}) approaches zero as $\hat y^*\sim 1/\sigma_1$, correctly accounting for the large
$\sigma_1$ asymptotics of our RBA result $R_{1/3}(\sigma_1) \sim \sigma_1^{-1/2}$
{\footnote{ Indeed one can argue that for events such that ${\cal A}_2(0)=\sigma_1 \gg1$
the macroscopic profile of the Airy process minus the parabola, ${\cal A}_2(\hat y) - \hat y^2$,
is changed away, in a region of
extension $\sim \sigma_1$, from minus the parabola $-\hat y^2$
into a triangular shape $\sim - \sigma_1^{1/2} |\hat y|$.
On each side near zero the problem thus becomes similar to finding the maximum of
a unit one sided Brownian with a negative drift
$\sim - \sigma_1^{1/2}$, which is attained at a distance $y^* \sim \sigma_1^{-1/2}$
near the origin. We thank I. Corwin for help in setting up this argument.}}

Note that, since the BR distribution has zero mean, (\ref{AiryExp}) implies
that the leading correction $O(\Delta^{-1/3})$ to the (unconditionned) first moment is
consistent with \eqref{asympt1}.

%
%

Our result for the variance (\ref{var1}) shows the absence of correction
of $O(\Delta^{-1/3})$, which implies that the following conditional covariance
vanishes
\bea
\overline{ \tilde {\cal A}_2(\hat X) {\cal C}_1 }^c|_{{\cal A}_2(0) = \sigma_1} \simeq 0
\eea
This is, again, in perfect agreement with fact that the Brownian
$\tilde B(\hat y)$ and $\tilde {\cal A}_2(\hat X)$ are not correlated, as
mentionned above, hence that
${\cal C}_1$ and $\tilde {\cal A}_2(\hat X)$ are also uncorrelated.
The leading correction to the variance is thus of the form
\bea
\langle \sigma^2 \rangle_{\sigma_1} = \langle \sigma^2 \rangle^c_{\mbox{\tiny GUE}}
+ \Delta^{-2/3} \left( \overline{ {\cal C}_1^2 }^c|_{{\cal A}_2(0) = \sigma_1}
+ \overline{ \tilde {\cal A}_2(\hat X) {\cal C}_2 }^c|_{{\cal A}_2(0) = \sigma_1}  \right)
\eea
and since the total coefficient of the $O(\Delta^{-2/3})$ was found in (\ref{var1})
to be {\it negative}, it means that there must be a non-zero (negative) covariance
of $\tilde {\cal A}_2(\hat X)$ with the next order correction ${\cal C}_2$. Note from
(\ref{AiryExp}) that the unconditionned variance of ${\cal C}_1$
can be bound between the sum and (absolute value of) the difference of the BR and GUE variances,
recalled in (\ref{cumGUE})-(\ref{cumBR}).

\subsection{Persistence of correlations} The correction term $\Delta^{-1/3} {\cal C}_1$ in (\ref{AiryExp}) is
at the origin of the property of ``persistence of correlations'' in the case of
the droplet initial condition. The latter is usually expressed as the property that
the two-time connected correlation of the KPZ heights $H_1,H_2$ at two large and well separated times
$t_1,t_2$, and same space point $X_1=X_2=0$,
do not decay to zero in the limit $t_2/t_1 \to +\infty$ \cite{Takeuchi,TakeuchiPersistence,FerrariSpohn2times,GueudreUnpub}. For the
cross-second moment one can write
\bea
&& \overline{H_1 H_2}^c \simeq_{\Delta \to + \infty} c ~ t_1^{2/3}  \label{persist}
\eea
where the constant $c$ is given as
\bea
\fl && c = \lim_{\Delta \to +\infty}  \overline{\sigma_1 (\sigma_1 + \Delta^{1/3} \sigma) }^c
= \overline{ {\cal A}_2(0) \max_{\hat y} [  {\cal A}_2(\hat y) - \hat y^2 + \sqrt{2} \tilde B(\hat y) ]}^c
\eea
a universal strictly positive constant of $O(1)$. More generally, the non-trivial
JPDF of ${\cal A}_2(0)$ and ${\cal A}_{\rm stat}(0) = \max_{\hat y} [  {\cal A}_2(\hat y) - \hat y^2 + \sqrt{2} \tilde B(\hat y) ]$ characterizes the property of persistence of correlations. { In particular the asymptotic value of the conditional correlation ratio (\ref{condcorrfun}) can be expressed in terms of the Airy process as
\bea
c_{\Delta=+\infty, \sigma_c}
= \frac{ \overline{{\cal A}_2(0)
\max_{\hat y} [  {\cal A}_2(\hat y) - \hat y^2 + \sqrt{2} \tilde B(\hat y) ]}^c|_{{\cal A}_2(0) > \sigma_c} }{ \overline{{\cal A}_2(0)^2}|_{{\cal A}_2(0) > \sigma_c} } \label{cc}
\eea
i.e. the ratio of two covariances conditioned on events such that ${\cal A}_2(0) > \sigma_c$.
We have performed a preliminary numerical simulation to measure this
quantity directly from (\ref{cc}) using a lattice DP model, which led to a value of $c \approx 0.6$ in
the range of values obtained using the RBA arguments (see previous Section)
and leave a more complete and precise determination of the function $c_{\Delta=+\infty, \sigma_c}$
to the future.}

In the DP language, it can be viewed as a measure of the
tendency of the two DP of lengths $t_1$ and $t_2$ to share a finite fraction $0<q<1$ of the
shorter DP path, since both must come through space coordinate $0$ at time $t=0$. This is not the case
for the flat initial condition, where correlation does decay to zero. In terms of
optimal paths this is particularly clear (since those tend to coalesce, see e.g. \cite{Pimentel}) and both behaviors
(for droplet and flat) were observed in a numerical simulation in \cite{GueudreUnpub}.

What we further find here is that the constant $c$ in \eqref{persist} (and the full JPDF mentioned above) is also {\it independent of the spatial separation of the points} in the KPZ regime $X \sim (\Delta t_1)^{2/3}$ (fixed $\hat X$, $\bar X=0$). One needs a much larger spatial separation $X \sim \Delta^{1/3} (\Delta t_1)^{2/3}$ to ``unbind'' the two
DP paths from each others, an effect which is visible in the numerical data of Fig. \ref{fig:correlations_integrated_X}.

%

\subsection{Small \texorpdfstring{$\Delta$}{Delta} limit}

To study the limit $\Delta \to 0$, we now write the equations \eqref{Airy1} in terms of the
variables $\hat X$ and $\hat Y$, defined in \eqref{defXY}, as
\bea \label{innewvariables}
&& \!\!\!\!\!\!\! \!\!\!\!\!\!\! \!\!\!  \sigma_1 =  {\cal A}_2(0) \\
&& \!\!\!\!\!\!\! \!\!\!\!\!\!\! \!\!\! \!\!\! \!\!\!  \!\!\! \!\!\!  \sigma  - \hat X^2  = \max_{\hat Y} [  \Delta^{-1/3}  [ {\cal A}_2(\Delta^{2/3} \hat Y) - {\cal A}_2(0)]
- \Delta \hat Y^2  +   \tilde {\cal A}_2(\hat X - \hat Y) -  (\hat X - \hat Y)^2 ] \nn
\eea
In the small $\Delta$ limit we can use \eqref{Brown} and expand the second line as
\bea
&&  \!\!\!\!\!\!\!  \zeta =\sigma - \hat X^2 = \max_{\hat Y}[ \sqrt{2} B(\hat Y) + \tilde {\cal A}_2(\hat X - \hat Y) -  (\hat X - \hat Y)^2] \label{minim}
+ O(\Delta^{1/3}) \nn \\
&& \!\!\!\!\!\!\! = \tilde {\cal A}_{\rm stat}(\hat X) + O(\Delta^{1/3})
\eea
an equality valid in law. Hence as $\Delta \to 0$, the one point PDF of $\sigma$ converges to the (extended) Baik-Rains
distribution $\partial_\sigma F_0(\sigma-\hat X^2;\hat X)$. In addition, since ${\cal A}_2(0)$ and $B(\tilde y)$ are uncorrelated,
the full JPDF of $\sigma_1,\sigma$ decouples in that limit into the product
of the GUE-TW distribution and of the BR distributions. This is perfectly consistent with what is found
in this paper from the RBA method in (\ref{ressmalldelta}) and in Appendix \ref{app:w}.
There this property was shown in the tail, but the above
argument shows that it holds for all couples $\sigma_1,\sigma$.

Note that using the RBA we obtain that in the double limit of large $\sigma_1$ and small $\Delta$ with fixed
$\hat w_{\sigma_1,\Delta}=\Delta^{1/3} \sigma_1^{1/2}$, the JPDF of $\sigma_1$ and $\sigma$ remains a
product of a GUE-TW distribution times a BR distribution in presence of a wedge of slope $\hat w_{\sigma_1,\Delta}$.
This means that conditioning in \eqref{innewvariables} to a large positive value of ${\cal A}_2(0)$ leads to
an effective wedge, i.e. (for say $\hat X=0$) an additional term $- \hat w_{\sigma_1,\Delta} |\hat Y|$ in
(\ref{minim}). It would be desirable to confirm this result using finer properties of the Airy process.

The two point connected correlation is easily estimated at small $\Delta$ \cite{TakeuchiCrossover,FerrariSpohn2times} from
\bea
\fl && \overline{H_1 H_2}^c = \frac{1}{2} (\overline{H_1^2}^c + \overline{H_2^2}^c
- \overline{H_{21}^2}^c ) = t_1^{2/3}
\big( \frac{1 + (1 + \Delta)^{2/3}}{2} \langle h_1^2 \rangle^c_{\mbox{\tiny GUE}}
- \frac{1}{2} \Delta^{2/3} \langle h^2 \rangle^c \big)\nn \\
\fl && ~~~~~~~~~~~~~~~~~~~~~~~~~~~~~~~~~~~~~~~~= t_1^{2/3}  \big( \langle h_1^2 \rangle^c_{\mbox{\tiny GUE}} - \frac{1}{2}  \Delta^{2/3} \langle h^2 \rangle^c_{\mbox{\tiny BR}} + O(\Delta)  \big)
\eea
where the only property used is that $h$ approaches the BR distribution for small $\Delta$
{ (which must be replaced by the extended BR in the case where $\hat X \neq 0$).}
Other such identities
can be obtained, e.g. from expanding $\overline{H_{21}^3}^c$ one obtains the time reversal antisymmetric correlation
\bea
\overline{H_1^2 H_2}^c - \overline{H_1 H_2^2}^c \simeq \frac{\Delta}{3} t_1^{2/3}  ( \langle h^3 \rangle^c_{\mbox{\tiny BR}} -
 \langle h^3 \rangle^c_{\mbox{\tiny GUE}} ) + O(\Delta^{4/3})
\eea
Note however that the statistical independence of $H_1$ and $H_{21}$ in the limit $\Delta \to 0$
is a stronger prediction, leading to the vanishing of all
$\lim_{\Delta \to 0} \overline{ H_{21}^p H_1^q}^c/(\Delta^{p/3} t_1^{(p+q)/3})$,
which could be checked in numerics and experiments.

\section{Conclusion} \label{sec:conclusion}

In summary we have obtained some exact results for the correlation functions at two different times in the
KPZ equation with droplet initial conditions, when both times are large, but their ratio $t_2/t_1$ is fixed.
We have used the replica Bethe ansatz method, which
has been very successful for the one-time problem, but until now, failed to produce correct and testable results
for the two-time
problem, the main difficulty being the calculation of the so-called form factors and the summation
of the resulting expressions over all eigenstates.

Our method could be described as ``catching the devil by the tail". We have restricted the summation over a simpler set of
states, enough to be able to perform the calculation with no further approximations,
but still rich enough to capture the exact tail of the two-time joint probability distribution function (JPDF). More precisely, calling $h_1$ the
scaled height at time $t_1$ and $h$ the {\it difference} between the height at time $t_2$ and the one at time $t_1$,
we have obtained the exact behavior of the JPDF, $P_{\Delta}(h_1,h)$, for large and positive $h_1$ and for any $h$. Furthermore it is expected to be quite good even for moderate values of $h_1$, not necessarily very large, in particular we expect very precise results in the whole region $h_1>-1$.

Our result for the tail is, therefore, rich enough to capture a notable physical aspect, namely that the JPDF factorizes into
two GUE-TW distributions when the ratio $t_2/t_2$ diverges, and into the product of a GUE-TW
and a Baik-Rains distribution when $t_2/t_1 \to 1^+$. Hence the feature of stationarity at two close times, which has been conjectured
and observed (to some extent) in numerics and experiments, is quite easily recovered.
To our knowledge this is the first time that this feature is shown through an
analytic calculation, and here we obtain it from the RBA method.
Our formula therefore provides the full crossover, in the large $h_1$ region, from BR stationarity
to GUE-TW as the time separation increases. It produces a number of results for the cumulants of $h$,
conditioned to a fixed
value, or interval of values, of $h_1$ that can be numerically evaluated. All these predictions being universal, we hope that
they will be tested soon in numerics and experiments.

Although this still leaves the formidable challenge of obtaining the complete JPDF $P_{\Delta}(h_1,h)$ as an open problem,
the partial summation carried here, and the consequent results, will certainly prove themselves as a useful guide for further progress.
In particular the constraint that stationarity must be recovered at close times emerges quite naturally within our approximation, and it remains to understand how it emerges more generally within the RBA. On the other hand,
extensions to other classes of initial conditions seem possible, and work in that direction is in progress. \\

%
\section*{Acknowledgments}

We thank P. Calabrese, I. Corwin, A. Borodin, K. Johansson, and K. Takeuchi for discussions.
This work is supported in part by LabEX ENS-ICFP:ANR-10-LABX-0010/ANR-10-IDEX-0001-02 PSL* (J.D.N.), and the National Science Foundation under Grant No. NSF PHY11-25915.

\newpage

\begin{appendices}

\section{Comment on previous works on the two-time problem }
\label{app:dots}
{\blue
We now discuss relations of the present work with the results reported in references \cite{dotsenko2times1,dotsenko2times2}. In formula (73) of reference \cite{dotsenko2times1} is presented an expression for the infinite time cumulative distribution $W(f_1, f_2,\Delta)$ of the two-time problem
with $X=0$. Using the following change of notation
\begin{align}\label{changevarDots}
& f_1 \Big|_{\text{Ref} \text{\cite{dotsenko2times1}}}= - \sigma_12^{2/3} \nn \\
& f_2 \Big|_{\text{Ref} \text{\cite{dotsenko2times1}}}= - 2^{2/3} \frac{\sigma_1 + \sigma \Delta^{1/3}}{(1 + \Delta)^{1/3}}
\end{align}
the JPDF  $ \partial_{f_1} \partial_{f_2} W(f_1,f_2,\Delta) $ obtained from formula (16) there,
corresponds to the JPDF computed in the present paper, defined in \eqref{gdef1}, is
\begin{equation}
\partial_{f_1}  \partial_{f_2} W(f_1, f_2,\Delta)\Big|_{\text{Ref} \text{\cite{dotsenko2times1}}} df_1 df_2 \to  P_{\Delta} (\sigma_1, \sigma ) d \sigma d\sigma_1
\end{equation}
while $\Delta$ is a common notation here and there.
The cumulative distribution $W(f_1, f_2,\Delta)$ is computed in principle by summing over all the number of replicas, i.e. all the possible string configurations for the state $| \gamma \rangle$ and the state $ | \mu \rangle$ (in our notations). However in Ref. \cite{dotsenko2times1} a substantial simplification is performed (not mentioned in the text), namely the author imposes on the starting formula (37) the following condition on the string configurations of the two states
\begin{equation}
n_s^\gamma \geq n_s^\mu\Big|_{\text{Ref} \text{\cite{dotsenko2times1}}}
\end{equation}
where in the notation of reference \cite{dotsenko2times1} $M_1 = n_s^\mu$ and $M=M_1 + M_2 = n_s^\gamma$, with $M_1,M_2 \geq 0$.
Therefore we claim that formula (73) in \cite{dotsenko2times1} cannot be considered \textit{a priori} as the exact expression of the cumulative two-time distribution
\begin{equation}\label{claimwrong}
\partial_{f_1}  \partial_{f_2} W(f_1, f_2,\Delta)  df_1 df_2 \Big|_{\text{Ref} \text{\cite{dotsenko2times1}}}  \neq P_{\Delta} (\sigma_1, \sigma) d \sigma d\sigma_1
\end{equation}
since all the terms with different types of string contents as $n_s^\gamma < n_s^\mu$ must also be taken into account.
In this work we focus on the cases
\begin{equation}
n_s^\gamma  =1 \leq n_s^\mu =\{1,2,3,\ldots, \infty \}
\end{equation}
and we show indeed that these are non-trivial terms which lead to a finite contribution to the probability distribution in the limit of large times, and for any $\Delta$.  We emphasize that the only contribution to the cumulative distribution function that is present in both works is the first diagonal contribution $1 = n_s^\gamma  = n_s^\mu$ which corresponds to the term $g^{(1,1)}_{\Delta,\lambda }(s_1, s)$ in this paper, and which indeed is equal (after implementing the change of variable \eqref{changevarDots}) to the contribution $M_1 = 1, M_2 = 0$ in formula (73) in \cite{dotsenko2times1}
\begin{equation}
W(f_1, f_2,\Delta) \Big|_{M_1=1,M_2= 0} = g^{(1,1)}_{\Delta} (- f_1, \frac{f_1 - f_2 (1 + \Delta)^{1/3}}{\Delta^{1/3}}  )
\end{equation}

In reference \cite{dotsenko2times2} another cumulative distribution is considered, however also in this case a similar approximation is done by imposing
\begin{equation}
n_s^\gamma =  n_s^\mu\Big|_{\text{Ref} \text{\cite{dotsenko2times2}}}
\end{equation}
where in principle all the contribution with different $n_s^\gamma ,  n_s^\mu \geq 0$ can contribute to the final formula (20) in \cite{dotsenko2times2}.  \\

A non-trivial check of our claim \eqref{claimwrong} can be performed in the limit $\Delta \to 0$.
The limit $\Delta \to 0$ of formula (73) in \cite{dotsenko2times1} was obtained in
Ref. \cite{dotsenko2times3}, see equation (19) there. However, this formula is
not the product of a GUE-TW distribution and a Baik-Rains distributions as predicted in the present paper (see equations \eqref{decoupleBR} and \ref{prod2}). The occurence of stationarity, hence of the
Baik-Rains distribution, in the limit $\Delta \to 0$, is an important physics requirement (also discussed in \cite{TakeuchiCrossover,FerrariSpohn2times})
which is missed by the neglect of important terms
in \cite{dotsenko2times1,dotsenko2times2,dotsenko2times3}.
}

\section{Calculation of one-string to multiple-string form factor} \label{app:ff}

\subsection{A magic identity for stationary KPZ}

In the RBA study of the KPZ equation with an initial condition given by a two-sided
Brownian \cite{SasamotoStationary}, with drifts $w_L$ and $w_R$ respectively, the following integrals
are needed
\begin{eqnarray} \label{Gla1}
\fl && G_{w}^L[\lambda_1,..,\lambda_p] = \int_{y_1<y_2< ..<y_p<0}
e^{\sum_{j=1}^p (w + i \lambda_i) y_j - \frac{1}{2} (2 j-1) y_j }
= \prod_{j=1}^{p} \frac{1}{j w + i \lambda_1 + .. + i \lambda_{j} - \frac{a j^2}{2} } \nn \\
\fl &&  G_{w}^R[\lambda_1,..,\lambda_p]  = \int_{0<y_1<y_2< ..<y_p}
e^{\sum_{j=1}^p (i \lambda_i-w) y_j + \frac{1}{2} (2 p+1-2 j) y_j }
= \prod_{j=1}^{p} \frac{1}{j w - i \lambda_p - .. - i \lambda_{p+1-j} - \frac{a j^2}{2} }
 \nonumber \\
 \fl &&
\end{eqnarray}
such that $G^R_{w}[\lambda_1,..,\lambda_p]  = G^L_{w}[- \lambda_p,.. ,-\lambda_1]$
(by convention for $p=0$, $G^R=G^L=1$).
In each line, the last identity is valid in the domain of parameters where the integral converge,
to which we will restrict from now on.
It was shown there \cite{SasamotoStationary} that the following ``magic" identity holds
\bea \label{magic}
\fl && \sum_{P \in S_n} A_P
\sum_{p=0}^n G_{w_L}^L[\lambda_{P_1},..,\lambda_{P_p}] G_{w_R}^R[\lambda_{P_{p+1}},..,\lambda_{P_n}]
= 2^{2 n}
\prod_{j=1}^n \frac{(w_L+w_R-j)}{(2 w_L -1 + 2 i \lambda_j)(2 w_R -1 - 2 i \lambda_j)} \nn \\
\fl &&
\eea
where the sum is over all permutations $P$ of the $n$ rapidities, and
the factors $A_P$ are given by (\ref{def1}). It is the existence of this identity which allows
for an explicit solution of the KPZ equation with stationary initial conditions, as
obtained in \cite{SasamotoStationary}.

\subsection{Application to the form factor}

Let us introduce the function
\be
\phi(y)= \theta(-y) e^{w_L y}
+\theta(y) e^{-w_R y}
\ee
The general form factor needed for the two-time JPDF problem is (upon a slight
generalization to $w_{L,R}$)
\bea
\fl && F^{n_2;n_1+n_2}_{\mu;\gamma} \equiv  \prod_{a=1}^{n_2} \int  dy_a \phi(y_a)
\psi_\mu^*(y_1,..,y_{n_2})
\psi_\gamma(x_1=0,..,x_{n_1}=0,y_{1},..,y_{n_2}) \\
\fl && = n_2! \sum_{P \in S_{n_2}} A^*_P \sum_{p=0}^{n_2} \int_{y_1<y_2..<y_p<0<y_{p+1}<..y_{n_2}}
\prod_{a=1}^{n_2} \phi(y_a)
 e^{- i \mu^*_{P_a} y_a}
\psi_\gamma(x_1=0,..,x_{n_1}=0,y_{1},..,y_{n_2}) \nn
\eea
where $A_P$ are given by (\ref{def1}) with $\lambda_\alpha \to \mu_\alpha$ and $n \to n_2$,
and complex conjugation is applied at the end to obtain $A_P^*$. In the second
line we have used the full symmetry of the wave-functions, chosen a particular order
for the $y_\alpha$, and inserted the explicit form (\ref{def1}) of the Bethe eingenfunction for
the state $|\mu \rangle$.

Consider now the case where $|\gamma\rangle$ is a single string $|\gamma\rangle = |q,N=n_1+n_2\rangle $. Recall that the wavefunction of a single string is given in the sector
$z_1 \leq .. \leq z_N$, by
\bea
\psi_\gamma(z_1,..,z_{N}) = N! e^{\sum_{\alpha=1}^N (i q + \frac{1}{2} (N+1-2 \alpha) ) z_\alpha }
\eea
Hence in the sector $y_1<..<y_p<x_1=..x_{n_1}=0<y_{p+1}<..y_{n_2}$ it becomes
\bea
\fl && \psi_\gamma(x_1=0,..,x_{n_1}=0,y_{1},..,y_{n_2})
= N! e^{\sum_{a=1}^p (i q + \frac{N+1}{2} ) y_a - a y_a }
e^{\sum_{a=p+1}^{n_2} (i q + \frac{N+1}{2} ) y_{a} - (a+n_1) y_{a} }
\eea

Putting all together we need to calculate
\bea
\fl && F^{n_2;n_1+n_2}_{\mu;\gamma} =
 n_2! N! \sum_{P \in S_{n_2}} A^*_P \sum_{p=0}^{n_2} \int_{y_1<y_2..<y_p<0<y_{p+1}<..y_{n_2}}
 e^{- \sum_{a=1}^{n_2} i \mu^*_{P_a} y_a } \\
\fl  && ~~~~~~~~~~~~
\times e^{\sum_{a=1}^p (i q + \frac{N+1}{2} ) y_a - (a - w_L) y_a }
e^{\sum_{a=p+1}^{n_2} (i q + \frac{N+1}{2} ) y_{a} - (a+n_1+w_R) y_{a} } \nn
\eea

On the other hand we can rewrite the functions defined above as
\bea
&& G_{w'_L}^L[\lambda_1,..,\lambda_p] = \int_{y_1<y_2< ..<y_p<0} e^{\sum_{a=1}^p i \lambda_a y_a }
e^{ \sum_{a=1}^p ( w'_L + \frac{1}{2} - a ) y_a } \\
&& G_{w'_R}^R[\lambda_1,..,\lambda_p]  = \int_{0<y_1<y_2< ..<y_p} e^{\sum_{a=1}^p i \lambda_a y_a }
e^{\sum_{a=1}^p (- w'_R + \frac{1}{2} + p - a) y_a } \\
&& G_{w'_R}^R[\lambda_{p+1},..,\lambda_{n_2}]  = \int_{0<y_{p+1}<y_2< ..<y_{n_2}}
e^{\sum_{a=p+1}^{n_2} i \lambda_a y_a } e^{\sum_{a=p+1}^n (- w'_R + \frac{1}{2} + n_2 - a) y_a } \nn \\
&&
\eea

Hence we can identify
\bea
\fl && ~~~~~ i \lambda_a + w'_L + \frac{1}{2} =- i \mu^*_a + w_L + i q + \frac{N+1}{2} \quad , \quad a=1,..,p \\
\fl && ~~~~~ i \lambda_a - w'_R + \frac{1}{2} + n_2 =- i \mu^*_a - w_R + i q + \frac{N+1}{2} - n_1 \quad , \quad a=p+1,..n_2
\eea
For instance, we can identify $\lambda_a \equiv - \mu^*_a$ and
$w'_L \equiv w_L + i q + N/2$ and $w'_R = w_R - i q - N/2 + n_1 + n_2= - i q + N/2$, and rewrite
\bea
\fl &&  ~~~~~~ F^{n_2;n_1+n_2}_{\mu;\gamma}  =
 n_2! N! \big[ \sum_{P \in S_{n_2}}  A_P \sum_{p=0}^{n_2}
G_{w'_L}^L[\lambda_{P_1},..,\lambda_{P_p}]
G_{w'_R}^R[\lambda_{P_{p+1}},..,\lambda_{P_{n_2}}] \big]_{\lambda_a \to - \mu^*_a}
\eea
where we have used that $A_P^*=A_P|_{\mu_a \to - \mu_a^*}$. We can now use the magic identity (\ref{magic}) and, remembering that $N=n_1+n_2$, we obtain our final result for the form factor with $n_s^\gamma=1$ and arbitrary eigenstate $|\mu \rangle$ as
\bea
\fl && F^{n_2;n_1+n_2}_{\mu;\gamma}  =  n_2! (n_1+n_2)!
2^{2 n_2} \\
\fl && ~~~~~~~~~~~ \times
\prod_{j=1}^{n_2} \frac{w_R+w_L+n_1+n_2 -j}{(2 w_L+2 i q + n_1+n_2 -1 - 2 i \mu^*_j)(2 w_R -2 i q + n_1+n_2  -1 + 2 i \mu^*_j)} \nn
\eea
If we now specify that $|\mu \rangle$ is also a string state with $|\mu \rangle=|{\bf p} , {\bf m}^\mu \rangle$
we can express the product in terms of Pocchammer symbols $(x)_n = x(x+1)..(x+ n-1)=\Gamma(x+n)/\Gamma(x)$ leading to
\begin{align}
& F^{n_2;n_1+n_2}_{{\bf p} , {\bf m}^\mu;q , n_1+n_2 } = n_2! (n_1+n_2)! (w_L+w_R + n_1)_{n_2}
\nn \\& \times
\prod_{j=1}^{n_s} \frac{1}{(w_L + i (q - p_j) + \frac{n_1+n_2-m_j}{2})_{m_j}
(w_R - i (q - p_j) + \frac{n_1+n_2-m_j}{2})_{m_j}}
\end{align}
which, for $w_L=w_R=w$, reduces to the formula (\ref{ffformula}) in the text. Finally note that
since $Re(w_{L,R}) \geq Re(w'_{L,R})$ in the above identifications, there is no additional
convergence problem with respect to the calculations in \cite{SasamotoStationary} (convergence
issues were discussed and resolved there).

\section{Calculation of \texorpdfstring{$\hat Z_{1, 1}(s_1,s) $}{FIXME}}
\label{sec:Z11}

In this Appendix we present an independent calculation of the contribution
of a single string both for the $\mu$ and $\gamma$ eigenstates. It
is related to the ``double tail" of the two times
JPDF, where both $h_1$ and $h_2-h_1$ are large.

\begin{align}
& \hat Z_{1, 1}(s_1,s)
= \sum_{n_1 \geq 1,n_2 \geq 1} \frac{(-1)^{n_1 + n_2}}{n_1! n_2!}
\int \frac{d p}{2\pi n_2}
\int \frac{d q}{2\pi (n_1+n_2)}  \nn  \\
& \times
e^{\frac{1}{12} n_2^3 \Delta t_1 - n_2 p^2 \Delta t_1 - \frac{1}{12}
(n_1+n_2)^3 t_1 - (n_1+n_2) q^2 t_1 - \lambda s_1 (n_1 + n_2) - \lambda \Delta^{1/3} n_2 s+
2 i \lambda^2 \Delta^{2/3} \tilde X p n_2} \nn \\&
 \frac{n_2! (n_1+n_2)! (2 w + n_1)_{n_2}}{(w + i (q - p) + \frac{n_1}{2})_{n_2}
(w - i (q - p) + \frac{n_1}{2})_{n_2}}
\end{align}
We now apply Mellin-Barnes in order to compute the sums over $n_1,n_2$
\begin{equation}
\sum_{n_1 \geq 1,n_2 \geq 1}  {(-1)^{n_1 + n_2}}{ } f(n_1,n_2)= \int_{-i \infty + \epsilon}^{+ i \infty + \epsilon}  \frac{dz_1}{2 i } \int_{-i \infty + \epsilon}^{+ i \infty + \epsilon} \frac{dz_2}{2  i } \frac{1}{\sin(\pi z_1) \sin(\pi z_2)}  f(z_1,z_2)
\end{equation}
where we assumed that there are no extra poles of $f(z_1,z_2) $ on the plane $\Re z_{1,2} > 0$. (Note that $\epsilon > 0$ since we exclude the pole in $z_{1,2}=0$).
We can rescale $z_1,z_2 \to \frac{z_1}{\lambda}, \frac{z_2}{\lambda}$ and take the infinite $\lambda$ limit
\begin{equation}
\lim_{\lambda \to \infty} \sum_{n_1 \geq 1,n_2 \geq 1}  {(-1)^{n_1 + n_2}}{ } f(n_1,n_2)= \int_{-i \infty + \epsilon}^{+ i \infty + \epsilon}  \frac{dz_1}{2 \pi i } \int_{-i \infty + \epsilon}^{+ i \infty + \epsilon} \frac{dz_2}{2 \pi i } \frac{1}{z_1 z_2} \lim_{\lambda \to \infty} f(\frac{z_1}{\lambda} , \frac{z_2}{\lambda})
\end{equation}
We also perform the change of variable $p \to p/(2 \lambda)$, $q \to q/(2 \lambda)$ and used the
rescaled slope $\tilde w = w \lambda \Delta^{1/3}$.

We obtain then in the limit $\lambda \to +\infty$:
\begin{align}\label{z11}
& \hat Z_{11}(s_1,s) =\int_{-i \infty + \epsilon}^{+ i \infty + \epsilon} \frac{dz_1}{2 \pi i } \int_{-i \infty + \epsilon}^{+ i \infty + \epsilon} \frac{dz_2}{2 \pi i }  \frac{1}{z_1 z_2} \frac{1}{(z_1  + z_2) z_2} \frac{z_1 +2 \frac{\tilde{w}}{\Delta^{1/3}}}{z_1 + z_2 + 2 \frac{\tilde{w}}{ \Delta^{1/3}} } \\&
\int \frac{dp}{4 \pi} \frac{dq}{4 \pi}  e^{\Delta z_2^3/3 -  \Delta z_2 p^2 + (z_1 + z_2)^3/3  - (z_1 + z_2) q^2  - (z_1 + z_2) s_1 - z_2 s \Delta^{1/3} + i \tilde X z_2 p \Delta^{2/3}} \nn \\&
\times \left( 1+\frac{ z_2}{ i \frac{p- q}{2} +\frac{z_1}{2}+ \frac{\tilde{w}}{\Delta^{1/3}}} \right) \left( 1+\frac{ z_2}{ -i \frac{p- q}{2}+\frac{z_1}{2}+ \frac{\tilde{w}}{\Delta^{1/3}}} \right)  \nn \\&= \int_0^\infty dv_1 dv_2 \int \frac{dp}{4 \pi} \frac{dq}{4 \pi}
\int \frac{dy_1}{\Delta^{1/3}} \int  {dy_2}{} \Ai(y_2 \Delta^{-1/3} + \Delta^{2/3}   p^2 + s + v_2 \Delta^{-1/3}  - i \tilde X p \Delta^{1/3} ) \nn  \\& \label{z11-2}  \times  \Ai( {y_1 +  q^2 + s_1   + v_1  }{ } ) \Big(\int_{-i \infty + \epsilon}^{+ i \infty + \epsilon} \frac{dz_1}{2 \pi i } \int_{-i \infty + \epsilon}^{+ i \infty + \epsilon} \frac{dz_2}{2 \pi i }  \frac{1}{z_1 z_2 } \\&     e^{ y_1 (z_1 + z_2) +  y_2 z_2} \left(
1+\frac{ z_2}{ i \frac{ p- q}{2}+ \frac{z_1}{2}+ \frac{\tilde{w}}{\Delta^{1/3}}}
\right)
\left(
1+\frac{ z_2}{ -i \frac{p- q}{2}+\frac{z_1}{2}+ \frac{\tilde{w}}{\Delta^{1/3}}}
\right)  \frac{z_1 + 2 \frac{\tilde{w}}{\Delta^{1/3}}}{z_1 + z_2 + 2 \frac{\tilde{w}}{\Delta^{1/3}}}         \Big) \nn
\end{align}
where in the last line we have used the Airy trick () and a shift of the variables
$y_1$ and $y_2$.

The limit $\tilde w \to 0^+$ can be taken by just setting $\tilde w=0$ in the
above complex integral over $z_1,z_2$, since they have a small positive real part.
This leads to some simplifications, and the integral is then easy to calculate,
by expanding the product in the last line of (\ref{z11-2}). This leads to
\begin{align}
& \hat Z_{11}(s_1,s)   = \int_0^\infty dv_1 dv_2 \int \frac{dp}{4 \pi} \frac{dq}{4 \pi}
\int \frac{dy_1}{\Delta^{1/3}} \int  {dy_2}{}  \Ai( {y_1 +  q^2 + s_1   + v_1  }{ } ) \nn \\& \times
 \Ai(y_2 \Delta^{-1/3} + \Delta^{2/3}   p^2 + s + v_2 \Delta^{-1/3}  - i \tilde X p \Delta^{1/3} )
 \nonumber \\& \times \Big(  \theta(y_1)\theta(  y_2)   -    4 \delta(y_1 + y_2) \theta(y_1)   \frac{\sin((q- p) y_1)}{(q-p)}        \Big)
\end{align}
In both terms we change $p \to p \Delta^{-1/3}$,
$v_2 \to v_2 \Delta^{1/3}$, and in the first one we also change
$y_2 \to y_2 \Delta^{1/3}$, while in the second one $y_2=-y_1$
because of the delta function. The sum can then be rewritten as
\begin{align} \label{Z11res}
&   \hat Z_{11}(s_1,s)  =    \left( \frac{1}{3 \pi} \int_0^\infty y^{3/2} \text{Ai}(y +   {s_1}{  }) \right) \left( \frac{1}{3 \pi} \int_0^\infty y^{3/2} \text{Ai}(y +  s ) \right)
  \\& \nn +\frac{1}{   \Delta^{1/3} } \int \frac{dq}{2 \pi  }   \int \frac{dp}{2 \pi  }  \int_{0}^{+\infty} dv_1 \int_{0 }^{+\infty} dv_2  \int_0^\infty dy  \\& \text{Ai}(- y \Delta^{-1/3} +  {     p^2 + v_2 }{ } + s   ) \text{Ai}(y +  {   q^2 + v_1 } + s_1 )   \frac{\sin[( {q}{ }-\frac{p}{\Delta^{1/3}})y  ]}{ {q}{ }-\frac{p}{\Delta^{1/3}}} \nn
\end{align}
We have used that
\bea
&&  \int_0^\infty dv_1 dy_1 \int \frac{dq}{4 \pi}  \Ai(y_1 +  q^2 + s_1  + v_1) =
\frac{1}{3 \pi} \int_0^\infty y^{3/2} \text{Ai}(y +  s_1) 
\eea

We can now check that this result can also be obtained from our more
general result (\ref{final0}) and ``expanding in the number of Airy functions"
containing $\sigma$, i.e. writing $\hat{g}^{(1)}_{\Delta }(s_1,s) = \hat{g}^{(1,1)}_{\Delta}(s_1,s) + \cdots$
where
\bea \label{g11res}
\fl  \hat{g}^{(1,1)}_{\Delta}(s_1,s) &=&
 \int_0^{+\infty} du K_\Ai(u+\sigma_1, u+\sigma_1)  \int_0^{+\infty} du K_\Ai(u+\sigma, u+\sigma) \\
\fl  & +& \Delta^{1/3}  \int_0^{+\infty} du K^{\Delta}_{\sigma_1}(u+\sigma_1, u+\sigma_1) \nn
\eea
should identify with $\hat Z_{11}(s_1,s)$. Using (\ref{id2})  is clear that the first
terms in both formula (\ref{Z11res} ) and (\ref{g11res}) are identical.
To check that the second terms are also the same, one first introduces the representation
$\frac{\sin(a y)}{a}
= \frac{1}{2} \int_{-y}^y du e^{i u a}$ with $a=q-\frac{p}{\Delta^{1/3}}$
in (\ref{Z11res} ), then uses (\ref{idAiry}) for both the integrals over
$p$ and $q$. Integration over $v_1$ and $v_2$ yields two Airy kernels,
and the change of variable $y=2^{-1/3} \Delta^{1/3} (y_1+y_2)$,
$u=2^{-1/3} \Delta^{1/3} (y_1-y_2)$ allows rewrite the
second term as
\bea
\fl && ~~~~~~~ \Delta^{1/3}  \int_0^{+\infty} dy_1 dy_2 K_\Ai(-y_1+\sigma, - y_2+\sigma)
K_\Ai(y_1 \Delta^{1/3} +\sigma, y_2 \Delta^{1/3} +\sigma) \\
\fl && ~~~~~~~~~~~~~~~~ =
\Delta^{1/3}  \int_0^{+\infty} du K^{\Delta}_{\sigma_1}(u+\sigma_1, u+\sigma_1)
\eea

\section{Small \texorpdfstring{$\Delta$}{Delta} limit and extended Baik-Rains distribution with finite drifts}
\label{app:w}

In this Section we first obtain a general formula for the generating function
for both times large, and for arbitrary $\Delta$ and arbitrary rescaled slope $\tilde w$ (also called drift).
Then we study its small $\Delta$ limit and show that the JPDF converges to
the product of the GUE-TW distribution and the extended Baik-Rains distribution with
a finite drift.

Let us start again from \eqref{eq:final_largetime} and keep the rescaled slope $\tilde{w}$ finite.
We can then rewrite the factors in the complex integrals as follows
\begin{align}
& \frac{1}{z_2 (z-z_2)} \frac{z - z_2 + \frac{2 \tilde{w}}{\Delta^{1/3}}}{z  + \frac{2 \tilde{w}}{\Delta^{1/3}}} e^{y z  + (4\Delta)^{1/3} z_2 (\kappa - \sigma)} = \frac{1}{z_2 (z-z_2)}  \left( 1- \frac{z_2}{z + \frac{2 \tilde{w}}{\Delta^{1/3}}} \right)
 e^{y z  + (4\Delta)^{1/3} z_2 (\kappa - \sigma)} \nn  \\&
  = \frac{1}{z_2 (z-z_2)}  \left( 1- \frac{1}{(4\Delta)^{1/3}}\frac{\partial_{\kappa}}{z + \frac{2 \tilde{w}}{\Delta^{1/3}}} \right)
 e^{y z  + (4\Delta)^{1/3} z_2 (\kappa - \sigma)}
\end{align}

We can then integrate by parts and move the derivative on the determinant and rewrite formula \eqref{eq:final_largetime} as
\begin{align}
&  \hat{g}^{(1)}_{\Delta }(s_1  , s) \nn \\ &
= \int \frac{d q}{4\pi   }  \int_0^\infty dv  \int  {dy}{ }   \int_{-\infty}^\infty  {d\kappa}{ } \Ai(  {y}{ } + v  + q^2 + 2^{2/3} \sigma_1 )   \nn \\&  \int_{-i\infty   + { 2\epsilon  } }^{i\infty   + 2\epsilon}
\frac{dz}{2 \pi i }   \int_{-i\infty  + \epsilon }^{i\infty  + \epsilon } \frac{dz_2}{2 \pi i }
     \frac{   e^{y z + (\kappa - \sigma)z _2 (4\Delta)^{1/3} }   }{z_2 (z - z_2) }
         \partial_{ \kappa}  \left( 1 + \frac{1}{(4\Delta)^{1/3}} \frac{\partial_{ \kappa}}{z + \frac{2 \tilde{w}}{\Delta^{1/3}}} \right) {\rm Det} [I - P_{ \kappa}\mathcal{F}^{\tilde{w}}_{ z,q} P_{ \kappa} ]
\end{align}
which is a general formula for finite slope $\tilde w$, where the kernel is given in (\ref{Fw})

Let us consider now the small $\Delta$ limit. In that limit the kernel in (\ref{Fw}) loses its dependence on $z$ and $q$ and becomes
\begin{align}
 \lim_{\Delta \to 0}  \mathcal{F}^{\tilde{w}}_{ z,q}(u_i,u_j)   & = B_{\hat{w}+ \hat X,\hat{w}- \hat X }(u_i,u_j)  \\
  \equiv  & \int_{0}^{\infty}dy
\Ai^{+}_{-}
\left( y+   u_i ,  \hat{w} + \hat X  , \hat{w} - \hat X   \right)     \Ai^{+}_{-}
\left( y+   u_j , \hat{w} - \hat X  , \hat{w}  + \hat X   \right) \nn
 \end{align}

Then we can integrate over $z$ and $z_2$ and obtain
\begin{align}
\hat g^{(1)}_{\Delta}(s_1  , s)  &
 \approx_{\Delta \to 0}       \int \frac{d q}{4\pi   }  \int_0^\infty dv \int  {dy}{ } \int_{-\infty}^\infty  {d\kappa}{ }  \Ai(  {y}{ } + v + q^2 + 2^{2/3} \sigma_1)\nn  \\& \times
\theta( y)  \theta(y +(4\Delta)^{1/3} (\kappa  - \sigma)  )  \partial_\kappa  \left( 1 + \frac{\partial_{\kappa}}{2 \hat{w}}\right) {\rm Det} [I - P_{\kappa}
B_{\hat{w}+ \hat X,\hat{w}- \hat X }  P_{\kappa} ]
\end{align}

This generating function is such that when we apply the differential operator $\partial_{\sigma_1 } \partial_{\sigma} - \frac{1}{\Delta^{1/3}} \partial^{2}_{\sigma}$
we obtain
\begin{equation}
\lim_{ \Delta \to 0} P^{(1)}_{\Delta}(\sigma_1,\sigma)  =  \partial_{\sigma_1}  F^{(1)}_2(\sigma_1)  \partial_{\sigma}   \left( 1 + \frac{\partial_{\sigma}}{2 \hat{w}}\right)  {\rm Det} [I - P_{\sigma }
B_{\hat{w}+ \hat X,\hat{w}- \hat X } P_{\sigma} ]
\end{equation}
where we have first shifted $y \to y - 2^{2/3} \sigma_1$, and then used that
\begin{equation}
[\partial_{\sigma_1} \partial_{\sigma} - \frac{1}{\Delta^{1/3}} \partial_{\sigma}^2]   \theta( y - 2^{2/3} \sigma_1)  \theta( y - 2^{2/3} \sigma_1 +(4\Delta)^{1/3} (\kappa  - \sigma)  ) = 2^{2/3} \delta(y - 2^{2/3} \sigma_1 ) \delta(\kappa - \sigma)
\end{equation}
as well as, using (\ref{idAiry}),
\begin{equation}
2^{2/3}  \int \frac{d q}{4\pi   }  \int_0^\infty dv     \Ai(  q^2  + v + 2^{2/3} \sigma_1 ) = K_\Ai(\sigma_1,\sigma_1)= \partial_{\sigma_1}  F^{(1)}_2(\sigma_1)
\end{equation}

Hence we obtain
\begin{equation}
\lim_{ \Delta \to 0} P^{(1)}_{\Delta}(\sigma_1,\sigma)  = F^{(1) \prime}_2(\sigma_1)
\ \partial_\sigma F_0(\sigma - \hat X^2 ; \hat X, \hat w)
\end{equation}
where
\bea
F_0(\sigma - \hat X^2 ; \hat X, \hat w) \equiv  
  \left( 1 + \frac{\partial_{\sigma}}{2 \hat{w}}\right)  {\rm Det} [I - P_{\sigma }B_{\hat{w}+ \hat X,\hat{w}- \hat X }  P_{\sigma} ]   \label{D9}
\eea
is the one-time, one-point CDF of the (scaled) height for the
KPZ equation with Brownian initial conditions with (scaled) slope $\hat w$ and
(scaled) position $\hat X$ \cite{SasamotoStationary}. It is the
so-called extended BR distribution, also discussed in (\ref{BR1BR2}).
In the limit $\hat X=0$ and $\hat{w} \to 0$, it becomes the CDF of
the Baik-Rains distribution.
\begin{equation}\label{bsigmabeforelimit}
F_0(\sigma) = \lim_{\hat{w}\to 0, \hat X=0} F_0(\sigma - \hat X^2 ; \hat X, \hat w)
\end{equation}
The derivation of the explicit distribution in this limit is recalled in the Appendix \ref{app:BR}.

\section{Baik-Rains distribution as a limit}
\label{app:BR}

In this Appendix we expand the formula (\ref{D9}) for small $\hat w$.
We recover results of Ref. \cite{SasamotoStationary} for $\hat w=0$ and obtain the
next order. We want to obtain explicit formula for the first two terms in the expansion
\bea \label{lim1}
\fl && ~~~~~ F_0(\sigma - \hat X^2; \hat X;\hat w) = F_0(\sigma - \hat X^2; \hat X)
+ \hat w [ \partial_{\hat w} F_0(\sigma - \hat X^2; \hat X;\hat w)]|_{\hat w=0}  + O(\hat w^2)
\eea
where we recall
\bea
F_0(\sigma - \hat X^2; \hat X;\hat w) =
  \left( 1 + \partial_\sigma \frac{1}{2 \hat{w}}\right)
{\rm Det} [I - P_{\sigma }B_{\hat{w}+ \hat X,\hat{w}- \hat X }   P_{\sigma} ]
\eea

Using the identity (\ref{id3}) we can rewrite
\bea \label{defBB}
B_{\hat{w}+ \hat X,\hat{w}- \hat X }(u_i,u_j) = K_\Ai(u_i,u_j) + 2 \hat w ~ \mathcal{B}_{ \hat{w}+\hat{X}}(u_i)
\mathcal{B}_{ \hat{w}-\hat{X}}(u_j)
\eea
where we recall that
\begin{equation} \label{Bw2}
\mathcal{B}_{\hat{w}}(u) = e^{\hat{w}^3/3 - u \hat{w}} - \int_0^\infty dy e^{\hat{w} y } \Ai(u  + y)
\end{equation}
Since the second term in \eqref{defBB} is a rank one projector, using (\ref{detproj2}), we can rewrite the Fredholm determinant in (\ref{D9}) as
\begin{align}
& {\rm Det} [I - P_{\sigma} B_{\hat{w}+ \hat X,\hat{w}- \hat X }  P_{\sigma} ]   = {\rm Det} [I -  P_{\sigma} K_{\Ai} P_{\sigma}]  \bigg(1 -
2 \hat{w } \int_{\sigma}^\infty du dv dv' \nn \\&  \times  \mathcal{B}_{ \hat{w }+ \hat X}(u) \mathcal{B}_{\hat{w }- \hat X}(v) [I -  P_\sigma K_{\Ai}  P_\sigma +  P_\sigma K_{\Ai}  P_\sigma ](u,v') [I - P_\sigma K_{\Ai}  P_\sigma ]^{-1}(v',v)  \bigg) \nn \\&
= F_2(\sigma) \big(1 - 2 \hat w {\rm \text{Tr}}[ (I - P_\sigma K_{\Ai}  P_\sigma)^{-1} (P_{\sigma} K_{\Ai} P_\sigma \mathcal{B}_{\hat{w}+\hat X})
\mathcal{B}^T_{\hat{w}-\hat X} ] \big) \nn  \\
& -   F_2(\sigma)   ( 2 \hat{w })
\int_\sigma^{+\infty} du \mathcal{B}_{\hat{w}+\hat X} (u) \mathcal{B}_{\hat{w}-\hat X}(u) \label{new1}
\end{align}


We now consider the limit $\hat w \to 0$.
Note that the first part $e^{\tilde{w}^3/3 - u \tilde{w}}$ leads to divergences in the integrals
over $u$ in the second term in (\ref{new1}) in the limit $\hat{w} \to 0$. From \eqref{Bw2} one obtains the following behavior as $\hat w \to 0$
\bea \label{seriesB}
\fl && ~~~~~ 2 \hat w \int_\sigma^{+\infty} du \mathcal{B}_{\hat{w}+\hat X} (u) \mathcal{B}_{\hat{w}-\hat X}(u)
= 1 - 2 \hat w \left(1 + {\cal L}_{\hat X}(\sigma) \right) + (2 \hat w)^2 {\cal M}_{\hat X}(\sigma) + O(\hat w^3)
\eea
where
\begin{align}
& \mathcal{L}_{\hat X}(\sigma) =
 - 1 + \sigma - \hat X^2 + 2 \int_\sigma^\infty du \int_{0}^\infty dy \cosh \big( \frac{\hat X^3}{3} - (u+y) \hat X \big)
   \Ai(u + y) \nn\\
 & - \int_\sigma^\infty du \int_{0}^\infty  dy_1 dy_2  e^{\hat X(y_1-y_2)}  \Ai(u + y_1)  \Ai(u + y_2)
 \label{defLX}
\end{align}
The next order correction is given by
\bea
\fl && \mathcal{M}_{\hat X}(\sigma) =  \frac{1}{2} (\sigma - \hat X^2)^2 +
\int_\sigma^\infty du \int_{0}^\infty dy (u-y)  \cosh \big( \frac{\hat X^3}{3} - (u+y) \hat X \big)
   \Ai(u + y) \\
   \fl && + \int_{\sigma}^{+\infty} du \int_{0}^\infty dy_1 dy_2 y_1 e^{\hat X(y_1-y_2)}  \Ai(y_1+u) \Ai(y_2+u) \nn
\eea
The functions $\mathcal{L}_{\hat X}$ and $\mathcal{M}_{\hat X}$ are the finite $\hat X$ generalization
of the functions $\mathcal{L}=\mathcal{L}_{0}$ and $\mathcal{M}=\mathcal{M}_{0}$ given in the text
in \eqref{funL} and \eqref{funM}.

By contrast, the first term in (\ref{new1}) behaves well in that limit, as the Airy kernel
makes the integrals convergent. We thus obtain
\bea
\fl && {\rm Det} [I - P_{\sigma} B_{\hat{w}+ \hat X,\hat{w}- \hat X }  P_{\sigma} ]
 \\
 \fl && = 2 \hat w F_2(\sigma) \Big( 1+ {\cal L}_{\hat X}(\sigma)
 -  {\rm \text{Tr}}[  (I - P_\sigma K_{\Ai}  P_\sigma)^{-1}  (P_{\sigma} K_{\Ai} P_\sigma \mathcal{B}_{\hat{w}+\hat X})
\mathcal{B}^T_{\hat{w}-\hat X} ]
 -  2 \hat w {\cal M}_{\hat X}(\sigma) \Big) \nn
 \\
 \fl && + O(\hat w^3) \nn
\eea
where a cancellation of the $O(1)$ terms has occured.
Let us apply now the operator $\left( 1 + \partial_\sigma \frac{1}{2 \hat{w}}\right)$
and first focus on the order zero term, i.e. the limit $\hat w \to 0$.
%
%
Then, using again (\ref{detproj2}) one obtains the (extended) Baik-Rains distribution
in terms of the Airy kernel
\begin{align} \label{BaikRainsExplicit}
 F_0(\sigma - \hat X^2;\hat X)
= \partial_{\sigma} \Big[ F_2(\sigma) \mathcal{L}_{\hat X}(\sigma) +  {\rm Det} [I -  P_{\sigma} (K_{\Ai}  +  (K_{\Ai} P_\sigma \mathcal{B}_{\hat X})
\mathcal{B}^T_{- \hat X} ) P_{\sigma} ]    \Big]
\end{align}
together with the two functions $\mathcal{B}_{\hat X}(u)$, defined in (\ref{Bw2}), and $\mathcal{L}_{\hat X}(\sigma)$
is defined in (\ref{defLX}). It generalizes Eq. (\ref{BR1}) to non-zero $\hat X$, in agreement with
\cite{SasamotoStationary} and
provides the one-point PDF of the ${\cal A}_{\rm stat}(\hat X)$ process (see
Appendix \ref{app:airy}).

One can push to the next order in $\hat w$. One finds
\bea
\fl && \partial_{\hat w} F_0(\sigma - \hat X^2; \hat X;\hat w)]|_{\hat w=0}
= 2   \Big( F_2(\sigma) {\cal L}_{\hat X}(\sigma) +  {\rm Det} [I -  P_{\sigma} (K_{\Ai}  +  (K_{\Ai} P_\sigma \mathcal{B}_{\hat X})
\mathcal{B}^T_{- \hat X} ) P_{\sigma} ] \Big) \nn \\
\fl && -    \partial_\sigma [ 2 F_2(\sigma) {\cal M}_{\hat X}(\sigma)
+ {\rm \text{Tr}}[  (I - P_\sigma K_{\Ai}  P_\sigma)^{-1}  (P_{\sigma} K_{\Ai} P_\sigma \partial_{\hat w} [\mathcal{B}_{\hat w+ \hat X})
\mathcal{B}^T_{\hat{w}-\hat X}]_{\hat w=0} ] \label{derBR}
\eea
which is used in the text in (\ref{p1br}) for $\hat X=0$ to compare to the effect of a small $\Delta$.

%

\section{Constrained summations}
\label{app:sums}

In this Appendix we will study constrained summations of the type
\begin{equation} \label{summa1}
F_n \equiv  \sum_{n_s=0}^\infty \frac{1}{n_s!} \prod_{j=1}^{n_s} \int du_j \sum_{m_j = 1}^{\infty}
\delta_{n,\sum_{j=1}^{n_s} m_j}  \det_{i,j=1}^{n_s} K(u_i,u_j;m_j)
\end{equation}
for kernels $K$ with some properties discussed below. In the absence of constraint, i.e.
without the factor $\delta_{n,\sum_{j=1}^{n_s} m_j}$, the
sum is simply a Fredholm determinant ${\rm Det}[I + \tilde K]$ with $\tilde K(u_i,u_j)=\sum_{m \geq 1}
K(u_i,u_j;m)$. For illustration, we will consider kernels of the form
\bea \label{kernG}
K(u_i,u_j;m) =\theta(u_i) \theta(u_j)  \int dy (-1)^m e^{\lambda m y} A(u_i,u_j;y)
\eea
so that the summation over the $m$ variable is of geometric type and
\bea \label{kernG2}
\tilde K(u_i,u_j)= - \theta(u_i) \theta(u_j)  \int dy \frac{1}{1 + e^{- \lambda y}} A(u_i,u_j;y)
\eea
but the method is more general.

Such constrained summations naturally occur in RBA calculations, where $m_j$ are string length variables
and $n$ the total number of replica, i.e. of particles in the associated LL problem. In the DP/KPZ problem
$n$ is the order of some moment of the partition sum/exponential of height field, averaged over the disorder/noise, respectively.
Two examples which we detail are:

(i) the single time DP/KPZ problem with point to point/droplet initial condition, studied e.g. in \cite{we}.
There $(-1)^n F_n$ is directly proportional to the moment of the partition sum $Z(t)=e^{h(0,t)}=Z_\eta(0,t|0,0)$
and it has the form (\ref{summa1}) and (\ref{kernG}) with
\bea
\fl && (-1)^n F_n = \frac{1}{n!} \overline{Z^n} \quad , \quad A(u_i,u_j;y) =
2^{-\frac{1}{3}} Ai(2^{\frac{1}{3}} u_i + 2^{-\frac{2}{3}} y)
Ai(2^{\frac{1}{3}} u_j + 2^{-\frac{2}{3}} y) \label{AA}
\eea
where $\lambda=(t/4)^{1/3}$.

(ii) for the present two-time problem, in Section (\ref{sec:calc}) we have encountered a sum over string lengths $\{ m_j \}_{j=1}^{n_s}$ constrained to have a fixed total sum $\sum_j m_j  = n_2$. There $n_2$ is the number of
particles/replica in the time interval $[t_1,t_2]$ and is associated to a moment of order $n_2$ of $Z_2$.
The associated $F_n$ has the form (\ref{summa1}), although it now involves additional external integration/sum
factors (e.g. over $q,n_1,n_2$), and the kernel has the form (\ref{kernG}), see e.g. Eq.(\ref{defKnnqz4}) ($K_{n_1,n_2,q,z}(u_i,u_j)$ there is the
kernel $\tilde K_z$ defined below in \eqref{Kzgeom}, with $\lambda \to \lambda \Delta^{1/3}$).

The general method is to rewrite the constrain as a countour integral around the pole $z = 0$ using
the identity
\begin{equation}
\delta_{n,\sum_{m_j}}  = \oint \frac{dz}{2 \pi i z} z^{n - \sum_j m_j}
\end{equation}
This allows to decouple the sums and one can then perform the sum over $m_j$ leading to the countour integral of a Fredholm determinant
\begin{equation} \label{Fnz}
F_n = \oint \frac{dz}{2 \pi i} z^{n - 1} {\rm Det}[ 1 +  \tilde{K}_z  ]
\end{equation}
where the Kernel $\tilde{K}_z$ is defined as a sum over $m$
\begin{equation}
\tilde{K}_z(u_i,u_j) = \sum_{m=1}^\infty z^{-m} K(u_i,u_j; m)
\end{equation}
so that $\tilde{K}_{z=1}=\tilde K$.

While the method is more general, for kernels of the form (\ref{kernG}) explicit formula can be obtained. Performing the geometric sum one obtains
\bea \label{Kzgeom}
\tilde K_z(u_i,u_j)= - \theta(u_i) \theta(u_j)  \int dy \frac{1}{1 + z e^{- \lambda y}} A(u_i,u_j;y)
\eea
so that the $z$-dependence is explicit. In particular the factor $\frac{1}{1 + z e^{- \lambda y}}$
has a pole for $z=- e^{\lambda y}$ on the negative real axis, which, after integration over $y$
results in a branch cut for $\tilde K_z$ along the real negative axis.
Using $\frac{1}{X- i 0^+} = PV \frac{1}{X} + i \pi \delta(X)$ we have, for $x \in \mathbb{R}$
\bea
\frac{1}{1 + z e^{- \lambda y}}|_{z=x\pm i 0^+} = PV \frac{1}{x + e^{-\lambda y}} \mp i \pi \delta(1+ x e^{-\lambda y})
\eea
hence we can write explicitly the Kernel just above and below the branch cut as
\bea
&& \tilde K_z|_{z=x\pm i 0^+} = \tilde B_x \pm i \pi \frac{1}{\lambda} \tilde P_x
\eea
in terms of two new real kernels
\bea
&& \tilde B_x(u_i,u_j) = - \theta(u_i) \theta(u_j) PV \int dy \frac{1}{1 + x e^{- \lambda y}} A(u_i,u_j;y) \\
&& \tilde P_x(u_i,u_j) =  \lambda \theta(u_i) \theta(u_j) \int dy ~ \delta(1 + x e^{- \lambda y}) A(u_i,u_j;y)
\eea
where we have inserted the factor $\lambda$ so that $\tilde P_x$ has a well defined large $\lambda$ limit
(see below).

To perform the $z$ contour integral in (\ref{Fnz}) we will choose the contour for the $z$ integration to enclose the negative real axis, infinitesimally close from above and below.
\footnote{we can either choose it from the start, or start from a large contour encircling zero, and assume that the only singularities of ${\rm Det} [ 1 +  \tilde{K}_z  ]$ arise from those of the factor $\frac{1} {1 + z e^{- \lambda y}} $, i.e. is a branch cut along the real negative axis.}
Hence we write, for small $\epsilon>0$
\begin{equation}
\oint  \frac{dz}{2 \pi i} f(z) = \frac{1}{2 \pi i } \left(\int_{-\infty}^{+\epsilon}  dx f(x - i \epsilon ) -
\int^{+\epsilon}_{- \infty} dx  f(x +  i \epsilon ) \right) + O(\epsilon)
\end{equation}
where $f(z) = z^{n - 1} {\rm Det}[ 1 +  \tilde{K}_z  ]$. Hence the final result involves the
difference of two FD
\bea
\fl && F_n =  (-1)^n \frac{1}{2 \pi i} \int_{-\infty}^{0}  dx (-x)^{n - 1} ( {\rm Det}[ 1 +  \tilde B_x
+ i \pi \frac{1}{\lambda} \tilde P_x   ] -
 {\rm Det}[ 1 +  \tilde B_x - i \pi \frac{1}{\lambda} \tilde P_x   ] )
\eea

Now we further specialize to the case where
\bea
A(u_i,u_j;y) = A(u_i,y) \bar A(u_j,y)
\eea
which is the case for both examples (i) and (ii) above, $A$ being Airy, or deformed Airy functions, respectively. It is also convenient
to choose the following parameterization for the real variable $x$
\bea
x= - e^{-\lambda f'}  = - e^{\lambda h'}
\eea
and denote $B_{f'}=\tilde B_x$ and $P_{f'}=\tilde P_x$.
In that case we have
\bea \label{appB}
\fl && ~~~~~~ B_{f'}(u_i,u_j)  =  PV \int dy A(u_i,y) \bar A(u_j,y)   \frac{1}{e^{-\lambda(y + f') } -1 }  \theta(u_i) \theta(u_j) \\
\fl && ~~~~~~ P_{f'}(u_i,u_j)   
= A(u_i,-f') \bar A(u_j,-f') \theta(u_i) \theta(u_j) \label{appP}
\eea
Since $P_{f'}=|U \rangle \langle V|$ is a rank one projector, we can use the matrix determinant
lemma (\ref{detproj2}) which implies
that the dependence in $P_f$ is only linear, hence we obtain the final result
\bea \label{appres1}
(-1)^n F_n =  \int_{-\infty}^{+\infty} df' e^{- \lambda n f'}
( Det[1+ B_{f'} + P_{f'}]- Det[1+ B_{f'}] )
\eea

It is interesting to note that:
\bea \label{kinfty}
\!\!\!\!\! \lim_{\lambda \to + \infty} B_{f'}(u_i,u_j)  = B_{f',\infty}(u_i,u_j)  = - \int dy A(u_i,y) \bar A(u_j,y)  \theta(y+f')
\eea
and we see that $P_{f'} = - \partial_{f'}  B_{f',\infty}$. Hence
in the large time limit
\bea \label{appres2}
\lim_{\lambda \to + \infty} (-1)^n F_n =  \int_{-\infty}^{+\infty} df' e^{- \lambda n f'} (- \partial_{f'}) Det[1+ B_{f',\infty}]
\eea

Let us now discuss the applications (i) and (ii).
For (i) one can read off from (\ref{appres1}) the one time PDF of the scaled free energy/height field for the
point to point DP/droplet KPZ at any finite time. Indeed since
$\overline{Z^n} = n! e^{\lambda n s} (-1)^n F_n$ for integer $n$, (\ref{appres1})
provides an analytic continuation of the moments $\overline{Z^n}$ for arbitrary complex n.
Furthermore it is equivalent to state that:
\bea
\ln Z = - G - \lambda f'
\eea
where $G$ is an independent unit Gumbel variable (i.e. of PDF $p(G)=e^{-G-e^{-G}}$),
and $f'$ is a random variable of PDF
\bea
P(f') = Det[1+ B_{f'} + P_{f'}]- Det[1+ B_{f'}]
\eea
where $B_{f'}$ and $P_{f'}$ are given by equations (\ref{appB}) and (\ref{appP})
with $A(u_i,y) \bar A(u_j,y)=A(u_i,u_j;y)$ given by (\ref{AA}). One recovers
exactly the result obtained in \cite{we} ($u$ there being $-f'$ here), by a completely
different method. In the large time limit one easily recovers the GUE-TW distribution
\bea
&& Det[1+ B_{f',\infty}] = F_2(- 2^{-2/3} f') \quad , \quad P(f') = 2^{-2/3} f_2(- 2^{-2/3} f')
\eea

The application (ii) to Section (\ref{sec:calc}) is now very similar, with $A(u_i,y)$ and $\bar A(u_j,y)$
given by deformed Airy functions. This leads to transform formula (\ref{startDet} )
into (\ref{Det2}) in the text, where $n=n_2$ and we use the notation $\kappa=-f'$,
with additional indices for the kernels.

\subsection{A simple example}\label{Section_constraint_simple}

We consider here a simple example to illustrate the method. Consider the following sum
\begin{equation}
S_\lambda= \sum_{n_1=0}^\infty \sum_{n_2=1}^\infty  (-1)^{n_1 + n_2} \frac{e^{ \lambda  y (n_1 + n_2)}}{\lambda(n_1+ n_2)} e^{ - \lambda s_1 n_1 -\lambda  s_2 n_2}
\end{equation}
Rewriting $\frac{1}{\lambda(n_1+ n_2)}=\int_0^{+\infty} e^{- u \lambda(n_1+ n_2)}$ the sums are decoupled and
can be performed exactly
\bea \label{initialsumsimple}
S_\lambda = - \int_0^\infty du \theta_\lambda(y - s_2 - u) \theta_\lambda(u+ s_1 - y)
\eea
where $\theta_\lambda(x)=1/(1+e^{-\lambda x})$. The limit $\lambda \to +\infty$ is then immediate
using that $\theta_{+\infty}(x)=\theta(x)$, the Heaviside step function. The same result can be
obtained using the Mellin-Barnes trick.

On the other hand let us now consider the sum written as a constrained sum
\bea\label{constrained_simple}
\fl && ~~~~~~ S_\lambda=\sum_{n_1=0}^\infty \sum_{n_2=1}^\infty  (-1)^{n_1} {e^{ \lambda  y n_1}}{ } e^{ - \lambda s_1 n_1  } \sum_{m=1}^\infty (-1)^m  \delta_{n_2,m} e^{- \lambda s_2 m + y \lambda m  }  \frac{1}{\lambda (n_1+ m)}
\eea
Although in this simple case we know the explicit result, we can pretend to not be able to resolve the constraint $\delta_{n_2,m}$ and therefore to choose a different trick to perform the sum.
Let us now use the contour integral representation of the Kronecker delta. This gives
\bea
\fl && S_\lambda= \sum_{n_1=0}^\infty \sum_{n_2=1}^\infty  (-1)^{n_1}  {e^{ \lambda  y n_1}}{}  e^{ - \lambda s_1 n_1  } \int_0^\infty du \sum_{m=1}^\infty (-1)^m \oint \frac{dz}{2 \pi i z} z^{n_2 - m } e^{- \lambda s_2 m + y \lambda m  - u \lambda (n_1 + m) } \nn \\
\fl && ~~~ =  \sum_{n_1=0}^\infty \sum_{n_2=1}^\infty  (-1)^{n_1 }  {e^{ \lambda  y n_1}}{ } e^{ - \lambda s_1 n_1  }  \int_0^\infty du \oint \frac{dz}{2 \pi i z} z^{n_2  } \frac{- e^{- \lambda u n_1}}{e^{-\lambda (y - s_2 - u ) } z + 1}
\eea
Now, to perform the integration we choose the countour $z$ encircling the negative axis
 \begin{equation}
\oint  \frac{dz}{2 \pi i} f(z) = \frac{1}{2 \pi i } \left(\oint_{-\infty}^{+\epsilon}  f(z - i \epsilon ) +   \oint_{+\epsilon}^{- \infty}  f(z +  i \epsilon ) \right)
\end{equation}
and we change variable $z = -e^{-\lambda f } \pm i \epsilon$, therefore we have
\begin{align}
= \sum_{n_1=0}^\infty & \sum_{n_2=1}^\infty  (-1)^{n_1 }  {e^{ \lambda  y n_1}}{ } e^{ - \lambda s_1 n_1  }  \lambda  \nn \\& \times  \int_0^\infty du \int_{-\infty}^{\infty}  \frac{df}{2 \pi i } (-1)^{n_2} e^{- \lambda f n_2  } \left( \frac{e^{\lambda (y - s_2 - u ) } e^{- \lambda u n_1}}{-1 + e^{\lambda (f + y - s_2 - u ) }  - i \epsilon   }- \frac{e^{\lambda (f + y - s_2 - u ) } e^{- \lambda u n_1}}{-1 + e^{\lambda (f + y - s_2 - u ) }  + i \epsilon   } \right) \nn \\
& = \sum_{n_1=0}^\infty \sum_{n_2=1}^\infty  (-1)^{n_1 }  {e^{ \lambda  y n_1}}{ } e^{ - \lambda s_1 n_1  }      \int_0^\infty du   e^{- u \lambda n_1}\int_{-\infty}^{\infty}  {df}{ }  e^{ - \lambda f n_2} \delta(f + y - s_2 - u ) \nn \\&
=  - \int_0^\infty du \frac{1}{1 + e^{\lambda(- y +u + s_2)}} \frac{1}{1 + e^{- \lambda (s_1 + u - y )}}
\end{align}
which gives the correct result \eqref{initialsumsimple}.
Hence we have verified that the procedure to compute a constrained summation as the one in \eqref{constrained_simple} leads to a correct result, at least in the case where one single constrained sum has to be performed.

\section{Useful identities}  \label{sec:useful}

\subsection{Determinant identities}

We recall two identities which are known for matrices but we assume extend
readily to operators. The so-called matrix determinant lemma
\bea \label{detproj2}
{\rm Det}( A + |U \rangle\langle V| ) = (1 + \langle V | A^{-1} | U \rangle) \, {\rm Det} A \,,
\eea
for any invertible operator $A$, and rank one projector $|U \rangle\langle V|$ (in quantum mechanics notation)
and the so-called Sherman-Morrison formula
\bea \label{form2}
( A + |U \rangle \langle V| )^{-1} = A^{-1} - \frac{A^{-1} |U \rangle\langle V| A^{-1} }{1 + \langle V | A^{-1} | U \rangle} .
\eea

\subsection{Derivatives of GUE-TW distribution}

The CDF of the TW-GUE, $F_2(\sigma)$, is written as a Fredholm determinant
\bea
F_2(\sigma)  = {\rm  Det }[ I- P_\sigma K_{\Ai} P_\sigma ] = {\rm  Det }[ I- P_0 K_{\Ai,\sigma} P_0]
\eea
where in the second form all integrations are on $[0,+\infty[$,
and we denote $K_{\Ai,\sigma}(u,v)\equiv K_\Ai(u+\sigma,v+\sigma)$. Using that
\bea
\partial_\sigma K_{\Ai,\sigma}(u,v) = - \Ai(u) \Ai(v)
\eea
Taking a derivative w.r.t. $\sigma$, one finds that
\bea  \label{derGUE}
\fl F_2'(\sigma) &=&  -  F_2(\sigma)  \text{Tr}[ P_0 \partial_{\sigma} K_{\Ai,\sigma} P_0
(1-P_{0} K_{\Ai,\sigma}P_{0} )^{-1} ] \\
\fl &=& F_2(\sigma)  \text{Tr}[(1-P_{0} K_{\Ai,\sigma}P_{0} )^{-1} \Ai_\sigma \Ai_\sigma^T ]  = {\rm  Det }[ I- P_\sigma (K_{\Ai} - \Ai \Ai^T) P_\sigma ]-F_2(\sigma) \nn
\eea
where we denote $\Ai_\sigma(u)=Ai(u+\sigma)$ and
we have used similar notation as in (\ref{b0}) to denote rank one projectors. On the second form
we can take another derivative
\bea \label{der2GUE}
\fl && F_2''(\sigma)  = F_2(\sigma) \text{Tr}[(1-P_{0} K_{\Ai,\sigma}P_{0} )^{-1} (\Ai'_\sigma \Ai_\sigma^T
+ \Ai_\sigma \Ai_\sigma^{\prime T})] \\
\fl && + F_2(\sigma) \text{Tr} \left[ (\partial_\sigma (1-P_{0} K_{\Ai,\sigma}P_{0} )^{-1}) \Ai_\sigma \Ai_\sigma^T \right]
+ F'_2(\sigma)  \text{Tr}[(1-P_{0} K_{\Ai,\sigma}P_{0} )^{-1} \Ai_\sigma \Ai_\sigma^T ] \nn
\eea
Now the last two terms exactly cancel using the second form in (\ref{derGUE}) for $F_2'(\sigma)$
as well as the identity
\bea
&& \text{Tr}\left[ (\partial_\sigma (1-P_{0} K_{\Ai,\sigma}P_{0} )^{-1}) A B^T \right] \\
&& = -
\text{Tr}[ (1-P_{0} K_{\Ai,\sigma}P_{0} )^{-1} A \Ai_\sigma^T ] \text{Tr}[ (1-P_{0} K_{\Ai,\sigma}P_{0} )^{-1} \Ai_\sigma B^T ] \nn
\eea
where $A B^T(u,v)=A(u) B(v)$ an arbitrary rank one projector. One thus obtains
\bea
\fl ~~~ F_2''(\sigma)  &=& 2 F_2(\sigma) \text{Tr}[(1-P_{0} K_{\Ai,\sigma}P_{0} )^{-1} \Ai'_\sigma \Ai_\sigma^T] \\
\fl ~~~ &=& 2 F_2(\sigma)  \text{Tr}[(1-P_{\sigma} K_{\Ai}P_{\sigma} )^{-1} \Ai' \Ai^T]
= 2 ( {\rm Det}[ 1 - P_{\sigma} (K_{\Ai}- \Ai' \Ai^T ) P_{\sigma} ] - F_2(\sigma) ) \nn
\eea
Interestingly, there is a similar form for the third derivative, although more involved, which
we do not use here.

\subsection{Asymptotics of Airy function and kernel}

Let us recall the asymptotics of the Airy function for large positive argument
\bea
&& Ai(y) = f_{Ai}(y) e^{- \frac{2}{3} y^{3/2}} \quad , \quad  f_{Ai}(y) = \frac{1}{2 \sqrt{\pi} y^{1/4}} \sum_{k=0}^\infty \frac{(-1)^k}{(\frac{2}{3} y^{3/2})^k} Ai_k \,,\\
&& Ai_k = 2^{-k} \frac{\Gamma(k+\frac{5}{6}) \Gamma(k+\frac{1}{6})}{2 \pi k!} \quad , \quad
Ai_k = \frac{(6k-1)(6 k-5)}{72 k} Ai_{k-1} \quad , \quad Ai_0=1 \nn \,\\&&
\end{eqnarray}
i.e. $Ai_0=1$, $Ai_1=5/72$. Using this one obtains the asymptotics of various functions needed in the text,
such as
\bea
\fl && K_\Ai(x,x)=\Ai'(x)^2 - x \Ai(x)^2 = e^{-\frac{4 x^{3/2}}{3}} \left(\frac{1}{8 \pi  x}-\frac{17}{192 \pi x^{5/2}
   }+O(x^{-7/2}) \right) \\
\fl && \int_{x}^{+\infty} dy K_\Ai(y,y) = \frac{1}{3} \left(2 x^2 \text{Ai}(x)^2-2
   x \text{Ai}'(x)^2-\text{Ai}(x)
   \text{Ai}'(x)\right)  \\
\fl    && = e^{-\frac{4 x^{3/2}}{3}}
   \left(\frac{1}{16 \pi  x^{3/2}}-\frac{35}{384 \pi
   x^3}+O(x^{-7/2})\right)
   \quad , \quad   \int_{x}^\infty dy \Ai(y ) \simeq \frac{e^{-\frac{2 x^{3/2}}{3}}}{2
   \sqrt{\pi } x^{3/4}}
\eea

\section{Higher orders in the large \texorpdfstring{$\Delta$}{Delta} expansion}
\label{app:higher}
Here we give some further results about the large $\Delta$ expansion. The function defined
in the text in (\ref{Pexp}) reads
\bea
\fl && C_1(\sigma_1,\sigma)= F_2''''(\sigma)  \int_{0}^{+\infty}  dy_1 dy_2 y_1 K_{\Ai}(y_1+\sigma_1, y_2+\sigma_1) \nn \\
\fl && - \left[\int_{0}^\infty dy y \Ai(y+\sigma_1 )\right]^2
\partial_\sigma \left( {\rm  Det } [ I- P_\sigma (K_{\Ai} - \Ai' {\Ai'}^T) P_\sigma ] - F_2(\sigma)\right) \\
\fl && - \left[\int_{0}^\infty dy y^2 \Ai(y+\sigma_1 )\right] \left[\int_{\sigma_1}^\infty dy \Ai(y )\right]
\partial_\sigma \left( {\rm  Det } [ I - P_\sigma (K_{\Ai} - \Ai'' \Ai^T) P_\sigma ] - F_2(\sigma)\right) \nn
\eea
Defining the moments
\bea
&& a_p=\int d\sigma \sigma^p \partial_\sigma \left( {\rm  Det } [ I- P_\sigma (K_{\Ai} - \Ai' {\Ai'}^T) P_\sigma ] - F_2(\sigma)\right) \\
&& b_p =  \int d\sigma \sigma^p
\partial_\sigma \left( {\rm  Det } [ I - P_\sigma (K_{\Ai} - \Ai'' \Ai^T) P_\sigma ] - F_2(\sigma)\right)
\eea
One thus obtains the expansion \eqref{condmomexp} for the conditional moments,
with the functions
\bea
\fl && R_{11}(\sigma_1)= \frac{ \int_{0}^{+\infty}  dy_1 dy_2 y_1 K_{\Ai}(y_1+\sigma_1, y_2+\sigma_1) }
{K_{\Ai}(\sigma_1,\sigma_1)} \quad  , ~
R_{12}(\sigma_1)= \frac{ \left[\int_{0}^\infty dy y \Ai(y+\sigma_1 )\right]^2 }{
K_{\Ai}(\sigma_1,\sigma_1)} \nn \\
\fl && R_{13}(\sigma_1)= \frac{ \left[\int_{0}^\infty dy y^2 \Ai(y+\sigma_1 )\right] \left[\int_{\sigma_1}^\infty dy \Ai(y )\right]}{K_{\Ai}(\sigma_1,\sigma_1)}
\eea
The coefficients $a_p,b_p$ (which obey $a_0=b_0=1$) can be
evaluated numerically. The three lowest ones are
\bea
&& a_p = \{-2.36145, 6.58327, -21.4239\} \quad , \quad p = 1,2,3  \\
&& b_p = \{-1.18072, 1.31663, -0.999958\} \quad , \quad p = 1,2,3
\eea

This allows to push by one order the results for the first three cumulants.
We obtain the mean as
\bea \label{mean1n}
\! \! \! \! \!   \langle \sigma \rangle_{\sigma_1}
&=& -1.77109 +  \frac{1}{\Delta^{1/3}}  R_{1/3}(\sigma_1) \\
&+ & \frac{1}{\Delta}  ( 2.36145 R_{11}(\sigma_1)
+ 1.18072 R_{12}(\sigma_1)) + O(\frac{1}{\Delta^{4/3}},e^{- \frac{4}{3} \sigma_1^{3/2}}) \nn
\eea
and the variance as
\bea
\label{var1more}
\! \! \! \! \!  \langle \sigma^2 \rangle^c_{\sigma_1} &=& 0.813195
- \frac{1}{\Delta^{2/3}}  ( R_{1/3}(\sigma_1)^2 - 2 R_{2/3}(\sigma_1) ) \nn \\
&+&  \frac{1}{\Delta} \big(  1.7814 R_{11}(\sigma_1) +
2.86569 R_{12}(\sigma_1))  \big)
+ O(\frac{1}{\Delta^{4/3}},e^{- \frac{4}{3} \sigma_1^{3/2}})
\eea
Finally, we obtain the expansion of the third cumulant as
\bea
\label{3cum}
\fl  \langle \sigma^3 \rangle^c_{\sigma_1} &=& 0.164325
+  \frac{1}{\Delta} \big(  - 6 R_{10}(\sigma_1) + 2.90614 R_{11}(\sigma_1)
+ 2.23476 R_{12}(\sigma_1) \nn \\
\fl &+& 2 R_{1/3}(\sigma_1)^3 - 6 R_{1/3}(\sigma_1) R_{2/3}(\sigma_1)
\big)
+ O(\frac{1}{\Delta^{4/3}},e^{- \frac{4}{3} \sigma_1^{3/2}})
\eea

\medskip

Finally, we give for completeness the function $C_{4/3}(\sigma_1,\sigma)$
which allows to obtain the next order $O(\Delta^{-4/3})$ as
\bea
\fl && C_{4/3}(\sigma_1,\sigma)= \frac{1}{3}\left( \int_{\sigma_1}^\infty dy y^3 \Ai(y)   \int_{\sigma_1}^\infty dy \Ai(y)   \right) \partial_{\sigma} \left((\det [1- P_\sigma K_{\Ai} P_\sigma + P_\sigma \Ai'''\Ai^T  P_\sigma ] -  F_2(\sigma) \right)\nn \\
\fl &&
  +   \left( \int_{0}^\infty dy y^2 \Ai(y+\sigma_1)   \int_{0}^\infty dy y \Ai(y+ \sigma_1)   \right) \partial_{\sigma} \left((\det [1- P_\sigma K_{\Ai} P_\sigma + P_\sigma \Ai''\Ai'^T  P_\sigma ] -  F_2(\sigma) \right) \nn \\
  \fl &&
  -   \Big[ \left(\int_{0} dy_1 dy_2 K_{\Ai}(y_1 + \sigma_1 , y_2 + \sigma_1) y_1 y_2 \right)\partial^2_\sigma \left(\det [1- P_\sigma K_{\Ai} P_\sigma + P_\sigma \Ai' \Ai'^T  P_\sigma ]- F_2(\sigma) \right)\nn  \\
  \fl &&  +  \left( \int_{0} dy_1 dy_2 K_{\Ai}(y_1 + \sigma_1 , y_2 + \sigma_1) y_1^2   \right) \partial^2_\sigma \left(\det [1- P_\sigma K_{\Ai} P_\sigma + P_\sigma \Ai''\Ai^T  P_\sigma ] -  F_2(\sigma) \right) \Big]
\eea

\section{A sum rule for the JPDF}\label{sumrule_appendix}

We here show that
\begin{equation} \label{sr}
\int_{-\infty}^{+\infty} d\sigma P_{\Delta}^{(1)}(\sigma_1, \sigma ) = F_2^{(1)}{}'(\sigma_1)
\end{equation}

We use the definition of the JPDF $P_{\Delta}^{(1)}$
\begin{equation}
P_{\Delta}^{(1)}(\sigma_1, \sigma )  = \partial_{\sigma_1} \partial_{\sigma}\hat{g}^{(1)}_{\Delta }(s_1,s) - \frac{1}{\Delta^{1/3}} \partial_{\sigma}^2  \hat{g}^{(1)}_{\Delta }(s_1,s)
\end{equation}
where $ \hat{g}^{(1)}_{\Delta }$ is given in equation \eqref{final0}
\begin{align}
  \hat{g}^{(1)}_{\Delta }(s_1,s)   & = 1  -  F^{(1)}_2(\sigma_1)  - F_2(\sigma)
  \text{Tr}[ P_{\sigma_1} K_\Ai ]
 + \Delta^{1/3}    {F_2(\sigma)}  \text{\text{Tr}} \left[ P_\sigma K^{\Delta}_{\sigma_1}   P_\sigma   (I-P_{\sigma} K_{\Ai}P_{\sigma} )^{-1}   \right]
\end{align}

Therefore we have
\begin{align}
& \int_{-\infty}^{+\infty} d\sigma \ P_{\Delta}^{(1)}(\sigma_1, \sigma )  \nn \\& = \partial_{\sigma_1} \left(   \hat{g}^{(1)}_{\Delta }(s_1, + \infty)  -   \hat{g}^{(1)}_{\Delta }(s_1,-\infty)  \right) - \frac{1}{\Delta^{1/3}} \left( \partial_{\sigma}\hat{g}^{(1)}_{\Delta }(s_1,s)|_{\sigma = +\infty}  -  \partial_{\sigma}\hat{g}^{(1)}_{\Delta }(s_1,s)|_{\sigma =-\infty}\right)
\end{align}
Now using that
\be
 \lim_{\sigma \to +\infty} F_2(\sigma)  =  1\quad , \quad \lim_{\sigma \to -\infty} F_2(\sigma)  = 0
\ee
and
\be
 \lim_{\sigma \to \pm \infty} F_2'(\sigma) =  0 \quad , \quad
 \lim_{\sigma \to +\infty}\text{\text{Tr}} \left[ P_\sigma K^{\Delta}_{\sigma_1}   P_\sigma   (I-P_{\sigma} K_{\Ai}P_{\sigma} )^{-1}   \right]  = 0
\ee
we conclude that
\begin{equation}
\left( \partial_{\sigma}\hat{g}^{(1)}_{\Delta }(s_1,s)|_{\sigma = +\infty}  -  \partial_{\sigma}\hat{g}^{(1)}_{\Delta }(s_1,s)|_{\sigma =-\infty}\right) = 0
\end{equation}
and
\begin{equation}
\partial_{\sigma_1} \left(   \hat{g}^{(1)}_{\Delta }(s_1, + \infty)  -   \hat{g}^{(1)}_{\Delta }(s_1,-\infty)  \right)  = - \partial_{\sigma_1}  \text{Tr}[ P_{\sigma_1} K_\Ai ] = F_2^{(1)}{}'(\sigma_1)
\end{equation}
which shows (\ref{sr}).

\section{Numerical details}\label{app:numerics}

\begin{figure}[!ht]
\centering
\includegraphics[scale=1.3]{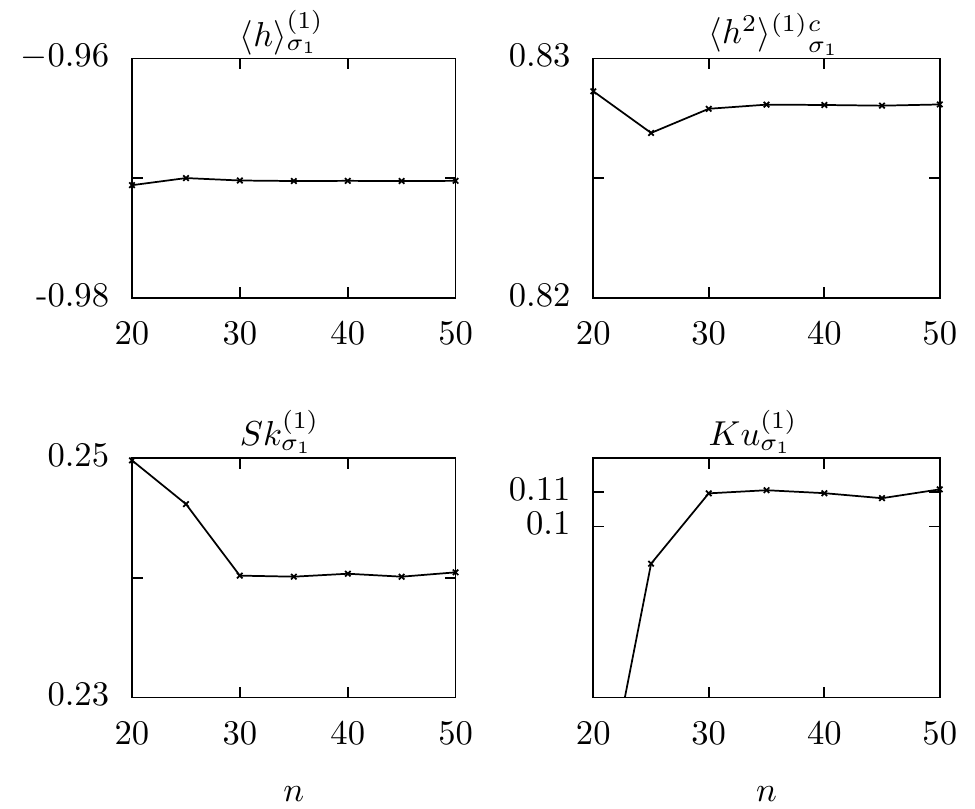}
\caption{Plot of the convergence of the lowest cumulants of the conditional probability distribution $P_{\Delta}^{(1)}(\sigma| \sigma_1)$, defined in \eqref{condcumint}, at fixed $\sigma_{1} = 1$ and $\Delta =1$ as a function of the number $n$  of points used for the discretization of the kernels of equation \eqref{jpdf0}.   }
\label{fig:convergence}
\end{figure}

For the numerical evaluation of equation \eqref{jpdf0} we chose a Gaussian quadrature discretization method to evaluate the inverse of the matrix $I- P_\sigma K_{\Ai} P_\sigma$ as well as the final trace and we numerically perform the derivatives necessary to compute the probability density function. This procedure is very slowly converging, as compared to the calculation of the usual GUE distribution (see \cite{Bornemann}), due to the definition of the kernel $K_{\sigma_1}^{\Delta}$ in equation \eqref{K4def0} which involves integration of the Airy functions on the negative axis, where they oscillate. On the other hand, by using two different representation of the kernel \eqref{K4def0} (equation \eqref{K4alt}), we obtain good convergence of the first cumulants already when the number of points $n$ in the discretization is around $n \sim 50$ (See Fig. \ref{fig:convergence}).

\end{appendices}

\newpage

\section*{References}

\bibliographystyle{iopart-num}
\bibliography{two_times_references}

\end{document}